\documentstyle[12pt]{article}
\def\beq{\begin{equation}}   \def\eeq{\end{equation}}
 
\newcommand{\gsim}{\lower.7ex\hbox{$
\;\stackrel{\textstyle>}{\sim}\;$}}
\newcommand{\lsim}{\lower.7ex\hbox{$
\;\stackrel{\textstyle<}{\sim}\;$}}

\newcommand{\ra}{\rightarrow}

\newcommand{\GeV}{\,\mbox{GeV}}
\newcommand{\MeV}{\,\mbox{MeV}}

\newcommand{\matel}[3]{\langle #1|#2|#3\rangle}

\begin{document}

\def\lsim{\mathrel{\rlap{\lower3pt\hbox{\hskip0pt$\sim$}}
    \raise1pt\hbox{$<$}}}         %less than or approx. symbol
\def\gsim{\mathrel{\rlap{\lower4pt\hbox{\hskip1pt$\sim$}}
    \raise1pt\hbox{$>$}}}         %greater than or approx. symbol

\begin{titlepage}
\renewcommand{\thefootnote}{\fnsymbol{footnote}}

\begin{flushright}
UND-HEP-97-BIG\hspace*{.2em}09\\
%\today\\
%hep-ph/9712475\\
%Version 2.0 \\ 
\end{flushright}
\vspace{.3cm}
\begin{center} \Large  
{\bf CP Violation -- An Essential Mystery in Nature's Grand Design}  
\footnote{Invited Lectures given at the XXV ITEP Winterschool 
of Physics, Feb. 18 - 27, 1997, Moscow, Russia, at 
`Frontiers in Contemporary Physics', May 11 - 16, 1997, Vanderbilt 
University, Nashville, U.S.A., and at the International School of 
Physics "Enrico Fermi", CXXXVII Course `Heavy Flavour Physics: 
A Probe of Nature's Grand Design', Varenna, Italy, July 8 - 18.}  
\end{center}
\vspace*{.3cm}
\begin{center} {\Large
I. Bigi \\ 
\vspace{.4cm}
{\normalsize 
{\it Dept.of Physics,
Univ. of Notre Dame du
Lac, Notre Dame, IN 46556, U.S.A.}\\
\vspace{.3cm}
e-mail address:\\
{\it bigi@undhep.hep.nd.edu} }}
\vspace*{.4cm}

{\Large{\bf Abstract}}\\
\end{center}
CP violation has so far been observed in one system only, namely in 
the decays of neutral kaons, and it can still be described by a single 
real quantity corresponding to a superweak scenario. 
In these lectures I describe why limitations on CP invariance are 
a particularly fundamental phenomenon and what experimental 
information is available. The KM ansatz constitutes the minimal implementation of CP violation: without requiring  
unknown degrees of freedom it can reproduce the known 
CP phenomenology. It unequivocally predicts large or even huge 
CP asymmetries of various kinds in the decays of beauty hadrons. 
New theoretical technologies will enable us in the foreseeable 
future to express at least some of these predictions in a 
quantitatively reliable fashion. There is tremendous potential 
for discovering New Physics in beauty transitions. Continuing efforts 
in strange decays and further dedicated searches for electric dipole moments and for CP asymmetries in charm decays are likewise essential for discovering crucial 
elements that still are missing in the puzzle that is  
Nature's Grand Design. 

\vspace*{.2cm}
\vfill
\noindent
\end{titlepage}
\addtocounter{footnote}{-1}

\newpage
\begin{quotation} 
\noindent 
Schl\" aft ein Lied in allen Dingen, \\ 
die da tr\" aumen fort und fort, \\
und die Welt hebt an zu singen, \\
findst Du nur das Zauberwort.\\

\bigskip 

\noindent 
Sleeps a song in all things \\ 
that dream on and on \\ 
and the world will start to sing \\ 
if you find the magic word. \\ 
\begin{flushright}
{\em J. v. Eichendorff} 
\end{flushright}
\end{quotation}

%%%%%%%%%%
\tableofcontents
%%%%%%%%%%%
%%%%%%%%%%%%%%%%%%%%%%%%%%
\section{Introduction}
%%%%%%%%%%%%%%%%%%%%%%%%%% 
Very few symmetries in nature are manifestly realized. Why do I 
think then that the breaking of CP invariance is very special -- 
more subtle, more fundamental and more profound than parity 
violation? 
\begin{itemize}
\item 
Parity violation tells us that nature makes a difference between 
"left" and "right" -- but not which is which! For the 
statement that neutrinos emerging from pion decays are 
left- rather than right-handed implies the use of 
positive instead of negative pions. "Left" and 
"right" is thus defined in terms of "positive"  
and "negative", respectively. This is like saying that 
your left thumb is on your right hand -- certainly 
correct, yet circular and thus not overly useful. 

On the other hand CP violation manifesting itself through 
\beq 
\frac{{\rm BR}(K_L \ra l^+ \nu \pi ^-)}
{{\rm BR}(K_L \ra l^- \bar \nu \pi ^+)} \simeq 1.006J\neq 1 
\eeq 
allows us to define "positive" and "negative" in 
terms of 
{\em observation} rather than {\em convention}, and 
subsequently likewise for "left" and "right". 
\item 
The limitation on CP invariance in the $K^0 - \bar K^0$ 
mass matrix 
\beq 
{\rm Im} M_{12} \simeq 1.1 \cdot 10^{-8} \; \; {\rm eV} \; \; 
\hat = \; \; \frac{{\rm Im} M_{12}}{m_K} \simeq 2.2 \cdot  
10^{-17} 
\eeq
represents the most subtle symmetry violation 
actually observed to date. 

\item 
CP violation constitutes one of the three essential ingredients in any 
attempt to understand the observed baryon number of the universe 
as a dynamically generated quantity rather than an initial 
condition \cite{DOLGOV2}. 

\item 
Due to CPT invariance -- on which I will not cast any doubt during   
these lectures -- CP breaking implies a violation of 
time reversal invariance 
\footnote{{\em Operationally} one defines time reversal as the 
reversal 
of {\em motion}: $\vec p \ra - \vec p$, $\vec j \ra - \vec j$ for 
momenta $\vec p$ and angular momenta $\vec j$.}. That nature 
makes an 
intrinsic distinction between past and future on the 
{\em microscopic} level that cannot be explained by statistical 
considerations is an utterly amazing observation.   

\item 
The fact that time reversal represents a very peculiar operation can 
be expressed also in a less emotional way, namely through 
{\em Kramers' Degeneracy} 
\cite{KRAMERS}. The time reversal operator $\bf T$ 
has to be {\em anti}-unitary; ${\bf T}^2$ then has eigenvalues 
$\pm 1$. Consider the sector of the Hilbert space with 
${\bf T}^2 = -1$ and assume the dynamics to conserve ${\bf T}$; 
i.e., the Hamilton operator $\bf H$ and ${\bf T}$ commute. It is 
easily shown that if $|E\rangle$ is an eigenvector of $\bf H$, so is 
${\bf T}|E\rangle$ -- with the {\em same} eigenvalue. Yet 
$|E\rangle$ and ${\bf T}|E\rangle$ are -- that is the main 
substance of this theorem -- orthogonal to each other! Each 
energy eigenstate in the Hilbert sector with ${\bf T}^2 = -1$ 
is therefore at least doubly degenerate. This degeneracy is realized 
in nature through {\em fermionic spin} degrees. Yet it is 
quite remarkable that the time reversal operator $\bf T$ already 
anticipates this option -- and the qualitative difference 
between fermions and bosons -- through ${\bf T}^2 = \pm 1$ -- 
{\em without} any explicit reference to spin! 

\end{itemize}

%%%%%%%%%%%
\subsection{General Description of Particle-Antiparticle Oscillations}
%%%%%%%%%%%%
A symmetry $\bf S$ can be manifestly realized in two different 
ways:    
\begin{itemize}
\item 
There exists a pair of degenerate states that transform into each 
other under $\bf S$. 
\item 
When there is an {\em un}paired state it has to be an eigenstate of 
$\bf S$. 
\end{itemize}
The observation of $K_L$ decaying into a $2\pi$ state -- which 
is CP even -- and a CP odd $3\pi$ combination therefore 
establishes CP violation only because $K_L$ and $K_S$ are 
{\em not} mass degenerate. 

In general, decay rates can exhibit CP violation in three different 
manners,  
namely through 
\begin{itemize}
\item 
the {\em existence} of a reaction, like 
$K_L \ra \pi \pi$, 
\item 
a {\em difference} in CP conjugate rates, like 
$K_L \ra l^- \bar \nu \pi ^+$ vs. 
$K_L \ra l^+  \nu \pi ^-$, 
\item 
a decay rate evolution that is {\em not a purely 
exponential} function of the proper time of decay; i.e., 
if one finds for a CP {\em eigenstate} $f$ 
\beq 
\frac{d}{dt} e^{\Gamma t}{\rm rate} 
(K_{neutral}(t) \ra f) \neq 0 
\eeq  
for all (real) values of $\Gamma$, then CP symmetry must be 
broken. This is easily proven: 
if CP invariance holds, the decaying state 
must be a CP eigenstate like the final state $f$; yet in that case 
the decay rate evolution must be purely exponential -- unless 
CP is violated. Q.E.D. 

\end{itemize} 
The whole formalism of particle-antiparticle oscillations is 
actually a straightforward application of basic quantum mechanics. 
I will describe it in terms of strange mesons; the generalization to 
any other flavour or quantum number (like beauty or charm) is 
obvious. In the absence of weak forces one has two mass degenerate 
and stable mesons $K^0$ and $\bar K^0$ carrying definite 
strangeness $+1$ and $-1$, respectively, since the strong and 
electromagnetic  
forces conserve this quantum number. The addition of the weak 
forces changes the picture qualitatively: strangeness is no longer 
conserved, kaons become unstable and the new mass eigenstates 
-- being linear superpositions of $K^0$ and $\bar K^0$ -- no longer 
carry definite strangeness. The violation of the quantum number 
strangeness has lifted the degeneracy: we have two physical 
states $K_L$ and $K_S$ with different masses and lifetimes: 
$\Delta m_K = m_L - m_K \neq 0 \neq \Delta \tau = 
\tau _L - \tau _S$. 

If CP is conserved in the $\Delta S=2$ transitions the mass 
eigenstates $K_1$ and $K_2$ have to be CP eigenstates as pointed out 
above: $|K_1\rangle = |K_+\rangle$, $|K_2\rangle = |K_-\rangle$,  
where ${\bf CP}|K_{\pm}\rangle \equiv \pm |K_{\pm}\rangle$. 
Using the phase convention 
\beq 
|\bar K^0 \rangle \equiv - {\bf C} |K^0\rangle 
\eeq 
the time evolution of a state that starts out as a $K^0$ is given by 
\beq 
|K^0(t)\rangle = 
\frac{1}{\sqrt{2}} e^{im_1t} e^{-\frac{\Gamma _1}{2}t} 
\left( |K_+\rangle + e^{i\Delta m_Kt} e^{-\frac{\Delta \Gamma}{2}t}
|K_-\rangle \right) 
\eeq 
The intensity of an initially pure $K^0$ beam traveling in vacuum  will then exhibit the 
following time profile: 
\beq 
I_{K^0}(t) = |\langle K^0|K^0(t)\rangle |^2 = 
\frac{1}{4} e^{-\Gamma _1 t } 
\left( 1 + e^{\Delta \Gamma _Kt} + 2e^{\frac{\Delta \Gamma _K}{2} t} 
{\rm cos}\Delta m_K t\right)  
\eeq 
The orthogonal state $|\bar K^0 (t)\rangle$ that was absent initially 
in this beam gets regenerated {\em spontaneously}: 
\beq 
I_{\bar K^0}(t) = |\langle \bar K^0|K^0(t)\rangle |^2 = 
\frac{1}{4} e^{-\Gamma _1 t } 
\left( 1 + e^{\Delta \Gamma _K t} - 2e^{\frac{\Delta \Gamma _K}{2} t} 
{\rm cos}\Delta m_K t\right)  
\eeq 
The oscillation rate expressed through $\Delta m_K$ and 
$\Delta \Gamma _K$ is naturally calibrated by the average 
decay rate $\bar \Gamma _K \equiv 
\frac{1}{2}( \Gamma _1 + \Gamma _2)$: 
\beq 
x_K \equiv \frac{\Delta m_K}{\bar \Gamma _K} \simeq 0.95 
\; \; \; , \; \; \; 
y_K \equiv \frac{\Delta \Gamma _K}{2\bar \Gamma _K} \simeq 1  
\eeq 
Two comments are in order at this point: 
\begin{itemize}
\item 
In any such binary quantum system there will be two lifetimes. 
The fact that they differ so spectacularly for neutral kaons  
-- $\tau (K_L) \sim 600 \cdot \tau (K_S)$ -- is 
due to a kinematical accident: the only available nonleptonic 
channel for the CP odd kaon is the 3 pion channel, for which 
it has barely enough mass.  
\item 
$\Delta m_K \simeq 3.7 \cdot 10^{-6}$ eV is often related 
to the kaon mass: 
\beq  
\frac{\Delta m_K}{m_K} \simeq 7 \cdot 10^{-15} 
\label{STRIKING} 
\eeq 
which is obviously a very striking number. Yet 
Eq.(\ref{STRIKING}) somewhat overstates the point. 
The kaon mass has nothing really to do with the 
$K_L-K_S$ mass difference 
\footnote{It would not be much more absurd to relate 
$\Delta m_K$ to the mass of an elephant!} and actually is  
measured relative to $\Gamma _K$. There is however 
one exotic application where it makes sense to state 
the ratio $\Delta m_K/m_K$, and that is in the context 
of antigravity where one assumes matter and antimatter 
to couple to gravity with the opposite sign. The gravitational 
potential $\Phi$  
would then produce a {\em relative} phase between $K^0$ and 
$\bar K^0$ of  2 $m_K\Phi t$. In the earth's potential this would 
lead to a gravitational oscillation time of 
$10^{-15}$ sec, which is much shorter than the lifetimes or 
the weak oscillation time; $K^0 - \bar K^0$ oscillations could 
then not be observed \cite{GOOD}. 
There are some loopholes in this argument -- 
yet I consider it still intriguing or at least entertaining.

\end{itemize}

%%%%%%%%%%%%%
\section{CP Phenomenology in $K_L$ Decays}
%%%%%%%%%%%%%%
%%%%%%%%%%%
\subsection{General Formalism}
%%%%%%%%%%%%%%%%%%%%

Oscillations become more complex once CP symmetry is broken in 
$\Delta S=2$ transitions, as seen from solving the (free) 
Schr\" odinger equation 
\beq 
i\frac{d}{dt} \left( 
\begin{array}{ll}
K^0 \\
\bar K^0
\end{array}  
\right)  = \left( 
\begin{array}{ll}
M_{11} - \frac{i}{2} \Gamma _{11} & 
M_{12} - \frac{i}{2} \Gamma _{12} \\ 
M^*_{12} - \frac{i}{2} \Gamma ^*_{12} & 
M_{22} - \frac{i}{2} \Gamma _{22} 
\end{array}
\right) 
\left( 
\begin{array}{ll}
K^0 \\
\bar K^0
\end{array}  
\right) 
\label{SCHROED} 
\eeq
CPT invariance imposes 
\beq 
M_{11}= M_{22} \; \; , \; \; \Gamma _{11} = \Gamma _{22} \; . 
\label{CPTMASS}
\eeq  
\begin{center} 
$\spadesuit \; \; \; \spadesuit \; \; \; \spadesuit $ \\ 
{\em Homework Problem \#1}: 
\end{center}
Which physical situation is 
described by an equation analogous to Eq.(\ref{SCHROED}) 
where however the two diagonal matrix elements differ 
{\em without} violating CPT? 
\begin{center} 
$\spadesuit \; \; \; \spadesuit \; \; \; \spadesuit $
\end{center} 
The mass eigenstates obtained through diagonalising this matrix 
are given by (for details see \cite{LEE,BOOK}) 
\beq 
|K_S\rangle \equiv |K_1\rangle = 
\frac{1}{\sqrt{|p|^2 + |q|^2}} \left( p |K^0 \rangle + 
q |\bar K^0\rangle \right) 
\eeq 
\beq 
|K_L\rangle \equiv |K_2\rangle = 
\frac{1}{\sqrt{|p|^2 + |q|^2}} \left( p |K^0 \rangle -  
q |\bar K^0\rangle \right) 
\eeq 
with 
\beq 
\frac{q}{p} = \sqrt{\frac{M_{12}^* - \frac{i}{2} \Gamma _{12}^*}
{M_{12} - \frac{i}{2} \Gamma _{12}}}
\eeq 
and eigenvalues 
\beq 
M_S - \frac{i}{2}\Gamma _S = M_{11} - \frac{i}{2} \Gamma _{11} 
- \sqrt{\left( M_{12} - \frac{i}{2} \Gamma _{12}\right) 
\left( M^*_{12} - \frac{i}{2} \Gamma ^*_{12}\right)} 
\eeq 
\beq 
M_L - \frac{i}{2}\Gamma _L = M_{11} - \frac{i}{2} \Gamma _{11} 
+ \sqrt{\left( M_{12} - \frac{i}{2} \Gamma _{12}\right) 
\left( M^*_{12} - \frac{i}{2} \Gamma ^*_{12}\right)} 
\eeq 
These states are conveniently expressed in terms of the 
CP eigenstates $|K_{\pm}\rangle$: 
\beq 
|K_S \rangle = \frac{1+q/p}{\sqrt{2(1+ |q/p|^2)}} 
\left( |K_+\rangle + \bar \epsilon |K_-\rangle\right) 
\eeq 
\beq 
|K_L \rangle = \frac{1+q/p}{\sqrt{2(1+ |q/p|^2)}} 
\left( |K_-\rangle + \bar \epsilon |K_+\rangle\right) 
\eeq 
where the parameter 
\beq 
\bar \epsilon \equiv \frac{1-q/p}{1+q/p} 
\eeq  
reflects the CP impurity in the state vector. 

A few comments -- some technical, some not -- 
might elucidate the situation: 
\begin{itemize}
\item 
If there is no relative phase between $M_{12}$ and 
$\Gamma _{12}$ 
\beq 
{\rm arg} \frac{M_{12}}{\Gamma _{12}} =0 
\eeq 
then $q/p =1$ and the state vectors conserve CP: 
$\bar \epsilon =0$. 
\item 
Yet for our later 
discussion one should take note that 
$q/p$ -- and therefore also $\bar \epsilon$ --  
{\em by itself cannot} be an observable. For a change in the 
phase {\em convention} adopted for defining $\bar K^0$ does 
not leave it invariant: 
\beq 
|\bar K^0 \rangle \ra e^{i\xi}|\bar K^0 \rangle \; 
\Longrightarrow \; 
(M_{12}, \Gamma _{12}) \ra e^{i\xi} 
(M_{12}, \Gamma _{12}) \; 
\Longrightarrow \; 
\frac{q}{p} \ra e^{-i\xi} \frac{q}{p} \; !
\eeq
On the other hand $|q/p|$ is independant of the phase 
convention and its deviation from unity is one measure 
of CP violation. 
\item 
On very general grounds -- without recourse to any model -- 
one can infer that CP violation in the neutral kaon system 
has to be small. CP invariance implies the two mass eigenstates 
$K_L$ and $K_S$  to be orthogonal -- as can be read off 
explicitely from the general expression 
\beq 
\langle K_L |K_S \rangle = \frac{1 -|q/p|^2} {1 +|q/p|^2}
\eeq
The Bell-Steinberger relation allows to place a bound on 
this scalar product from inclusive decay rates 
\cite{LEE,BOOK}: 
\beq 
\langle K_L |K_S \rangle \leq \sqrt{2} 
\sum _f \sqrt{\frac{\Gamma _L^f\Gamma _S^f}{\Gamma _S^2} } 
\leq \sqrt{2} \sqrt{\frac{\Gamma _L}{\Gamma _S}} 
\simeq 0.06 
\label{BELL} 
\eeq 
There is no input from any CP measurement. What is essential, 
though, is the huge lifetime ratio. 
\item 
There are actually two processes underlying the transition 
$K_L \ra 2\pi$: $\Delta S=2$ forces generate the mass eigenstates 
$K_L$ and $K_S$ whereas $\Delta S=1$ dynamics drive the decays 
$K \ra 2\pi$. Thus CP violation can enter in two a priori independant 
ways, namely through the $\Delta S=2$ and the $\Delta S=1$ 
sector. This distinction can be made explicit in terms of the 
transition amplitudes: 
\beq 
\eta _{+-} \equiv 
\frac{A(K_L \ra \pi ^+ \pi ^-)}{A(K_S \ra \pi ^+ \pi ^-)} \equiv 
\epsilon _K+ \epsilon ^{\prime} \; , \; 
\eta _{00} \equiv 
\frac{A(K_L \ra \pi ^0 \pi ^0)}{A(K_S \ra \pi ^0 \pi ^0)} \equiv 
\epsilon _K- 2\epsilon ^{\prime}  
\eeq
The quantity $\epsilon _K$ describes the CP violation common to the 
$K_L$ decays; it thus characterizes the decaying {\em state} and is 
referred to as {\em CP violation in the mass matrix} or 
{\em superweak CP violation}; $\epsilon ^{\prime}$ on the other 
hand differentiates between different channels and thus 
characterizes {\em decay} dynamics; it is called {\em direct 
CP violation}.   
\item  
{\em Maximal} parity and/or charge conjugation violation 
can be defined by saying there is no right-handed neutrino 
and/or left-handed antineutrino, respectively. Yet 
{\em maximal} CP violation {\em cannot} be defined in an analogous 
way: for the existence of the right-handed antineutrino which is the 
CP conjugate to the left-handed neutrino is already required by 
CPT invariance. 

\end{itemize} 

%%%%%%%%%%%%
\subsection{Data}
%%%%%%%%%   
The data on CP violation in neutral kaon decays are as follows: 
\begin{enumerate} 
\item 
{\em Existence} of $K_L \ra \pi \pi$:
\beq 
\begin{array}{l}
{\rm BR}(K_L \ra \pi ^+ \pi ^-) = (2.067 \pm 0.035) \cdot 10^{-3} \\ 
{\rm BR}(K_L \ra \pi ^0 \pi ^0) = (0.936 \pm 0.020) \cdot 10^{-3} \\ 
\end{array}
\label{ETADATA}
\eeq 
\item 
Search for {\em direct} CP violation: 
\beq 
\frac{\epsilon ^{\prime}}{\epsilon _K} \simeq 
{\rm Re} \frac{\epsilon ^{\prime}}{\epsilon _K} = 
\left\{   
\begin{array}{ll}
(2.3 \pm 0.65) \cdot 10^{-3} & NA\; 31 \\
(1.5 \pm 0.8) \cdot 10^{-3} & PDG \; '96\;  average \\
(0.74 \pm 0.52 \pm 0.29) \cdot 10^{-3} & E\; 731 \\
\end{array} 
\right. 
\label{DIRECTCPDATA} 
\eeq 
\item 
Rate {\em difference} in semileptonic decays: 
\beq 
\delta _l\equiv 
\frac{\Gamma (K_L \ra l^+ \nu \pi ^-) - 
\Gamma (K_L \ra l^- \bar \nu \pi ^+)}
{\Gamma (K_L \ra l^+ \nu \pi ^-) + 
\Gamma (K_L \ra l^- \bar \nu \pi ^+)}  = 
(3.27 \pm 0.12)\cdot 10^{-3} \; , 
\label{SLDIFFDATA} 
\eeq
where an average over electrons and muons has been taken. 
\item 
{\em T violation}: 
\beq 
\frac{\Gamma (K^0 \Rightarrow \bar K^0) - 
\Gamma (\bar K^0 \Rightarrow K^0)} 
{\Gamma (K^0 \Rightarrow \bar K^0) + 
\Gamma (\bar K^0 \Rightarrow K^0)} = 
(6.3 \pm 2.1 \pm 1.8) \cdot 10^{-3} \; \; \; 
CPLEAR 
\label{CPLEAR}
\eeq 
from a third of their data set \cite{CPLEARBLOCH}. 
It would be premature to claim this asymmetry has been 
established; yet it represents an intriguingly direct test of time 
reversal violation and is sometimes referred to as the Kabir 
test . It requires tracking the flavour identity of 
the {\em decaying} meson as a $K^0$ or $\bar K^0$ through its 
semileptonic decays -- $\bar K^0 \ra l^- \bar \nu \pi ^+$ vs. 
$K^0 \ra l^+  \nu \pi ^-$ -- and also of the {\em initially 
produced} kaon. The latter is achieved through correlations 
imposed by associated production. The CPLEAR collaboration 
studied low energy proton-antiproton annihilation 
\beq 
p \bar p \ra K^+ \bar K^0 \pi ^- \; \; vs. \; \; 
p \bar p \ra K^-  K^0 \pi ^+ \; ; 
\eeq 
the charged kaon reveals whether a $K^0$ or a $\bar K^0$ was 
produced in association with it. In the future the CLOE collaboration 
will study T violation in $K^0 \bar K^0$ production at DA$\Phi$NE: 
\beq 
e^+ e^- \ra \phi (1020) \ra K^0 \bar K^0 
\eeq  
\end{enumerate}

%%%%%%%%%%%%%
\subsection{Phenomenological Interpretation}
%%%%%%%%%%%%%%%
%%%%%%%%%%%
\subsubsection{Semileptonic Transitions}
%%%%%%%%%%%%%%%%%
CPT symmetry imposes constraints well beyond the equality of 
lifetimes for particles and antiparticles: certain {\em sub}classes of 
decay rates have to be equal as well. For example one finds 
\beq 
\Gamma (\bar K^0 \ra l^- \bar \nu \pi ^+) = 
\Gamma (K^0 \ra l^+ \nu \pi ^-)  
\eeq 
The rate asymmetry in semileptonic decays listed in 
Eq.(\ref{SLDIFFDATA}) thus reflects pure superweak CP violation: 
\beq 
\delta _l = \frac{ 1 - |q/p|^2}{1+|q/p|^2} 
\eeq
From the measured value of $\delta _l$ one then obtains 
\beq 
\left| \frac{q}{p}\right| = 1 + (3.27 \pm 0.12) \cdot 10^{-3} 
\eeq  
Since one has for the $K^0 - \bar K^0$ system specifically 
\beq 
\left| \frac{q}{p}\right| \simeq 
1 + \frac{1}{2} {\rm arg}\frac{M_{12}}{\Gamma _{12}} 
\eeq 
one can express this kind of CP violation through a phase: 
\beq 
\Phi (\Delta S=2) \equiv {\rm arg}\frac{M_{12}}{\Gamma _{12}} = 
(6.54 \pm 0.24)\cdot 10^{-3} 
\label{PHI2}
\eeq
The result of the Kabir test, Eq.(\ref{CPLEAR}), yields: 
\beq 
\Phi (\Delta S=2) = (6.3 \pm 2.1 \pm 1.8)\cdot 10^{-3}\; ,  
\eeq  
which is of course consistent with Eq.(\ref{PHI2}). 

Using the measured value of $\Delta m_K/\Delta \Gamma _K$ one 
infers 
\beq 
\frac{M_{12}}{\Gamma _{12}} = - (0.4773 \pm 0.0023)
\left[ 1 - i (6.54 \pm 0.24)\cdot 10^{-3})\right] 
\eeq  
%%%%%%%%%%%%%%%
\subsubsection{Nonleptonic Transitions}
%%%%%%%%%%%%%%%%%
From Eq.(\ref{ETADATA}) one deduces 
\beq 
\begin{array}{l} 
|\eta _{+-}| = (2.275 \pm 0.019)\cdot 10^{-3} \\ 
|\eta _{00}| = (2.285 \pm 0.019)\cdot 10^{-3} \\ 
\end{array}
\eeq 
As mentioned before the ratios $\eta _{+-,00}$ are sensitive also 
to direct CP violation generated by a phase between the 
decay amplitudes $A_{0,2}$ for $K_L\ra (\pi \pi )_I$, where the 
subscript $I$ denotes the isospin of the $2\pi$ system: 
\beq 
\Phi (\Delta S=1) \equiv {\rm arg}\frac{A_2}{A_0} 
\eeq
One finds 
\beq 
\eta _{+-} \simeq  \frac{i \tilde x}{2\tilde x+i}
\left[ \Phi (\Delta S=2)  + 2 \omega \Phi (\Delta S=1) \right] \; , 
\eeq  
with  
\beq 
\tilde x \equiv  \frac{\Delta m_K}{\Delta \Gamma _K} = 
\frac{\Delta m_K}{\Gamma (K_S)} 
=\frac{1}{2} x_K \simeq 0.477 \; \; , \; \; 
\omega \equiv \left| \frac{A_2}{ A_0}\right| \simeq 0.05
\eeq
where the second quantity represents the observed
enhancement of $A_0$ for which a name -- "$\Delta I=1/2$ rule" -- 
yet no quantitative dynamical explanation has been found. 
Equivalently one can write   
\beq 
\frac{\epsilon ^{\prime}}{\epsilon _K} \simeq 2 \omega 
\frac{\Phi (\Delta S=1)}{\Phi (\Delta S=2)}
\eeq 
The data on $K_L \ra \pi \pi$ can thus be expressed as 
follows \cite{WINSTEIN} 
\beq 
\begin{array}{l} 
\Phi (\Delta S=2) = (6.58 \pm 0.26) \cdot 10^{-3} \\ 
\Phi (\Delta S=1) = (0.99 \pm 0.53) \cdot 10^{-3}  
\end{array}
\eeq  
%%%%%%%%%%
\subsubsection{Resume}
%%%%%%%%%%%
The experimental results can be summarized as follows: 
\begin{itemize}
\item 
The decays of neutral kaons exhibit unequivocally CP violation 
of the superweak variety, which is expressed through the 
angle $\Phi (\Delta S=2)$. The findings from 
semileptonic and nonleptonic transitions concur to an impressive 
degree. 
\item 
Direct CP violation still has not been established. 
\item 
A theorist might be forgiven for mentioning that the evolution of the 
measurements over the last twenty odd 
years has not followed the straight line this brief summary 
might suggest to the uninitiated reader. 
\end{itemize} 

%%%%%%%%%%%%%%%%%%%
\section{Theoretical Implementation of CP Violation}
%%%%%%%%%%%%%%%

%%%%%%%%%%
\subsection{Some Historical Remarks}
%%%%%%%%%%%%
Theorists can be forgiven if they felt quite pleased with the state 
of their craft in 1964:
\begin{itemize}
\item 
The concept of (quark) families had emerged, at least in a 
rudimentary form. 
\item 
Maximal parity and charge conjugation violations had been found in 
weak charged current interactions, yet CP invariance apparently 
held. Theoretical pronouncements were made ex cathedra why this 
had to be so!
\item 
{\em Pre}dictions of the existence of two kinds of neutral 
kaons with different lifetimes and masses had been 
confirmed by experiment \cite{PAIS}. 
\end{itemize}
That same year the reaction $K_L \ra \pi ^+ \pi ^-$ was 
discovered \cite{FITCH}! Two things should be noted here. 
The Fitch-Cronin experiment had predecessors: rather than 
being an isolated effort it was the culmination of a whole 
research program. Secondly there was at least one theoretical 
voice, namely that of Okun \cite{OKUN}, who in 1962/63 had listed a 
dedicated search for $K_L \ra \pi \pi$ as one of the most important 
unfinished tasks. Nevertheless for the vast majority of the community the 
Fitch-Cronin observation came as a shock and caused considerable 
consternation among theorists. Yet -- to their credit -- these data 
and their consequence, namely that CP invariance was broken, 
were soon accepted as facts. This was phrased -- though 
{\em not explained} -- in terms of the Superweak Model 
\cite{WOLFSW} later that same year. 

In 1970 the renormalizability of the $SU(2)_L\times U(1)$ 
electroweak gauge theory was proven. I find it quite amazing 
that it was still not realized that the physics known at that time 
could not produce CP violation. As long as one had to struggle 
with infinities in the theoretical description one could be forgiven 
for not worrying unduly about a tiny quantity like 
BR$(K_L \ra \pi ^+ \pi ^-) \simeq 2.3 \cdot 10^{-3}$. Yet no such 
excuse existed any longer once a renormalizable theory had been 
developed! The existence of the Superweak Model somewhat 
muddled the situation in this respect: for it provides merely 
a classification of the dynamics underlying CP violation rather 
than a dynamical description itself. 

The paper by Kobayashi and Maskawa \cite{KM}, 
written in 1972 and published in 1973, was the first 
\begin{itemize}
\item 
to state clearly that the 
$SU(2)_L\times U(1)$ gauge theory even with two complete 
families \footnote{Remember this was still before the $J/\psi$ 
discovery!} is necessarily CP-invariant and 
\item 
to list the possible extensions that could generate CP 
violation; among them -- as one option -- was the three (or more) 
family scenario now commonly referred to as the KM ansatz. They 
also discussed the impact of right-handed currents and of a 
non-minimal Higgs sector. 
\end{itemize}

%%%%%%%%%%%
\subsection{The Minimal Model: The KM Ansatz}
%%%%%%%%%
Once a theory reaches a certain degree of complexity, many potential 
sources of CP violation emerge. Popular examples of such a scenario 
are provided by models implementing supersymmetry or its 
local version, supergravity; hereafter both are referred to as 
SUSY. In my lectures I will however focus on the minimal 
theory that can support CP violation, namely the Standard Model 
with three families. All of its dynamical elements have been 
observed -- except for the Higgs boson, of course. 

%%%%%%%%%%%
\subsubsection{Weak Phases like the Scarlet Pimpernel}
%%%%%%%%%%

Weak interactions at low energies are described by four-fermion 
interactions. The most general expression for spin-one 
couplings are 
$$ 
{\cal L}_{V/A} = \left( \bar \psi _1 \gamma _{\mu}
(a + b \gamma _5)\psi _2\right) 
\left( \bar \psi _3 \gamma _{\mu}
(c + d \gamma _5)\psi _4\right) + 
$$
\beq 
+ \left( \bar \psi _2 \gamma _{\mu}
(a^* + b^* \gamma _5)\psi _1\right) 
\left( \bar \psi _4 \gamma _{\mu}
(c^* + d^* \gamma _5)\psi _3\right)
\eeq 
Under CP these terms transform as follows: 
$$  
{\cal L}_{V/A} \stackrel{CP}{\Longrightarrow} 
CP {\cal L}_{V/A} (CP)^{\dagger} = 
\left( \bar \psi _2 \gamma _{\mu}
(a + b \gamma _5)\psi _1\right) 
\left( \bar \psi _4 \gamma _{\mu}
(c + d \gamma _5)\psi _3\right) + 
$$ 
\beq 
+ \left( \bar \psi _1 \gamma _{\mu}
(a^* + b^* \gamma _5)\psi _2\right) 
\left( \bar \psi _3 \gamma _{\mu}
(c^* + d^* \gamma _5)\psi _4\right)
\eeq 
If $a,b,c,d$ are real numbers, one obviously has 
${\cal L}_{V/A}= CP {\cal L}_{V/A} (CP)^{\dagger} $ and CP 
is conserved. Yet CP is {\em not necessarily} broken if these 
parameters are complex, as we will explain specifically 
for the Standard Model. 

{\em Weak Universality} arises naturally whenever the weak 
charged current interactions are described through a 
{\em single} non-abelian gauge group -- $SU(2)_L$ in the case 
under study. For the single {\em self}-coupling of the gauge bosons 
determines also their couplings to the fermions; 
one finds for the quark couplings to the charged $W$ bosons: 
\beq 
{\cal L}_{CC} = 
g \bar U_L^{(0)}\gamma _{\mu}D_L^{(0)} W^{\mu} + 
\bar U_R^{(0)} {\bf M}_U U_L^{(0)} + 
\bar D_R^{(0)} {\bf M}_D D_L^{(0)} + h.c. 
\eeq   
where $U$ and $D$ denote the up- and down-type quarks, 
respectively: 
\beq 
U= (u,c,t) \; \; \; , \; \; \; \; D=(d,s,b) 
\eeq 
and ${\bf M_U}$ and ${\bf M_D}$ 
their 3$\times$3 mass matrices. In general 
those will not be diagonal; to find the physical states, one has to 
diagonalize these matrices: 
\beq 
{\bf M}^{diag}_U = {\bf K}^U_R{\bf M}_U ({\bf K}^U_L)^{\dagger} \; \; , \; \; 
{\bf M}^{diag}_D = {\bf K}^D_R{\bf M}_D ({\bf K}^D_L)^{\dagger}
\eeq 
\beq 
U_{L,R} = {\bf K}^U_{L,R} U^{(0)}_{L,R} \; \; , \; \; 
D_{L,R} = {\bf K}^D_{L,R} D^{(0)}_{L,R}
\eeq  
with ${\bf K}_{L,R}^{U,D}$ representing four unitary 3$\times$3 matrices. 
The coupling of these physical fermions to $W$ bosons is then given 
by 
\beq 
{\cal L}_{CC} = 
g \bar U_L({\bf K}_L^U)^{\dagger}{\bf K}_L^D\gamma _{\mu}D  W^{\mu} + 
\bar U_R {\bf M}^{diag}_U U_L + 
\bar D_R {\bf M}^{diag}_D D_L + h.c. 
\eeq   
and the combination $({\bf K}_L^U)^{\dagger}{\bf K}_L^D 
\equiv {\bf V}_{CKM}$ 
represents the KM matrix, which obviously has to be unitary 
like $K^U$ and $K^D$. Unless the 
up- and down-type mass matrices are aligned in flavour 
space (in which case they would be diagonalized by the 
same operators ${\bf K}_{L,R}$) one has ${\bf V}_{CKM} \neq 1$.  

In the neutral current sector one has 
\beq 
{\cal L}_{NC} = g^{\prime} 
\bar U_L^{(0)} \gamma _{\mu}U_L^{(0)}Z_{\mu} = 
g^{\prime} 
\bar U_L \gamma _{\mu}U_LZ_{\mu}
\eeq 
and likewise for $U_R$ and $D_{L,R}$; i.e. {\em no} 
flavour changing neutral currents are generated, let alone 
new phases. CP violation thus has to be embedded into the 
charged current sector. 

If ${\bf V}_{CKM}$ is real (and thus orthogonal), CP symmetry is 
conserved in the weak interactions. Yet the occurrance of 
complex matrix elements does not {\em automatically} signal 
CP violation. This can be seen through a straightforward 
(in hindsight at least) algebraic argument. A unitary 
$N\times N$ matrix contains $N^2$ independant real 
parameters; $2N-1$ of those can be eliminated through 
re-phasing of the $N$ up-type and $N$ down-type fermion 
fields (changing all fermions by the {\em same} phase obviously 
does not affect  ${\bf V}_{CKM}$). Hence there are $(N-1)^2$ 
real physical parameters in such an $N \times N$ matrix. 
For $N=2$, i.e. two families, one recovers a familiar result, 
namely there is just one mixing angle, the Cabibbo 
angle. For $N=3$ there are four real physical parameters, 
namely three (Euler) angles -- and one phase. It is the latter 
that provides a gateway for CP violation. For $N=4$ Pandora's 
box opens up: there would be 6 angles and 3 phases. 

PDG suggests a "canonical" parametrization for the $3\times 3$ CKM 
matrix: 
$$ 
{\bf V}_{CKM} = 
\left(  
\begin{array}{ccc} 
V(ud) & V(us) & V(ub) \\
V(cd) & V(cs) & V(cb) \\
V(td) & V(ts) & V(tb) 
\end{array} 
\right) 
$$
\beq 
= \left( 
\begin{array}{ccc} 
c_{12}c_{13} & s_{12}c_{13} & s_{13}e^{-i \delta _{13}}  \\
- s_{12}c_{23} - c_{12}s_{23}s_{13}e^{i \delta _{13}} &
c_{12}c_{23} - s_{12}s_{23}s_{13}e^{i \delta _{13}} & 
c_{13}s_{23} \\
s_{12}s_{23} - c_{12}c_{23}s_{13}e^{i \delta _{13}} &
- c_{12}s_{23} - s_{12}c_{23}s_{13}e^{i \delta _{13}} &
c_{13}c_{23} 
\end{array}
\right) 
\label{PDGKM} 
\eeq
where 
\beq 
c_{ij} \equiv {\rm cos} \theta _{ij} \; \; , \; \;  
s_{ij} \equiv {\rm sin} \theta _{ij}
\eeq 
with $i,j = 1,2,3$ being generation labels. 

This is a completely general, yet not unique parametrisation: a 
different set of 
Euler angles could be chosen; the phases can be shifted around 
among the matrix elements 
by using a different phase convention. In that sense one can refer to 
the KM phase as the Scarlet Pimpernel: "Sometimes here, sometimes 
there, sometimes everywhere!"  

Using just the observed hierarchy 
\beq 
|V(ub)| \ll |V(cb)| \ll |V(us)| , |V(cd)| \ll 1
\label{HIER}
\eeq 
one can, as first realized by Wolfenstein, expand 
${\bf V}_{CKM}$ in powers of the Cabibbo angle $\theta _C$: 
\beq 
{\bf V}_{CKM} = 
\left( 
\begin{array}{ccc} 
1 - \frac{1}{2} \lambda ^2 & \lambda & 
A \lambda ^3 (\rho - i \eta + \frac{i}{2} \eta \lambda ^2) \\
- \lambda & 1 - \frac{1}{2} \lambda ^2 - i \eta A^2 \lambda ^4 & 
A\lambda ^2 (1 + i\eta \lambda ^2 ) \\ 
A \lambda ^3 (1 - \rho - i \eta ) \\
& - A\lambda ^2 & 1 
\end{array}
\right) 
+ {\cal O}(\lambda ^6) 
\label{WOLFKM}
\eeq 
where 
\beq 
\lambda \equiv {\rm sin} \theta _C
\eeq
For such an expansion in powers of $\lambda$ to be self-consistent, 
one has to require that $|A|$, $|\rho |$ and $|\eta |$ are of order 
unity. Numerically we obtain  
\beq 
\lambda = 0.221 \pm 0.002  
\label{KMNUm1} 
\eeq 
from $|V(us)|$,  
\beq 
A = 0.81 \pm 0.06 
\label{KMNUM2}
\eeq 
from $|V(cb)| \simeq 0.040 \pm 0.002|_{exp} \pm 0.002|_{theor}$ 
and 
\beq 
\sqrt{\rho ^2 + \eta ^2} \sim 0.38 \pm 0.11 
\label{KMNUM3} 
\eeq
from $|V(ub)| \sim (3.2 \pm 0.8)\cdot 10^{-3}$. 

We see that the CKM matrix is a very special unitary matrix: 
it is almost diagonal, it is almost symmetric and the matrix 
elements get smaller the more one moves away from the 
diagonal. 
Nature most certainly has encoded a profound message in 
this peculiar pattern. Alas -- we have not succeeded yet in 
deciphering it! I will return to this point at the end of my 
lectures.  

%%%%%%%%%%%%%%%
\subsubsection{Unitarity Triangles}
%%%%%%%%%%%%%%

The qualitative difference between a two and a three family scenario 
can be seen also in a less abstract way.  
Consider $\bar K^0 \ra \pi ^+ \pi ^-$; it can proceed through 
a tree-level process 
$[s\bar d] \ra [d \bar u][u\bar d]$ , in which case its 
weak couplings are 
given by $V(us)V^*(ud)$. Or it can oscillate first to $K^0$ 
before decaying; i.e., on the quark level it is the 
transition $[s\bar d] \ra [d\bar s] \ra [d \bar u][u\bar d]$ 
controlled by 
$\left( V(cs)\right) ^2 \left( V^*(cd)\right) ^2 V^*(us)V(ud)$. 
At first sight it would seem that those two combinations of 
weak parameters are not only different, but should also exhibit a 
relative phase. Yet the latter is not so -- if there are two families 
only! In that case the four quantities $V(ud)$, $V(us)$, $V(cd)$ 
and $V(cs)$ have to form a unitary $2\times 2$ which leads to the 
constraint 
\beq 
V(ud)V^*(us) + V(cd)V^*(cs) = 0 
\label{UNIT2FAM}
\eeq 
Using Eq.(\ref{UNIT2FAM}) twice one gets  
$$  
\left( V(cs) V^*(cd)\right) ^2 V^*(us)V(ud) = 
- |V(cd)V(cs)|^2 V(cs)V^*(cd) = 
$$ 
\beq 
= |V(cd)V(cs)|^2 V^*(ud) V(us) \; ; 
\eeq  
 i.e., the two combinations $V^*(ud)V(us)$ and 
$\left( V(cs)\right) ^2 \left( V^*(cd)\right) ^2 V^*(us)V(ud)$ are 
actually parallel to each other with {\em no} 
relative phase. A penguin 
operator with a charm quark as the internal fermion 
line generates another contribution to 
$K_L \ra \pi ^-\pi ^+$, this one controlled by $V(cs)V^*(cd)$. Yet 
the unitarity condition Eq.(\ref{UNIT2FAM}) forces this contribution 
to be antiparallel to $V^*(ud)V(us)$; i.e., again no relative phase. 

The situation changes fundamentally for three families: the weak 
parameters $V(ij)$ now form a $3\times 3$ matrix and the condition 
stated in  Eq.(\ref{UNIT2FAM}) gets extended: 
\beq 
V(ud)V^*(us) + V(cd)V^*(cs) + V(td)V^*(ts) = 0 
\label{UNIT3FAM}
\eeq 
{\em This is a triangle relation in the complex plane.} 
There emerge now 
relative phases between the weak parameters and the 
loop diagrams with internal charm and top quarks can generate 
CP asymmetries. 

Unitarity imposes altogether nine algebraic conditions on the 
matrix elements of ${\bf V}_{CKM}$, of which six are triangle relations 
analogous to Eq.(\ref{UNIT3FAM}). 
There are several nice features about this representation in terms of 
triangles; I list four now and others later: 
\begin{enumerate}
\item 
The {\em shape} of each triangle is independant of the phase 
convention adopted for the quark fields. Consider for example 
Eq.(\ref{UNIT3FAM}): changing the phase of any of the 
up-type quarks will not affect the triangle at all. Under 
$|s\rangle \ra |s\rangle e^{i \phi _s}$ the whole triangle will 
rotate around the left end of its base line by an angle 
$\phi _s$ -- yet the shape of the triangle -- in contrast 
to its orientation in the complex plane -- remains the same! 
The angles inside the triangles are thus observables;  
choosing an orientation for the triangles is then a matter 
of convenience. 
\item 
It is easily shown that all six KM triangles possess 
the same area. 
Multiplying Eq.(\ref{UNIT3FAM}) by the phase 
factor $V^*(ud)V(us)/|V(ud)V(us)|$, which does not change the 
area, yields 
\beq 
|V(ud)V(us)| + \frac{V^*(ud)V(us)V(cd)V^*(cs)}{|V(ud)V(us)|} + 
\frac{V^*(ud)V(us)V(td)V^*(ts)}{|V(ud)V(us)|} = 0 
\eeq 
$$  
{\rm area (triangle \; of \; Eq.(\ref{UNIT3FAM})})  = 
\frac{1}{2} |{\rm Im}V(ud)V(cs)V^*(us)V^*(cd)| = 
$$ 
\beq 
= \frac{1}{2} |{\rm Im}V(ud)V(ts)V^*(us)V^*(td)|
\eeq
Multiplying Eq.(\ref{UNIT3FAM}) instead by the phase 
factors $V^*(cd)V(cs)/|V(cd)V(cs)|$ or $V^*(td)V(ts)/|V(td)V(ts)|$ 
one sees that the area of this triangle can be expressed in other ways 
still. Among them is 
\beq 
{\rm area (triangle \; of \; Eq.(\ref{UNIT3FAM})})  = 
\frac{1}{2} |{\rm Im}V(cd)V(ts)V^*(cs)V^*(td)|  
\eeq 
Due to the unitarity relation 
\beq 
V^*(cd)V(td) + V^*(cb)V(tb) = - V^*(cs)V(ts) 
\label{UNIT3FAM2}
\eeq
one has  
\beq 
{\rm area ( triangle \; of \; Eq.(\ref{UNIT3FAM})})  = 
\frac{1}{2} |{\rm Im}V(cd)V(tb)V^*(cb)V^*(td)|  
\eeq 
-- yet this is exactly the area of the triangle defined by 
Eq.(\ref{UNIT3FAM2})! 
This is the re-incarnation of the original 
observation that there is a {\em single irreducible}  
weak phase for three families. 
\item 
In general one has for the area of these triangles 
$$  
A_{CPV}(\rm every \; triangle) = \frac{1}{2} J 
$$ 
\beq  
J = {\rm Im}V^*(km) V(lm)V(kn)V^*(ln) = 
{\rm Im}V^*(mk) V(ml)V(nk)V^*(nl) 
\label{KMAREA}
\eeq
irrespective of the indices $k,l,m,n$; $J$ is obviously re-phasing 
invariant. 
\item  
If there is a representation of $V_{CKM}$ where 
all phases were confined to a $2\times 2$ 
sub-matrix exactly rather than approximately, then one can  
rotate all these phases away; i.e., CP is conserved in such a 
scenario! Consider again the triangle described by 
Eq.(\ref{UNIT3FAM}): it can always be rotated such that its 
baseline -- $V(ud)V^*(us)$  -- 
is real. Then Im$V(td)V^*(ts)$ = - Im$V(cd)V^*(cs)$ holds. 
If, for example, there were no phases in the third row and column, 
one would have Im$(V(td)V^*(ts)) =0$ and therefore 
Im$V(cd)V^*(cs) =0$ as well; i.e., $V(ud)V^*(us)$ and 
$V(cd)V^*(cs)$ were real relative to each other; 
therefore $J=0$, i.e. all six triangles had zero area meaning 
there are no relative weak phases! 
\end{enumerate}  

%%%%%%%%%%%
\subsection{Evaluating $\epsilon _K$ and $\epsilon ^{\prime}$} 
%%%%%%%%%
In calculating observables in a given theory -- in the case under 
study $\epsilon _K$ and $\epsilon ^{\prime}$ 
within the KM Ansatz -- one is faced with the `Dichotomy of the Two 
Worlds', namely 
\begin{itemize}
\item 
one world of {\em short}-distance physics where even the strong 
interactions can be treated {\em perturbatively} in terms of 
quarks and gluons and in which theorists like to work, and 
\item 
the other world of {\em long}-distance physics where one has 
to deal with hadrons the behaviour of which is controlled 
by {\em non}-perturbative dynamics and where, by the way, 
everyone, including theorists, lives. 
\end{itemize}
Accordingly the calculational task is divided into two 
parts, namely first determing the relevant 
transition operators in the short-distance world and then 
evaluating their matrix elements in the hadronic world. 

%%%%%%%%%%%%%%%%%
\subsubsection{$\Delta S=2$ Transitions}
%%%%%%%%%%%%%%
Since the {\em elementary} interactions in the 
Standard Model can change strangeness at most by one unit, 
the $\Delta S=2$ amplitude driving $K^0 - \bar K^0$ 
oscillations is obtained by iterating the 
basic $\Delta S=1$ coupling: 
\beq 
{\cal L}_{eff} (\Delta S=2) = {\cal L}(\Delta S=1) \otimes 
{\cal L}(\Delta S=1) 
\eeq  
There are actually two ways in which the $\Delta S=1$ transition 
can be iterated: 

{\bf (A)} 
The resulting $\Delta S=2$ transition is described by 
a {\em local} operator. The celebrated box diagram makes 
this connection quite transparent. The contributions that do 
{\em not} depend on the mass of the internal quarks cancel against 
each other due to the GIM mechanism. Integrating over the internal 
fields, namely the $W$ bosons and the top and charm quarks 
\footnote{The up quarks act merely as a subtraction term here.} 
then yields a convergent result: 
$$  
{\cal L}_{eff}^{box}(\Delta S=2, \mu ) = 
\left( \frac{G_F}{4\pi }\right) ^2 \cdot 
$$ 
\beq  
\cdot \left[  \xi _c^2 E(x_c) \eta _{cc} + 
\xi _t^2 E(x_t) \eta _{tt} + 
2\xi _c \xi _t E(x_c, x_t) \eta _{ct}
 \right] \cdot  [\alpha _S(\mu ^2)]^{-6/27} 
\left( \bar s \gamma _{\mu}(1- \gamma _5) d\right) ^2 
\label{LAGDELTAS2}
\eeq  
with $\xi _i$ denoting combinations of KM parameters 
\beq 
\xi _i = V(is)V^*(id) \; , \; \; i=c,t \; ; 
\eeq 
$E(x_i)$ and $E(x_c,x_t)$ reflect the box loops with equal and 
different internal quarks, respectively \cite{INAMI}:  
\beq 
E(x_i) = x_i 
\left(   
\frac{1}{4} + \frac{9}{4(1- x_i)} - \frac{3}{2(1- x_i)^2} 
\right) 
- \frac{3}{2} \left( \frac{x_i}{1-x_i}\right) ^3 
{\rm log} x_i 
\eeq 
$$  
E(x_c,x_t) = x_c x_t 
\left[ \left( 
\frac{1}{4} + \frac{3}{2(1- x_t)} - \frac{3}{4(1- x_t)^2} \right) 
\frac{{\rm log} x_t}{x_t - x_c} + (x_c \leftrightarrow x_t) - 
\right. 
$$ 
\beq 
\left. - \frac{3}{4} \frac{1}{(1-x_c)(1- x_t)} \right]  
\eeq
\beq 
x_i = \frac{m_i^2}{M_W^2} 
\eeq  
and $\eta _{ij}$ containing the QCD radiative corrections from 
evolving the effective Lagrangian from $M_W$ down to 
the internal quark mass. The factor $[\alpha _S(\mu ^2)]^{-6/27}$ 
reflects the fact that a scale 
$\mu$ must be introduced at which the four-quark operator 
$\left( \bar s \gamma _{\mu}(1- \gamma _5) d\right) ^2 $ is 
defined. This dependance on the auxiliary variable 
$\mu$ drops out when one takes the matrix element of this 
operator (at least when one does it correctly).  
Including next-to-leading log 
corrections one finds (for $m_t \simeq 180$ GeV) \cite{BURAS}: 
\beq 
\eta _{cc} \simeq 1.38 \pm 0.20 \; , \; \; 
\eta _{tt} \simeq 0.57 \pm 0.01 \; , \; \; 
\eta _{cc} \simeq 0.47 \pm 0.04 
\eeq                       

{\bf (B)} 
However there is also a {\em non}-local $\Delta S=2$ 
operator generated from the iteration of ${\cal L}(\Delta S=1)$. 
It presumably provides a major contribution to $\Delta m_K$. 
Yet for $\epsilon _K$ it is not sizeable within the KM ansatz 
\footnote{This can be inferred from the observation that 
$|\epsilon ^{\prime}/\epsilon _K|\ll 0.05$} and will be 
ignored here. 

Even for a local four-fermion operator it is non-trivial to 
evaluate an on-shell matrix element 
between hadron states since that is 
clearly controlled by non-perturbative dynamics. Usually one 
parametrizes this matrix element as follows: 
$$  
\matel{\bar K^0}{(\bar s \gamma _{\mu}(1-\gamma _5)d) 
(\bar s \gamma _{\mu}(1-\gamma _5)d)}{K^0} = 
$$ 
\beq 
= \frac{4}{3} B_K 
\matel{\bar K^0}{(\bar s \gamma _{\mu}(1-\gamma _5)d)}{0} 
\matel{0}{(\bar s \gamma _{\mu}(1-\gamma _5)d)}{K^0} = 
\frac{4}{3} B_K f_K^2m_K
\label{BAGFACT} 
\eeq 
The factor $B_K$ is -- for historical reasons of no consequence now -- 
often called the bag factor; $B_K = 1$ is referred to as 
{\em vacuum saturation} or {\em factorization ansatz} since it 
corresponds to a situation where inserting the vacuum intermediate 
state into Eq.(\ref{BAGFACT}) reproduces the full result 
after all colour contractions of the quark lines have been included. 
Several theoretical techniques have been employed to estimate the 
size of $B_K$; their findings are listed in 
Table \ref{TABLEBAG}.  
\begin{table}
\begin{tabular} {|l|l|}
\hline   
Method & $B_K$ \\ 
\hline 
\hline 
Large $N_C$ Expansion & $\frac{3}{4}$\\
\hline 
Large $N_C$ Chiral Pert. with loop correction & $0.66 \pm 0.1$\\
\hline 
Lattice QCD & $0.84 \pm 0.2$ \\
\hline 
\end{tabular}
\centering
\caption{Values of $B_K$ from various theoretical techniques} 
\label{TABLEBAG} 
\end{table} 
These results, which are all consistent with each other and with 
several phenomenological studies as well, can be summarized as 
follows:  
\beq 
B_K \simeq 0.8 \pm 0.2 
\label{BAG} 
\eeq 
Since the size of this matrix element is determined 
by the strong interactions, one indeed expects $B_K \sim 1$. 

We have assembled all the ingredients now for calculating 
$\epsilon _K$. The starting point is given by 
\footnote{The exact expression is 
$|\epsilon _K|  = \frac{1}{\sqrt{2}}
\left| \frac{{\rm Im}M_{12}}{\Delta m_K} - \xi _0\right| $ 
where $\xi _0$ denotes the phase of the 
$K^0 \ra (\pi \pi )_{I =0}$ isospin zero amplitude; its 
contribution is 
numerically irrelevant.}: 
\beq 
|\epsilon _K|  
\simeq 
\frac{1}{\sqrt{2}}
\left| \frac{{\rm Im}M_{12}}{\Delta m_K} \right| 
\eeq 
The CP-odd part Im$M_{12}$ is obtained from 
\beq 
{\rm Im} M_{12} = 
{\rm Im}\matel{K^0}{{\cal L}_{eff}(\Delta S=2)}{\bar K^0} 
\eeq 
whereas for $\Delta m_K$ one inserts the experimental value,   
since the long-distance contributions to $\Delta m_K$ are not under 
theoretical control. One then finds 
$$  
|\epsilon _K|_{KM} \simeq |\epsilon _K|_{KM}^{box} \simeq  
$$ 
$$ 
\simeq \frac{G_F^2}{6 \sqrt{2} \pi ^2} 
\frac{M_W^2m_K f_K^2 B_K}{\Delta m_K} 
\left[ {\rm Im}\xi _c^2 E(x_c) \eta _{cc} + 
 {\rm Im}\xi _t^2 E(x_t) \eta _{tt} + 
2{\rm Im}(\xi _c\xi _t) E(x_c,x_t) \eta _{ct} \right] 
$$ 
$$  
\simeq 1.9 \cdot 10^4 B_K \left[ {\rm Im}\xi _c^2 E(x_c) \eta _{cc} + 
 {\rm Im}\xi _t^2 E(x_t) \eta _{tt} + 
2{\rm Im}(\xi _c\xi _t) E(x_c,x_t) \eta _{ct} \right] 
\simeq 
$$ 
\beq 
\simeq 
7.8 \cdot 10^{-3} \eta B_K (1.3 - \rho ) 
\label{EPSKM} 
\eeq 
where I have used the numerical values for the KM parameters 
listed above and 
$x_t \simeq 5$ corresponding to $m_t = 180$ GeV. 

To reproduce the observed value of $|\epsilon _K|$ one needs 
\beq 
\eta \simeq \frac{0.3}{B_K} \frac{1}{1.3 - \rho } 
\label{EPSREP} 
\eeq
For a given $B_K$ one thus obtains another $\rho - \eta $ 
constraint. Since $B_K$ is not precisely known 
\footnote{Some might argue that this is an understatement.} 
one has a fairly broad band in the $\rho - \eta $ plane 
rather than a line. Yet I find it quite remarkable and very 
non-trivial that Eq.(\ref{EPSREP}) {\em can} be 
satisfied since 
\beq 
\frac{0.3}{B_K} \sim 0.3 \div 0.5 
\eeq 
without stretching any of the parameters or bounds, in particular 
\beq 
\sqrt{\rho ^2 + \eta ^2} \sim 0.38 \pm 0.11 \; .   
\eeq  
 While this does of course not amount to a {\em pre}diction, 
one should keep in mind for proper perspective 
that in the 1970's and early 
1980's values like $|V(cb)| \sim 0.04$ and 
$|V(ub)|  \sim 0.004$ would have seemed quite unnatural; claiming 
that the top quark mass had to be 180 GeV would have been 
outright preposterous even in the 1980's! Consider a 
scenario with $|V(cb)| \simeq 0.04$ and $|V(ub)| \simeq 0.003$, 
yet $m_t \simeq 40$ GeV; in the mid 80's this would have appeared 
to be quite natural (and there had even been claims that top quarks 
with a mass of $40\pm 10$ GeV had been discovered). In that 
case one would need 
\beq 
\eta \sim \frac{0.75}{B_K} 
\label{EPS40} 
\eeq 
to reproduce $|\epsilon _K|$. Such a large value for $\eta$ would hardly 
be compatible with what we know about $|V(ub)|$ 
\footnote{For some time it was thought that $B_K \simeq 0.3 \div 
0.5$ was the best estimate. This would make satisfying 
Eq.(\ref{EPS40}) completely out of the question!}. 
\begin{center} 
$\spadesuit \; \; \; \spadesuit \; \; \; \spadesuit $ \\ 
{\em  Homework Problem \# 2}: 
\end{center}
Eq.(\ref{EPSKM}) suggests that 
a non-vanishing value for $\epsilon _K$ is  generated from the 
box diagram with internal charm quarks only -- 
Im$\xi _c^2\; E(x_c) = - \eta A^2 \lambda ^6 E(x_c) \neq 0$ -- 
{\em without} top quarks. How does this match up with the 
statement that the intervention of three families 
is needed for a CP asymmetry to arise? 
\begin{center} 
$\spadesuit \; \; \; \spadesuit \; \; \; \spadesuit $
\end{center} 

%%%%%%%%%%%%%%%%%
\subsubsection{$\Delta S=1$ Decays}
%%%%%%%%%%%%%%
At first one might think that no {\em direct} CP asymmetry 
can arise in $K \ra \pi \pi$ decays since it requires the interplay 
of three quark families. Yet upon further reflection one realizes 
that a one-loop diagram produces the so-called Penguin 
operator which changes isopin by half a unit only, 
is {\em local} and contains a CP {\em odd} component since it involves virtual charm 
and top quarks. With direct CP violation thus being  
of order $\hbar$, i.e. a pure quantum 
effect, one suspects already at this point that it will be reduced 
in strength.  

The quantity $\epsilon ^{\prime}$ is suppressed relative to 
$\epsilon _K$  due to two other reasons: 
\begin{itemize}
\item 
The GIM factors are actually quite different for $\epsilon _K$ and 
$\epsilon ^{\prime}$: in the former case they are of the type 
$(m_t^2 - m_c^2)/M_W^2$, in the latter log$(m_t^2/m_c^2)$. Both 
of these expressions vanish for $m_t=m_c$, yet for the realistic 
case $m_t \gg m_c$ they behave very differently: $\epsilon _K$ 
is much more enhanced by the large top mass than 
$\epsilon ^{\prime}$. This means of course that 
$|\epsilon ^{\prime}/\epsilon _K|$ is a rather steeply decreasing 
function of $m_t$. 
\item 
There are actually two classes of Penguin operators contributing 
to $\epsilon ^{\prime}$, namely strong as well as electroweak 
Penguins. The latter become relevant  
since they are more enhanced than the former for very heavy top 
masses due to the coupling of the longitudinal virtual $Z$ boson 
(the re-incarnation of one of the original Higgs fields) to 
the internal top line. Yet electroweak and strong Penguins contribute 
with the opposite sign! 
\end{itemize}
CPT invariance together with the measured $\pi \pi$ phase shifts 
tells us that the two complex quantities $\epsilon ^{\prime}$ and 
$\epsilon _K$ are almost completely real to each other; i.e., 
their ratio is practically real: 
\beq 
\frac{\epsilon ^{\prime}}{\epsilon _K} \simeq 2 \omega 
\frac{\Phi (\Delta S=1)}{\Phi (\Delta S=2)}
\label{EPSPOVEREPSTH} 
\eeq 
where, as defined before,  
\beq 
\omega \equiv \frac{|A_2|}{|A_0|} \simeq 0.05 \; \; , \; \; 
\Phi (\Delta S=2) \equiv {\rm arg}\frac{M_{12}}{\Gamma _{12}} 
\; \; , \Phi (\Delta S=1) \equiv {\rm arg} \frac{A_2}{A_0}
\eeq
Eq.(\ref{EPSPOVEREPSTH}) makes two points obvious:
\begin{itemize}
\item 
Direct CP violation -- $\epsilon ^{\prime} \neq 0$ -- 
requires a relative phase between the isospin 0 and 2 amplitudes; 
i.e., $K \ra (\pi \pi )_0$ and $K \ra (\pi \pi )_2$ have to exhibit 
different CP properties. 
\item 
The observable ratio $\epsilon ^{\prime} / \epsilon _K$ is 
{\em artifically reduced} by the 
enhancement of the $\Delta I =1/2$ amplitude, as 
expressed through $\omega$. 
\end{itemize} 
Several $\Delta S=1$ transition operators contribute to 
$\epsilon ^{\prime}$ and their renormalization has to be treated 
quite carefully. Two recent detailed analyses yield 
\cite{BURASPRIME,CIUCHINIPRIME}
\beq 
-2.1 \cdot 10^{-4} \leq \frac{\epsilon ^{\prime}}{\epsilon _K} \leq 
13.3 \cdot 10^{-4} 
\label{BURAS1} 
\eeq 
\beq 
\frac{\epsilon ^{\prime}}{\epsilon _K} =  
(4.6 \pm 3.0 \pm 0.4)\cdot 10^{-4} 
\label{CIUCHINI} 
\eeq 
These results are quite consistent with each other and show  
\begin{itemize}
\item  
that the KM ansatz leads to a prediction typically in the range 
{\em below} $10^{-3}$, 
\item 
that the value could happen to be zero or 
even slightly negative and 
\item 
that large theoretical uncertainties persist due to cancellations among 
various contributions. 
\end{itemize} 
This last (unfortunate) point can be illustrated also by comparing 
these predictions with older ones made before top quarks were 
discovered and their mass measured; those old predictions 
\cite{FRANZINI} are 
very similar to Eqs.(\ref{BURAS1},\ref{CIUCHINI}), once the now 
known value of $m_t$ has been inserted. 

Two new experiments running now -- NA 48 at CERN and KTEV at 
FNAL -- and one expected to start up soon -- CLOE at DA$\Phi $NE -- 
expect to measure $\epsilon ^{\prime}/\epsilon _K$ with a 
sensitivity of $\simeq \pm 2\cdot 10^{-4}$. Concerning their future 
results one can distinguish four scenarios: 
\begin{enumerate}
\item 
The `best' scenario: 
$\epsilon ^{\prime}/\epsilon _K \geq 2 \cdot 10^{-3}$. One would 
then 
have established unequivocally direct CP violation of a strength that 
very probably reflects the intervention of new physics beyond the 
KM ansatz. 
\item 
The `tantalizing' scenario: 
$1\cdot 10^{-3}\leq \epsilon ^{\prime}/\epsilon _K 
\leq 2 \cdot 10^{-3}$. It would be tempting to interprete this 
discovery of direct CP violation as a sign for new physics -- yet 
one could not be sure! 
\item 
The `conservative' scenario: 
$\epsilon ^{\prime}/\epsilon _K \simeq 
{\rm few}\cdot10^{-4} > 0$. This strength of direct CP violation could 
easily be accommodated 
within the KM ansatz -- yet no further constraint would 
materialize. 
\item 
The `frustrating' scenario: 
$\epsilon ^{\prime}/\epsilon _K \simeq 0$ within errors! 
No substantial conclusion could be drawn then concerning the 
presence or absence of direct CP violation, and the allowed KM 
parameter space would hardly shrink. 
\end{enumerate}

%%%%%%%%%%%%
\section{`Exotica'}
%%%%%%%%%%%%%
In this section I will discuss important possible manifestations 
of CP and/or T violation that are exotic only in the sense that 
they are unobservably small with the KM ansatz.  

%%%%%%%%%%
\subsection{$K_{3\mu}$ Decays}
%%%%%%%%%%%%%
In the reaction  
\beq 
K^+ \ra \mu ^+ \nu \pi ^0
\eeq 
one can search for a transverse polarisation of the emerging 
muons: 
\beq 
P_{\perp}^{K^+}(\mu ) \equiv 
\langle \vec s (\mu) \cdot 
(\vec p (\mu ) \times \vec p (\pi ^0))\rangle  
\eeq
where $\vec s$ and $\vec p$ denote spin and momentum, 
respectively. 
The quantity $P_{\perp}(\mu )$ constitutes a {\em T-odd}  
correlation: 
\beq 
\left. 
\begin{array}{l}
\vec p \Rightarrow - \vec p \\ 
\vec s \stackrel{T}{\Rightarrow} - \vec s
\end{array} 
\right\} \leadsto 
P_{\perp}(\mu ) \stackrel{T}{\Rightarrow} - P_{\perp}(\mu ) 
\eeq   
Once a {\em non-}vanishing value has been observed for a 
parity-odd correlation one has unequivocally found a manifestation 
of parity violation. From $P_{\perp}^{K^+}(\mu ) \neq 0$ one can 
deduce that T is violated -- yet the argument is more subtle as can 
be learnt from the following homework problem. 
\begin{center} 
$\spadesuit \; \; \; \spadesuit \; \; \; \spadesuit $ \\ 
{\em Homework Problem \#3}: 
\end{center}
Consider 
\beq 
K_L \ra \mu ^+ \nu \pi ^- 
\eeq 
Does $P_{\perp}^{K_L}(\mu ) \equiv 
\langle \vec s(\mu ) \cdot (\vec p(\mu ) \times \vec p(\pi ^-)) 
\rangle \neq 0$ necessarily imply that T invariance does not hold in 
this reaction?
\begin{center}  
$\spadesuit \; \; \; \spadesuit \; \; \; \spadesuit $ \\ 
\end{center}
Data on $P_{\perp}^{K^+}(\mu )$ are still consistent with zero 
\cite{SCHMIDT}: 
\beq 
P_{\perp}^{K^+}(\mu ) = ( -1.85 \pm 3.60) \cdot 10^{-3} \; ; 
\label{YALE}
\eeq 
yet being published in 1981 they are ancient by the standards of our 
disciplin. 

On general grounds one infers that 
\beq 
P_{\perp}^{K^+}(\mu ) \propto {\rm Im} \frac{f_-^*}{f_+} 
\label{POLTH} 
\eeq 
holds where $f_-\, [f_+]$ denotes the chirality changing 
[conserving] decay amplitude. Since $f_-$ practically vanishes 
within the Standard Model, one obtains a fortiori 
$P_{\perp}^{K^+}(\mu )|_{KM} \simeq 0$. 

Yet in the presence of charged Higgs fields one has 
$f_- \neq 0$. CPT implies that 
$P_{\perp}^{K^+}(\mu ) \neq 0$ represents CP violation as 
well, and actually one of the {\em direct} variety. A rather 
model independant guestimate on how large such an effect 
could be is obtained from the present bound on 
$\epsilon ^{\prime}/\epsilon _K$: 
\beq 
P_{\perp}^{K^+}(\mu ) \leq 20 \cdot 
(\epsilon ^{\prime}/\epsilon _K) \cdot \epsilon _K \leq 
10^{-4} 
\eeq 
where the factor 20 allows for the `accidental' reduction of 
$\epsilon ^{\prime}/\epsilon _K$ by the $\Delta I=1/2$ rule: 
$\omega \simeq 1/20$. This bound is a factor of 100 larger 
than what one could obtain within KM.  It could actually be bigger 
still since there is a loophole in this generic 
argument: Higgs couplings to leptons could be strongly 
enhanced through a large ratio of vacuum expectation values $v_1$ 
relative to $v_3$, where $v_1$ controls the couplings to 
up-type quarks and $v_3$ to leptons. 
Then 
\beq 
P_{\perp}^{K^+}(\mu )|_{Higgs} \leq {\cal O}(10^{-3}) 
\eeq 
becomes conceivable with the Higgs fields as heavy as 
80 - 200 GeV 
\cite{GARISTO}. Such Higgs exchanges would be quite insignificant 
for $K_L \ra \pi \pi $! 

Since $K_{\mu 3}$ studies provide such a unique 
window onto Higgs dynamics, I find it mandatory to probe for 
$P_{\perp}(\mu ) \neq 0$ in a most determined way. 
It is gratifying to note that an on-going KEK experiment will be 
sensitive to $P_{\perp}(\mu )$ down to the $10^{-3}$ level -- 
yet I strongly feel one should not stop there, but push 
further down to the $10^{-4}$ level. 
%%%%%%%%%%
\subsection{Electric Dipole Moments}
%%%%%%%%%%%
Consider a system -- such as an elementary particle or  
an atom -- in a weak external electric field $\vec E$. The 
energy shift of this system due to the electric field 
can then be expressed through an expansion in powers 
of $\vec E$ \cite{BERN}: 
\beq 
\Delta E = \vec d \cdot \vec E + d_{ij}E_i E_j + 
{\cal O}(|\vec E|^3) 
\label{ESHIFTDIP}
\eeq 
where summation over the indices $i,j$ is understood. The 
coefficient $\vec d$ of the term linear in $\vec E$ is called 
electric dipole moment or sometimes permanent 
electric dipole moment (hereafter referred to as EDM) whereas that 
of the quadratic 
term is often named an {\em induced} dipole moment. 

For an elementary object one has 
\beq 
\vec d = d \vec j 
\eeq 
where $\vec j$ denotes its total angular momentum since that 
is the only available vector. Under time reversal one finds  
\beq 
\begin{array}{l}
\vec j \; \stackrel{T}{\Rightarrow} \; - \vec j \\ 
\vec E \; \stackrel{T}{\Rightarrow} \vec E \; . 
\end{array} 
\eeq 
Therefore 
\beq 
{\rm T \; invariance}  \leadsto d = 0 \; ; 
\eeq 
i.e., such an electric dipole moment has to vanish, unless T is 
violated (and likewise for parity). 

The EDM is at times confused with an induced electric dipole 
moment objects can possess due to their internal structure. To 
illustrate that consider an atom with two {\em nearly degenerate} 
states of opposite parity: 
\beq 
{\bf P}|\pm \rangle = \pm |\pm \rangle \; , \; 
{\bf H}|\pm \rangle = E_{\pm} |\pm \rangle \; , \; E_+ < E_- \; , \; 
\frac{E_- - E_+}{E_+} \ll 1 
\eeq 
Placed in a constant external electric field $\vec E$ the states 
$|\pm \rangle $ will mix to produce new energy eigenstates; 
those can be found by diagonalising the matrix of the 
Hamilton operator: 
\beq 
H = 
\left( 
\begin{array}{ll}
E_+ & \Delta \\ 
\Delta & E_- 
\end{array} 
\right) 
\eeq 
where $\Delta = \vec d_{ind} \cdot \vec E$ with  
$\vec d_{ind}$ being the transition matrix element between 
the $|+\rangle$ and $|-\rangle$ states induced by the electric 
field. The two new energy eigenvalues are 
\beq 
E_{1,2} = \frac{1}{2} (E_+ + E_-) \pm 
\sqrt{\frac{1}{4} (E_+ - E_-)^2 + \Delta ^2} 
\eeq 
For $E_+ \simeq E_-$ one has 
\beq 
E_{1,2} \simeq \frac{1}{2} (E_+ + E_-) \pm |\Delta | \; ; 
\eeq 
i.e., the energy shift appears to be linear in $\vec E$: 
\beq 
\Delta E = E_2 - E_1 = 2 |\vec d_{ind} \cdot \vec E| 
\label{FAKELINEAR} 
\eeq   
Yet with $\vec E$ being sufficiently small one arrives at 
$4(\vec d_{ind}\cdot \vec E)^2 \ll (E_+ - E_-)^2$ and therefore 
\beq 
E_1 \simeq E_- + \frac{(\vec d_{ind}\cdot \vec E)^2}{E_- - E_+} \; , \; 
E_2 \simeq E_+ - \frac{(\vec d_{ind}\cdot \vec E)^2}{E_- - E_+} \; ; 
\eeq
i.e., the induced energy shift is {\em quadratic} in $\vec E$ 
rather than  
linear and therefore does {\em not} imply T violation! The distinction 
between an EDM and an induced electric dipole moment is somewhat 
subtle -- yet it can be established in an unequivocal way by 
probing for a linear Stark effect with weak electric fields. A more 
careful look at Eq.(\ref{FAKELINEAR}) already indicates that. 
For the energy shift stated there does not change under 
$\vec E \Rightarrow - \vec E$ as it should for an EDM which also 
violates parity! 

The data for neutrons read: 
\beq 
d_n = \left\{   
\begin{array}{l} 
(-3 \pm 5)\cdot 10^{-26} \; ecm \; \; \; \; \; {\rm ILL} \\ 
(2.6 \pm 4 \pm 1.6)\cdot 10^{-26} \; ecm \; \; \; \; \; {\rm LNPI}  
\end{array}
\right. 
\label{EDMNEUT}
\eeq 
These numbers and the experiments leading to them are very 
impressive: 
\begin{itemize} 
\item 
One uses neutrons emanating from a reactor and 
subsequently cooled down 
to a temperature of order $10^{-7}$ eV. This is comparable to the 
kinetic energy a neutron gains when dropping 1 m in the 
earth's gravitational field. 
\item 
Extrapolating the ratio between the neutron's radius 
-- $r_N \sim 10^{-13}$ cm -- with its EDM of no more than 
$10^{-25}$ ecm to the earth's case, one would say that it 
corresponds to a situation where one has searched for a displacement 
in the earth's mass distribution of order $10^{-12} \cdot r_{earth} 
\sim 10^{-3} {\rm cm} = 10$ microns!  
\end{itemize} 

A truly dramatic increase in sensitivity for the 
{\em electron's} EDM has 
been achieved over the last few years: 
\beq 
d_e = (-0.3 \pm 0.8) \cdot 10^{-26} \; \; e \, cm 
\label{EDMEL}
\eeq 
This quantity is searched for through measuring electric dipole 
moments of {\em atoms}. At first this would seem to be a 
losing proposition theoretically: for according to Schiff's 
theorem an atom when placed inside an external electric 
field gets deformed in such a way that the electron's EDM is 
completely shielded; i.e., $d_{atom} = 0$. This theorem 
holds true in the nonrelativistic limit, yet is vitiated by relativistic 
effects. Not surprisingly the latter are particularly large for 
heavy atoms; one would then expect the electron's 
EDM to be only partially shielded: $d_{atom} = S\cdot d_e$ with 
$S < 1$. Yet amazingly -- and highly welcome of 
course -- the electron's EDM can actually get magnified by two to three 
orders of magnitude in the atom's electric dipole moment; for 
Caesium one has \cite{BERN} 
\beq 
d_{Cs} \simeq 100 \cdot d_e 
\eeq
This enhancement factor is the theoretical reason behind the 
greatly improved sensitivity for $d_e$ as expressed through 
Eq.(\ref{EDMEL}); the other one is experimental, namely the great 
strides made by laser technology applied to atomic physics. 

The quality of the number in Eq.(\ref{EDMEL}) can be illustrated 
through a comparison with the electron's magnetic moment. 
The electromagnetic form factor $\Gamma _{\mu}(q)$ 
of a particle like the electron evaluated at momentum 
transfer $q$ contains two tensor terms: 
\beq 
d_{atom} =  
\frac{1}{2m_e}\sigma _{\mu \nu} q^{\nu}\left[ i F_2(q^2) + 
F_3(q^2) \gamma _5 \right]  + ... 
\eeq  
In the nonrelativistic limit one finds for the EDM: 
\beq 
d_e = - \frac{1}{2m_e} F_3(0) 
\eeq 
On the other hand one has 
\beq 
\frac{1}{2}(g-2) = \frac{1}{e} F_2(0) 
\eeq 
The {\em precision} with which $g-2$ is known for the 
electron -- $\delta [(g-2)/2] \simeq 10^{-11}$ -- 
(and which represents one of the great success stories of 
field theory) corresponds to an {\em uncertainty} in the electron's 
{\em magnetic} moment 
\beq 
\delta \left[ \frac{1}{2m_e} F_2(0)\right] \simeq 
2\cdot 10^{-22} \; \; e\, cm
\eeq 
that is several orders of magnitude larger than the bound on its 
EDM! 

Since the EDM is, as already indicated above, described by a 
dimension-five operator in the Lagrangian 
\beq 
{\cal L}_{EDM} = - \frac{i}{2} d 
\bar \psi \sigma _{\mu \nu}\gamma _5 \psi F^{\mu \nu} 
\eeq
with $F^{\mu \nu}$ denoting the electromagnetic field strength 
tensor, one can calculate $d$ within a given theory of CP violation 
as a finite quantity.  Within the KM ansatz one finds that the 
neutron's EDM is zero for all practical purposes 
\footnote{I ignore here the Strong CP Problem, which is 
discussed in the next section.}:  
\beq 
\left. d_N\right| _{KM} < 10^{-30} \; \; e\, cm 
\eeq
and likewise for $d_e$. Yet again that is due to very specific features 
of the KM mechanism and the chirality structure of the Standard 
Model. In alternative models -- where CP violation enters 
through {\em right}-handed currents or a non-minimal 
Higgs sector (with or without involving SUSY) -- one finds 
\beq 
\left. d_N\right| _{New\; Physics} \sim 
10^{-27} - 10^{-28} \; \; e\, cm 
\eeq  
as reasonable benchmark figures. 

%%%%%%%%%%%%%%%%%%%%%%%%%%%%%%%%%%%%%%%%%%%
\section{The Strong CP Problem}
%%%%%%%%%%%%%%%%%%%%%%%%%%%%%%%%%%%%%%%%%%%%%%%%
%%%%%%%%%%%
\subsection{The Problem}
%%%%%%%%%%

It is often listed among the attractive features of QCD that it `naturally' conserves 
baryon number, flavour, parity and CP. Actually the last two points 
are not quite true,   
which had been overlooked for 
some time 
although it can be seen in different ways 
\cite{PECCEI}. Consider 
\beq 
{\cal L}_{eff} = \sum _q \bar q \left( i \not{D} - m_q\right) q 
- \frac{1}{4} G \cdot G + 
\frac{\theta g_S^2}{32 \pi ^2} 
G\cdot \tilde G 
\label{QCDCP} 
\eeq 
where $D_{\mu}$, 
$G$ and 
$\tilde G$ denote the covariant derivative, the 
gluon field strength tensor and its dual, respectively:
\beq 
D_{\mu} = \partial _{\mu} + i g_S A^i_{\mu}t^i 
\eeq  
\beq 
G_{\mu \nu} \equiv G_{\mu \nu}^i t^i \; \; , \; \; 
G_{\mu \nu} ^i = \partial _{\mu} A^i_{\nu} - 
\partial _{\nu} A^i_{\mu} + g_S if_{ijk}A^j_{\mu}A^k_{\nu} \; \; 
, \; \; \tilde G_{\mu \nu} \equiv \frac{i}{2} 
\epsilon _{\mu \nu \alpha \beta} G_{\alpha \beta} 
\eeq  
\beq 
G\cdot G \equiv G_{\mu \nu} G_{\mu \nu} \; \; , 
\; \; 
G\cdot \tilde G \equiv G_{\mu \nu} \tilde G_{\mu \nu} 
\eeq
In adding the operator $G \cdot \tilde G$ 
to the usual QCD Lagrangian we have followed a general 
tenet of quantum field theory:  any Lorentz scalar gauge invariant 
operator of dimension four has to be included in the Lagrangian 
unless there is a specific reason -- in particular a symmetry 
requirement -- that enforces its absence. For otherwise 
radiative corrections will resurrect such an operator with 
a (logarithmically) divergent coefficient! 

Such an operator exists also in an 
abelian gauge theory like QED where the field strength tensor 
takes on a simpler form: $F_{\mu \nu} = 
\partial _{\mu} A_{\nu} - \partial _{\nu} A_{\mu}$.  One 
then finds 
\beq 
F_{\mu \nu} \tilde F_{\mu \nu} = 
\partial _{\mu} K^{QED}_{\mu} \; , 
\; \; K^{QED}_{\mu} = 2 \epsilon _{\mu \alpha \beta \gamma } 
A_{\alpha}  \partial _{\beta} A_{\gamma} \; ; 
\eeq 
i.e., this extra term can be reduced to a total 
derivative which is usually dropped without further ado 
as physically irrelevant. 

For nonabelian gauge theories one obtains 
\beq 
G_{\mu \nu}\tilde G_{\mu \nu} = \partial _{\mu} K_{\mu} \; , 
\; \; K_{\mu} = 2 \epsilon _{\mu \alpha \beta \gamma } 
\left( A_{\alpha}  \partial _{\beta} A_{\gamma} + 
\frac{2}{3} i g_S A_{\alpha}A_{\beta} A_{\gamma} 
\right) \; . 
\label{KTOPO} 
\eeq 
The extra term is still a total derivative and our first reaction 
would be to just drop it for that very reason. 
Alas this time we would be wrong in doing so!  
Let us recapitulate the usual argument. If a term in the 
Lagrangian can be expressed as a total divergence 
like $\partial _{\mu}K_{\mu}$ than its contribution to the 
{\em action} which determines the dynamics can be 
expressed as the integral of the current $K$ over a surface 
at infinity. Yet with physical observables having to 
vanish rapidly at infinity to yield finite values for energy etc., 
such integrals are expected to yield zero. The field strength indeed 
goes to zero at infinity -- but not necessarily the gauge potentials 
$A_{\mu}$! The field configuration at large 
space-time distances has to approach that of a 
ground state for which $G_{\mu \nu} = 0$ holds. Yet the 
latter property 
does not suffice to define the ground state {\em uniquely}: 
it 
still allows ground states to differ by pure gauge configurations 
which obviously satisfy $G_{\mu \nu} = 0$. 
This is also true for abelian gauge 
theories, yet remains without dynamical significance. The structure  
of nonabelian gauge theories on the other hand is much more 
complex and they possess an infinity of {\em inequivalent}  
states defined by $G_{\mu \nu}=0$ 
\cite{REBBI}. Their differences can be expressed 
through topological characteristics of their gauge field 
configurations.  To be more precise: these states can be characterised 
by an integer, the so-called {\em winding number}; accordingly they 
are denoted by $|n \rangle$. They are {\em not} gauge 
invariant. Not surprisingly 
then transitions between states 
$|n_1\rangle$ and $|n_2\rangle $ with $n_1 \neq n_2$ 
can take place. The net change $\Delta n$ in winding number between 
$t=-\infty$ and $t=\infty$ is described by their $K$ charge, 
the space integral of the zeroth component of the current 
$K_{\mu}$ defined in Eq.(\ref{KTOPO}).  
A gauge invariant state is constructed as a 
linear superposition of the states $|n \rangle$ labeled by a real 
parameter $\theta$  
\beq 
|\theta \rangle = \sum _n e^{-i n\theta }|n\rangle \; . 
\eeq 
One easily shows that for a 
gauge invariant operator $O_{g.inv.}$ 
$\matel{\theta}{O_{g.inv.}}{\theta ^{\prime}}=0$ 
holds if $\theta \neq \theta ^{\prime}$. 
We thus see that the state space of QCD consists of 
{\em disjoint} sectors built up from ground states $|\theta \rangle $.  

For vacuum-to-vacuum transitions one then finds 
\beq 
\langle \theta _+|\theta _-\rangle = 
\sum _{n,m} e^{i\theta (m - n)}\langle m_+|n_-\rangle = 
\sum _{\Delta n}e^{i \Delta n \theta}
\sum _n \langle (n+\Delta n)_+| n_- \rangle \;  , 
\; \; \Delta n = m - n \; , 
\eeq 
which can be reformulated in the path integral formalism 
\beq 
\langle \theta _+|\theta _-\rangle = 
\sum _{\Delta n} \sum _{\rm fields} 
e^{i \int d^4x {\cal L}_{eff}} 
\delta (\Delta n - \frac{g_S^2}{16 \pi ^2} \int d^4x G \cdot 
\tilde G) \; ;  
\eeq 
i.e., the $G \cdot \tilde G$ term in Eq.(\ref{QCDCP}) acts as a 
Lagrangian multiplier implementing the change in winding 
number $\Delta n$. 

This is easily generalized to any transition amplitude, and 
the situation can be summarized as follows: 
\begin{itemize}
\item 
There is an infinity set of {\em inequivalent} groundstates in QCD 
labeled by a real parameter $\theta$. 
\item 
The dependance of observables on $\theta$ can be determined by 
employing the {\em effective} Lagrangian of 
Eq.(\ref{QCDCP}). 

\end{itemize} 

The problem with this additional term in the Lagrangian is 
that $G\cdot \tilde G$ -- in contrast to $G \cdot G$ -- 
violates both parity and time reversal invariance!   
This is best seen by expressing $G_{\mu \nu}$ and its dual 
through the colour electric and colour magnetic fields $\vec E$ 
and $\vec B$, respectively: 
\begin{eqnarray} 
G\cdot G \propto |\vec E|^2 +|\vec B|^2 
&\stackrel{{\bf P}, {\bf T}}{\Longrightarrow}& 
|\vec E|^2 +|\vec B|^2 \\  
G \cdot \tilde G \propto 2\vec E \cdot \vec B 
&\stackrel{{\bf P}, {\bf T}}{\Longrightarrow}& 
- 2\vec E \cdot \vec B
\end{eqnarray}   
since  
\begin{eqnarray} 
\vec E \stackrel{{\bf P}}{\Longrightarrow} - \vec E 
\; \; \; &,& \; \; \;      
\vec B \stackrel{{\bf P}}{\Longrightarrow}  \vec B \\    
\vec E \stackrel{{\bf T}}{\Longrightarrow}  \vec E  
\; \; \; &,& \; \; \;      
\vec B \stackrel{{\bf T}}{\Longrightarrow} - \vec B \; ; 
\end{eqnarray}  
i.e., for $\theta \neq 0$ neither parity nor time reversal invariance 
are fully conserved by QCD. This is the {\em Strong CP Problem}. 

The problem which resides in gluodynamics spreads into 
the quark sector through the `chiral'  anomaly 
\cite{ABJANOM}: 
\beq 
\partial _{\mu} J_{\mu}^5 = 
\partial _{\mu} \sum _q \bar q_L \gamma _{\mu}q_L = 
\frac{g_S^2}{32\pi ^2} G \cdot \tilde G \neq 0 \; ; 
\label{TRIANOM} 
\eeq   
i.e., the axial current of massless quarks, which is conserved 
{\em classically}, ceases to be so on the quantum level 
\footnote{This is why it is called an anomaly.}. This 
chiral anomaly is also called the `triangle' anomaly because it 
is produced by a diagram with a triangular fermion loop.  

There are two further aspects to the anomaly expressed in 
Eq.(\ref{TRIANOM}): 
\begin{itemize}
\item 
The anomaly actually solves one long standing puzzle of 
{\em strong} dynamics, the `U(1) Problem': 
In the limit of massless 
$u$ and $d$ quarks QCD would appear to have a {\em global} 
$U(2)_L \times U(2)_R$ invariance. While the vectorial part 
$U(2)_{L+R}$ is a manifest symmetry, the axial part 
$U(2)_{L-R}$ is spontaneously realized leading to the emergence 
of four Goldstone bosons. In the presence of quark masses those 
bosons acquire a mass as well. The pions readily play the part, 
but the $\eta$ meson does not 
\footnote{An analogous discussion can be given with $s$ quarks included. The spontaneous breaking of the global 
$U(3)_{L-R}$ symmetry leads to the existence of nine 
Goldstone bosons, yet the $\eta ^{\prime}$ meson is far 
too heavy for this role.}! Yet from the anomaly one infers that 
due to quantum corrections the axial $U(1)_{L-R}$ was never there 
in the first place even for massless quarks: therefore only 
three Goldstone bosons are predicted, the pions! 
\item 
On the other hand the anomaly aggravates the 
Strong CP Problem 
when electroweak dynamics are included. For the quarks 
acquire their masses from the Higgs mechanism driving 
the phase transition 
\beq 
SU(2)_L \times U(1) \leadsto U(1)_{QED} 
\eeq 
The resulting quark mass ${\cal M}_{quark}$ matrix cannot be expected to be diagonal and Hermitian ab initio; 
it will have to be diagonalized through  
chiral rotations of the quark fields: 
\beq 
{\cal M}_{quark}^{diag} = 
U_R^{\dagger} {\cal M}_{quark} U_L
\eeq  
Exactly because of the axial anomaly this induces an additional 
term in the Lagrangian of the Standard Model: 
\beq 
{\cal L}_{SM,eff} = {\cal L}_{QCD} + {\cal L}_{SU(2)_L \times U(1)} 
+ \frac{\bar \theta g_S^2}{32 \pi ^2} G \cdot \tilde G 
\eeq 
where 
\beq 
\bar \theta = \theta _{QCD} + \Delta \theta _{EW}\; , 
\; \;  \Delta \theta _{EW} = {\rm arg \; det} U_R^{\dagger} U_L 
\eeq 
Since the electroweak sector has to contain sources of 
CP violation other than $G \cdot \tilde G$, the second term 
in $\bar \theta$ has no a priori reason to vanish. 
\end{itemize}

%%%%%%%%%
\subsection{The Neutron Electric Dipole Moment}
%%%%%%%%%%%
Since the gluonic operator $G \cdot \tilde G$ does not change 
flavour one suspects right away that its most noticeable impact 
would be to generate an electric dipole moment (EDM) 
for neutrons. 
This is indeed the case, yet making this connection more concrete 
requires a more sophisticated argument. In the context of the Strong CP Problem one views the neutron 
EDM -- $d_N$ -- as due to the photon coupling to a virtual 
proton or pion in a fluctuation of the neutron: 
$ 
n \; \Longrightarrow \; p^* \pi ^* \; \Longrightarrow  \; n
$.  
Of the two effective pion nucleon couplings in this one-loop process 
one is produced by ordinary strong forces and conserves 
P, T and CP; 
the other one is induced by $G\cdot \tilde G$. The first 
estimate was obtained by Baluni 
\cite{BALUNI} in a nice paper using bag model 
computations of the transition amplitudes between 
the neutron and its excitations: 
$d_N \simeq 2.7 \cdot 10^{-16} \; \bar \theta \; e cm$. 
In \cite{CREWTHER} 
chiral perturbation theory instead was employed: 
$d_N \simeq 5.2 \cdot 10^{-16} \; \bar \theta \; e cm$. 
More recent estimates yield values in roughly the same range: 
$d_N \simeq (4\cdot 10^{-17} \div 2\cdot 10^{-15}) \bar \theta 
\; e\; cm$ \cite{PECCEI}. Hence 
\beq 
d_N \sim {\cal O}(10^{-16} \bar \theta ) \; \; e \; cm 
\eeq 
and 
one infers  
\beq 
d_N \leq 1.1 \cdot 10^{-25} \; \; e \; cm \; \;  \; 95\% \; C.L. 
\; \; \; \leadsto \; \; \; 
\bar \theta < 10^{-9 \pm 1} 
\label{THETABOUND}
\eeq 
Although $\theta _{QCD}$ is a QCD parameter it might not necessarily 
be of order unity; nevertheless its truly tiny size begs for 
an explanation. The only kind of explanation that is usually 
accepted as `natural' in our community is one based on 
symmetry. Such an explanation has been put forward based 
on a so-called Peccei-Quinn symmetry. Yet before we start speculating too wildly, we want 
to see whether there are no more mundane explanations.   

%%%%%%%%%%%%%
\subsection{Are There Escape Hatches?}
%%%%%%%%%%%%
One could argue that the Strong CP Problem is fictitious 
using one of two lines of reasoning: 
\begin{itemize}
\item 
Being the coefficient of a dimension-four operator 
$\bar \theta$  
can in general \footnote{Exceptions will be mentioned below.} 
be renormalized to any value, including zero. 
This is technically correct; however $\theta \leq {\cal O}(10^{-9})$ 
is viewed as highly `unnatural':  
\begin{itemize} 
\item 
A priori there is no reason why $\theta _{QCD}$ and 
$\Delta \theta _{EW}$ should practically vanish. 
\item 
Even if $\theta _{QCD} = 0 = \Delta \theta _{EW}$ were set 
{\em by fiat} quantum corrections to $\Delta \theta _{EW}$ are 
typically much larger than $10^{-9}$ and ultimately actually 
infinite. 
\item 
To expect that $\theta _{QCD}$ and $\Delta \theta _{EW}$ 
cancel as to render $\bar \theta$ sufficiently tiny would 
require fine tuning of a kind which would have to strike even 
a skeptic as unnatural. For $\theta _{QCD}$ reflects dynamics 
of the strong sector and $\Delta \theta _{EW}$ that of the 
electroweak sector. 
\item 
In models where CP symmetry is realized in a spontaneous 
fashion one has $\theta _{tree} = 0$ and 
\beq 
\bar \theta = \delta \theta _{ren} 
\eeq 
turns out to be a finite and calculable quantity that has no 
apparent reason to be smaller than, say, $10^{-4}$. 
\end{itemize}
\item 
A more respectable way out is provided by the 
following observation: if one of the quark masses 
vanishes   
the resulting chiral invariance would remove any 
$\bar \theta$ dependance of observables by rotating it 
-- through the anomaly -- into the quark mass matrix with its zero 
eigenvalue. However most authors argue quite forcefully that 
neither the up quark nor a fortiori the down quark mass can 
vanish \cite{PECCEI}: 
\beq 
m_d (1\; \GeV) > m_u (1\; {\GeV}) \simeq 5\; \MeV 
\eeq 
where the notation shows that one has to use the running mass 
evaluated at a scale of 1 GeV. 
\end{itemize}

%%%%%%%%%%
\subsection{Peccei-Quinn Symmetry}
%%%%%%%%%%%
As just argued $\bar \theta \leq {\cal O}(10^{-9})$ 
could hardly come about accidentally; an organizing principle had 
to arrange various contributions and corrections in such a way as to 
render the required cancellations.  There is the general 
philosophy that such a principle has to come from an underlying 
symmetry. We have already sketched such an approach: a global chiral invariance allows to rotate the 
dependance on $\bar \theta$ away; we failed however in our 
endeavour because this symmetry is broken by $m_q \neq 0$. Is 
it possible to invoke some other variant of chiral symmetry for 
this purpose even if it is spontaneously broken?  
One 
particularly intriguing ansatz is to re-interprete a physical 
quantity that is conventionally taken to be a constant as a 
dynamical degree of freedom that relaxes itself to a certain 
(desired) value in response to forces acting upon it. One early 
example is provided by the original Kaluza-Klein theory 
\cite{KALUZA} 
invoking a six-dimensional `space'-time manifold: two compactify 
dynamically and thus lead to the quantization of electric and 
magnetic charge. 

Something similar has been suggested by 
Peccei and Quinn \cite{PQ}. They augmented the Standard Model 
by a global $U(1)_{PQ}$ symmetry -- now referred to as 
the Peccei-Quinn symmetry -- that is axial. The spontaneous breaking of this symmetry gives rise to a Goldstone boson -- named the axion -- 
with zero mass on the Lagrangian level. Goldstone couplings 
to other fields usually have to be derivative, i.e. involve 
$\partial _{\mu}a(x)$, but not $a(x)$ directly. Since 
$U(1)_{PQ}$ is axial it exhibits a triangle anomaly again; 
this is implemented in the effective Lagrangian by having a 
term linear in the axion field coupled to $G\cdot \tilde G$: 
\beq 
{\cal L}_{eff} = {\cal L}_{SM} + \frac{g_S^2\bar \theta}{32\pi ^2} 
G \cdot \tilde G + \frac{g_S^2}{32\pi ^2}\frac{\xi}{\Lambda _{PQ}} 
aG \cdot \tilde G - \frac{1}{2}\partial _{\mu} a \partial _{\mu} a 
+ {\cal L}_{int}(\partial _{\mu} a,\psi ) 
\label{LAGAXION}
\eeq
The size of the parameters $\Lambda _{PQ}$ and 
$\xi$ and the form of 
${\cal L}_{int}(\partial _{\mu} a,\psi )$ describing the 
(purely derivative) coupling of the axion field to other fields 
$\psi$ 
depend on how the Peccei-Quinn symmetry is specifically 
realized. 

The term $a G \cdot \tilde G$ represents an {\em explicit} 
breaking of the $U(1)_{PQ}$ symmetry. This gives 
rise to an axion mass. Yet the primary role of 
$a G \cdot \tilde G$ is to make the $ G \cdot \tilde G$ term 
disappear from the effective Lagrangian. $U(1)_{PQ}$ 
invariance being realized {\em spontaneously} means 
that $a(x)$ acquires a vacuum expectation value (=VEV) 
$\langle a\rangle$; the physical axion excitations  
are then described by the shifted field 
$ 
a_{phys}(x) = a(x) - \langle a \rangle 
$  
and one rewrites Eq.(\ref{LAGAXION}) as follows: 
\beq 
{\cal L}_{eff} = {\cal L}_{SM} + \frac{g_S^2}{32\pi ^2} 
\bar \theta  G \cdot \tilde G - 
\frac{1}{2}\partial _{\mu} a_{phys} \partial _{\mu} a_{phys}  
+\frac{g_S^2}{32\pi ^2}\frac{\xi}{\Lambda _{PQ}} 
a_{phys} G \cdot \tilde G 
+ {\cal L}_{int}(\partial _{\mu} a_{phys},\psi ) 
\eeq 
where now 
\beq 
\bar \theta  = \theta _{QCD} + 
{\rm arg\, det} U_R^{\dagger} U_L - 
\frac{\langle a\rangle }{f_a} \; , 
\; \; f_a = \frac{\Lambda _{PQ}}{\xi}\; ; 
\eeq 
i.e., the size of the 
coefficient of the $G \cdot \tilde G$ operator is 
determined by the VEV of the axion field. 

The term $a G \cdot \tilde G$ generates an effective potential 
for the axion field; its minimum defines the ground state: 
\beq 
\langle \frac{\partial V_{eff}}{\partial a}\rangle \equiv 
- \frac{\xi }{\Lambda _{PQ}} \frac{g_S^2}{32 \pi ^2} 
\langle G  \cdot \tilde G \rangle = 0 
\eeq 
That means that the term $a G \cdot \tilde G$ -- the reincarnation 
of the anomaly -- singles out one of the previously degenerate 
$\bar \theta$ states as the true ground state. This is not 
surprising since $a G \cdot \tilde G$ is {\em not} invariant 
under $U(1)_{PQ}$. It is a pleasant surprise, though, that 
for this lowest energy state $G\cdot \tilde G$ settles into a 
{\em vanishing} expectation value thus banning 
the Strong CP Problem {\em dynamically}. 

%%%%%%%%%
\subsection{The Dawn of Axions -- and Their Dusk?}
%%%%%%%%%%%
Rather than ending here the story contains another twist or 
two. The breaking of $U(1)_{PQ}$ gives rise to a Nambu-Goldstone 
boson. Actually the axion is, as already mentioned, a 
pseudo-Nambu-Goldstone 
boson; for it acquires a mass due to the anomaly: 
\beq 
m_a^2 \sim \frac{G \cdot \tilde G}{\Lambda _{PQ}^2} 
\sim {\cal O}\left(  \frac{\Lambda _{QCD}^4}{\Lambda _{PQ}^2} 
\right) 
\eeq 
Since one expects on general grounds 
$\Lambda _{PQ} >> \Lambda _{QCD}$ one is dealing with a 
very light boson. The question is how light would the 
axion be. 

The electroweak scale $v_{EW} = 
\left( \sqrt{2}G_F\right) ^{-\frac{1}{2}} \simeq 250$ GeV 
provides the discriminator for two scenarios: 
\begin{itemize} 
\item 
\beq 
\Lambda _{PQ} \sim v_{EW} 
\eeq 
In that case axions can or even should be seen in accelerator based 
experiment. Such scenarios are referred to as {\em visible} axions. 
\item 
\beq 
\Lambda _{PQ} \gg v_{EW} 
\eeq 
Such axions could not be 
found in accelerator based experiments; therefore they are 
called {\em invisible} scenarios. Yet that does not mean that 
they necessarily escape detection! For they could be of 
great significance for the formation of stars, whole galaxies 
and even the universe. 

\end{itemize} 

%%%%%%%%%
\subsubsection{Visible Axions}
%%%%%%%%%%%
The simplest scenario involves two $SU(2)_L$ doublet Higgs fields 
that possess opposite hypercharge 
\footnote{In the Standard Model the Higgs doublet and its 
charge conjugate fill this role.}. 
They also carry a $U(1)$ 
charge {\em in addition} to the hypercharge; this second (and 
global) $U(1)$ is identified with the PQ symmetry, and the axion is 
its pseudo-Nambu-Goldstone boson. 
The anomaly  induces a mass for the axion: 
\beq 
m_a \simeq \frac{m_{\pi}F_{\pi}}{v}N_{fam} 
\left(  x + \frac{1}{x} \right)  
\frac{\sqrt{m_um_d}}{(m_u + m_d)} \simeq 
25 \; N_{fam} \left(  x + \frac{1}{x} \right) \; \; {\rm KeV} 
\eeq 
where $N_{fam}$ denotes the number of families. 
Such an axion is almost certainly much lighter than the 
pion. Depending on the axion's mass two cases have to be 
decided:
\begin{itemize}
\item 
If $m_a > 2 m_e$, the axion decays very rapidly into electrons and 
positrons: 
\beq 
\tau (a \ra e^+ e^-) \simeq 4 \cdot 10^{-9}
\left( \frac{1\; {\rm MeV}}{m_a} \right) 
\frac{x^2 \; {\rm or} \; 1/x^2}{\sqrt{1- \frac{4m_e^2}{m_a^2}}} \; \; {\rm sec} 
\eeq 
\item 
If on the other hand $m_a < 2 m_e$ then the axion decays 
fairly slowly into two photons:  
\beq 
\tau (a \ra \gamma \gamma ) \sim 
{\cal O}\left( \frac{100 \; {\rm KeV}}{m_a}  
\right) \; \; {\rm sec} 
\eeq 
\end{itemize} 
I have presented here a very rough sketch of scenarios 
with visible axions since we can confidently declare that they 
have been  
ruled out experimentally.  They have been looked for in beam 
dump experiments -- without success. Yet the more telling 
blows have come from searches in rare decays:
\begin{itemize}
\item 
For long-lived axions -- $m_a < 2 m_e$ -- one expects a 
dominating contribution to $K^+ \ra \pi ^+ + \; nothing$ from 
\beq 
K^+ \ra \pi ^+ + a 
\label{KTOAXION}
\eeq 
with the axion decayig well outside the detector. For the two-body 
kinematics of Eq.(\ref{KTOAXION}) one has very tight bounds from 
published data \cite{ADLER}: 
\beq 
BR(K^+ \ra \pi ^+ X^0) < 5.2 \cdot 10^{-10} \; \; \; 90\%\; C.L.
\eeq
for $X^0$ being a practically massless and noninteracting particle. 
Theoretically one would expect:
\beq 
\left. BR(K^+ \ra \pi ^+ a) \right|_{theor} \sim 
3\cdot 10^{-5} \cdot \left( x + 1/x \right) ^{-2} 
\label{KTOAXTH}
\eeq
Although Eq.(\ref{KTOAXTH}) does not represent a precise prediction 
the discrepancy between expectation and observation is 
conclusive. 
\item 
One arrives at the same conclusion that 
{\em long-lived visible} axions 
do not exist from the absence of {\em quarkonia} decay 
into them: neither $J/\psi \ra a\gamma$ nor 
$\Upsilon \ra a \gamma$ has been seen. 
\item 
The analysis is a bit more involved for 
{\em short-lived} axions -- $m_a > 2 m_e$. Yet again 
their absence has been established through a combination 
of experiments. Unsuccessful searches for 
\beq 
\pi ^+ \ra a e^+ \nu 
\eeq  
figure prominently in this endeavour. Likewise the 
{\em absence} of axion driven 
{\em nuclear de-excitation} has been 
established on a level that appears to be conclusive 
\cite{NUCLEAR}. 
%%%%%%%%%
\subsubsection{Invisible Axions}
%%%%%%%%
Invisible axion scenarios involve a complex 
scalar field $\sigma$ that 
(i) is an $SU(2)_L$ {\em singlet}, 
(ii) 
carries a PQ charge and 
(iii)  
possesses a huge VEV $\sim \Lambda _{PQ} >> v_{ew}$. 
\end{itemize}
The reasons underlying those requirements should be obvious; 
they can be realized in two distinct (sub-)scenarios: 
\begin{enumerate}
\item 
Only presumably very heavy new quarks carry a PQ charge. This 
situation is referred to as {\em KSVZ} axion \cite{KSVZAX}. The minimal 
version can do with a single $SU(2)_L$ Higgs doublet. 
\item 
Also the known quarks and leptons carry a PQ charge. Two 
$SU(2)_L$ Higgs doublets are then required in addition to 
$\sigma$. The fermions do not couple directly to 
$\sigma$, yet become sensitive to PQ breaking through 
the Higgs potential. This is referred to as {\em DFSZ} axion 
\cite{DFSZAX}. 
\end{enumerate} 
From current algebra one infers for the axion mass in either 
case 
\beq 
m_a \simeq 0.6 \; {\rm eV} \cdot \frac{10^7 \; {\rm GeV}}{f_a} 
\eeq 
The most relevant coupling of such axions is to two photons 
\beq 
{\cal L}( a \ra \gamma \gamma ) = 
- \tilde g_{a\gamma \gamma} \frac{\alpha}{\pi} 
\frac{a(x)}{f_a} \vec E \cdot \vec B \; ,  
\eeq 
where $\tilde g_{a\gamma \gamma}$ is a model dependant 
coefficient of order unity. 

Axions with such tiny masses have lifetimes easily in excess 
of the age of the universe. Also their couplings to other 
fields are so minute that they would not betray their 
presence -- hence their name {\em invisible} axions -- 
under ordinary circumstances! 
Yet in astrophysics and cosmology more favourable 
extra-ordinary conditions can arise. 

Through their couplings to electrons axions would provide 
a cooling mechanism to {\em stellar evolution}. Not 
surprisingly their greatest impact occurs for the lifetimes of 
red giants and the supernovae like SN 1987a. The actual 
bounds depend on the model -- whether it is a 
KSVZ or DFSZ axion -- but relatively mildly only. 
Altogether astrophysics tells us that {\em if} 
axions exist one has 
\beq 
m_a < 3 \cdot 10^{-3} \; \; {\rm eV} 
\eeq 

{\em Cosmology} on the other hand provides us with  a 
{\em lower} bound through a very intriguing line of 
reasoning. At temperatures $T$ above $\Lambda _{QCD}$ 
the axion is massless and all values of 
$\langle a(x)\rangle$ are equally likely. For 
$T \sim 1$ GeV the anomaly induced potential 
turns on driving $\langle a(x)\rangle$ 
to a value as to yield $\bar \theta =0$ at the new potential 
minimum. The energy stored previously as {\em latent heat} 
is then released into axions oscillating around its new VEV. 
Precisely because of the invisible axion's couplings are so 
immensely suppressed the energy cannot be dissipated into 
other degrees of freedom. We are then dealing with a fluid of 
axions. Their typical momenta is the inverse of their 
correlation length which in turn cannot exceed their horizon; 
one finds 
\beq 
p_a \sim \left( 10^{-6}\; {\rm sec}\right) ^{-1} 
\sim 10^{-9}\; {\rm eV} 
\eeq 
at $T \simeq 1$ GeV; i.e., the axions despite their minute 
mass form a very {\em cold} fluid and actually represent 
a candidate 
for cold dark matter. Their contribution to the density of the 
universe relative to its critical value is \cite{SIKIVIE}
\beq 
\Omega _a = \left(  \frac{0.6 \cdot 10^{-5}\; {\rm eV}}
{m_a}\right) ^{\frac{7}{6}}\cdot 
\left(  \frac{200\; {\rm MeV}}
{\Lambda _{QCD}}\right) ^{\frac{3}{4}}\cdot 
\left(  \frac{75 \; {\rm km/sec\, \cdot Mpc}}
{H_0}\right) ^2 \; ; 
\eeq 
$H_0$ is the present Hubble expansion rate. For axions not to overclose the universe one thus has to require: 
\beq 
m_a \geq 10^{-6} \; {\rm eV} 
\eeq
or 
\beq 
\Lambda _{PQ} \leq 10^{12} \; {\rm GeV} 
\eeq  
This means also that we might be existing in a bath 
of cold axions still making up a significant fraction of the 
matter of the universe. 

Ingenious suggestions have been made to search for such 
cosmic background axions. The main handle one has on them 
is their coupling to two photons. They can be detected 
by stimulating the conversion 
\beq 
{\rm axion} \; \;  \stackrel{\vec B}
{\longrightarrow} \; \;  {\rm photon} 
\eeq 
in a strong magnetic field $\vec B$: 
the second photon 
which is virtual in this process effects the interaction 
with the inhomogeneous magnetic field in the cavity.  The 
available microwave technology allows an impressive 
experimental sensitivity. No signal has been found yet, but the search continues and soon should reach a level where one has a 
good chance to see a signal \cite{LLNL}. 

%%%%%%%%
\subsection{The Pundits' Judgement}
%%%%%%%%%
The story of the Strong CP Problem is a particularly intriguing one. 
We -- like most though not all of our community -- find the 
theoretical arguments persuasive that there is a problem that 
has to be resolved. The inquiry has been based on an impressive 
arsenal of theoretical reasoning and has inspired fascinating 
experimental undertakings. 

Like many modern novels the problem -- if its is indeed 
one -- has not found any resolution. On he other hand it has 
the potential to lead the charge towards a new paradigm 
in high energy physics.

%%%%%%%%%%%%%%%%
\section{Summary on the CP Phenomenology with Light Degrees of 
Freedom}
%%%%%%%%%%%%%%%%%
To summarize our discussion up to this point: 
\begin{itemize} 
\item 
The following data represent the most sensitive probes:  
\beq 
{\rm BR}(K_L \ra \pi ^+ \pi ^-) = 2.3 \cdot 10^{-3} \neq 0 
\eeq  
\beq 
\frac{{\rm BR}(K_L \ra l^+ \nu \pi ^-)}
{{\rm BR}(K_L \ra l^- \nu \pi ^+)} \simeq 1.006 \neq 1 
\eeq 
\beq 
{\rm Re} \frac{\epsilon ^{\prime}}{\epsilon _K} = 
\left\{ 
\begin{array}{l} 
(2.3 \pm 0.7) \cdot 10^{-3} \; \; NA\, 31 \\ 
(0.6 \pm 0.58 \pm 0.32 \pm 0.18) \cdot 10^{-3} \; \; 
E\, 731 
\end{array}  
\right. 
\eeq
\beq 
{\rm Pol}_{\perp}^{K^+}(\mu ) = (-1.85\pm 3.60) \cdot 10^{-3} 
\eeq
\beq 
d_N < 12 \cdot 10^{-26} \; \; e\, cm 
\eeq   
\beq 
d_{Tl} = (1.6 \pm 5.0) \cdot 10^{-24} \; \; e\, cm 
\stackrel{theor.}{\leadsto} 
d_e = (-2.7 \pm 8.3) \cdot 10^{-27} \; \; e\, cm
\eeq 
\item 
An impressive amount of experimental ingenuity, acumen and 
commitment 
went into producing this list. We know that CP violation 
unequivocally exists in nature; it can be 
characterized by a {\em single} non-vanishing quantity: 
\beq 
{\rm Im} M_{12} \simeq 1.1 \cdot 10^{-8} \; eV \neq 0 
\eeq  
\item 
The `Superweak Model' states that there just happens to 
exist a $\Delta S=2$ interaction that is fundamental or effective -- 
whatever the case may be -- generating Im$M_{12} =$ 
Im$M_{12}|_{exp}$ while $\epsilon ^{\prime} =0$. It 
provides merely  
a {\em classification} for possible dynamical implementations 
rather than such a dynamical implementation itself.  

\item 
The KM ansatz allows us to incorporate CP violation into the 
Standard Model. Yet it does not regale us with an understanding. 
Instead it relates the origins of CP violation to central 
mysteries of the Standard Model: Why are there families? 
Why are there three of those?  What is underlying 
the observed pattern in the fermion masses? 
\item 
Still the KM ansatz succeeds in {\em accommodating} the data in an 
unforced way: $\epsilon _K$ emerges to be naturally 
small, $\epsilon ^{\prime}$ naturally tiny (once the huge 
top mass is built in), the EDM's for neutrons [electrons] 
naturally (tiny)$^2$ [(tiny)$^3$] etc. 

\end {itemize}

%%%%%%%%%%%
\section{CP Violation in Beauty Decays -- The KM Perspective}
%%%%%%%%%%%%
The KM predictions for strange decays and electric dipole 
moments given above will be subjected to sensitive tests in the 
foreseeable future. Yet there is one question that most naturally 
will come up in this context: "Where else to look?" 
I will show below that on very general grounds one has to 
conclude that the decays of beauty hadrons provide by far the 
optimal lab. Yet first I want to make some historical remarks. 

%%%%%%%%%%%
\subsection{The Emerging Beauty of $B$ Hadrons}
%%%%%%%%%%%%
%%%%%%%%%%
\subsubsection{Lederman's Paradise Lost -- and Regained!}
%%%%%%%%%%%%
In 1970 Lederman's group studying the Drell-Yan process 
\beq 
p p \ra \mu ^+ \mu ^- X
\eeq 
at Brookhaven observed a shoulder in the di-muon mass 
distribution around 3 GeV. 1974 saw the `Octobre Revolution' when 
Ting et al. and Richter et al. found a narrow resonance -- the $J/\psi$ 
-- with a mass of 3.1 GeV at Brookhaven and SLAC, respectively, and 
announced it. In 1976/77 Lederman's group working at 
Fermilab saw 
a structure -- later referred to as the Oops-Leon -- around 6 GeV, 
which then disappeared. In 1977 Lederman et al. discovered 
three resonances in the mass range of 9.5 - 10.3 GeV, the 
$\Upsilon$, $\Upsilon ^{\prime}$ and $\Upsilon ^{\prime \prime}$! 
That shows that persistence can pay off -- at least sometimes and for  
some people. 

%%%%%%%%%%
\subsubsection{Longevity of Beauty}
%%%%%%%%%
The lifetime of weakly decaying beauty quarks can be related 
to the muon lifetime 
\beq  
\tau (b) \sim \tau (\mu ) \left( \frac{m_{\mu}}{m_b} \right) ^5 
\frac{1}{9} \frac{1}{|V(cb)|^2} \sim 
3 \cdot 10^{-14} \left|  \frac{{\rm sin}\theta _C}{V(cb)}\right| ^2  
\; {\rm sec} 
\eeq  
for a $b$ quark mass of around  5 GeV; the factor 1/9 reflects 
the fact that the virtual $W ^-$ boson in $b$ quark decays can 
materialize as a $d \bar u$ or $s \bar c$ in three colours each and 
as three lepton pairs. I have ignored phase space corrections here. 
Since the $b$ quark has to decay outside its own family one would 
expect $|V(cb)| \sim {\cal O}({\rm sin}\theta _C) = 
|V(us)|$. Yet starting in 1982 data showed a considerably longer 
lifetime 
\beq 
\tau ({\rm beauty}) \sim 10^{-12} \; {\rm sec} 
\eeq 
implying 
\beq 
|V(cb)| \sim {\cal O}({\rm sin}^2\theta _C) \sim 0.05 
\eeq 
The technology to resolve decay vertices for objects of such 
lifetimes happened to have just been developed -- for charm 
studies! 

%%%%%%%%%%%
\subsubsection{The Changing Identity of Neutral $B$ Mesons}
%%%%%%%%%%%%
Speedy $B_d - \bar B_d$ oscillations were discovered by ARGUS in 
1986: 
\beq 
x_d \equiv \frac{\Delta m(B_d)}{\Gamma (B_d)} \simeq 
{\cal O}(1)
\eeq 
These oscillations can then be tracked like the decays. This 
observation was also the first evidence that top quarks had to be 
heavier than orginally thought, namely $m_t \geq M_W$. 

%%%%%%%%%%%
\subsubsection{Beauty Goes to Charm (almost always)}
%%%%%%%%%%%
It was soon found that $b$ quarks exhibit a strong preference 
to decay into charm rather than up quarks  
\beq 
\left| \frac{V(ub)}{V(cb)} \right| ^2 \ll 1 
\eeq 
establishing thus the hierarchy 
\beq 
|V(ub)|^2 \ll |V(cb)|^2 \ll |V(us)|^2 \ll 1
\eeq 

%%%%%%%%%
\subsubsection{Resume}
%%%%%%%%
We will soon see how all these observations form crucial inputs 
to the general message that big CP asymmetries should emerge in 
$B$ decays and that they (together with interesting rare decays)  
are within reach of experiments. It is for this reason that I strongly 
feel that the only appropriate name for this quantum number is 
{\em beauty}! A name like bottom would not do it justice. 

%%%%%%%%%%%%
\subsection{The KM Paradigm of Huge CP Asymmetries}
%%%%%%%%%%%% 
%%%%%%%%%%%%%%%
\subsubsection{Large Weak Phases!}
%%%%%%%%%%%%%%
The Wolfenstein representation expresses the CKM matrix as an 
expansion: 
\beq 
V_{CKM} = 
\left( 
\begin{array}{ccc} 
1 & {\cal O}(\lambda ) & {\cal O}(\lambda ^3) \\ 
{\cal O}(\lambda ) & 1 & {\cal O}(\lambda ^2) \\ 
{\cal O}(\lambda ^3) & {\cal O}(\lambda ^2) & 1 
\end {array} 
\right) 
\; \; \; , \; \; \; \lambda = {\rm sin}\theta _C 
\eeq 
The crucial element in making this expansion meaningful is the 
`long' lifetime of beauty hadrons of around 1 psec. That number 
had to change by an order of magnitude -- which is out of the 
question -- to invalidate the conclusions given below for the 
size of the weak phases. 

The unitarity condition yields 6 triangle relations: 
\beq 
\begin{array}{ccc} 
V^*(ud)V(us) + &V^*(cd)V(cs) + &V^*(td) V(ts) = 
\delta _{ds}= 0 \\
{\cal O}(\lambda ) & {\cal O}(\lambda ) & {\cal O}(\lambda ^5) 
\end{array} 
\label{TRI1} 
\eeq 
\beq 
\begin{array}{ccc} 
V^*(ud)V(cd) + &V^*(us)V(cs) + &V^*(ub) V(cb) = 
\delta _{uc}= 0 \\
{\cal O}(\lambda ) & {\cal O}(\lambda ) & {\cal O}(\lambda ^5) 
\end{array} 
\label{TRI2} 
\eeq 
\beq 
\begin{array}{ccc} 
V^*(us)V(ub) + &V^*(cs)V(cb) + &V^*(ts) V(tb) = 
\delta _{sb}= 0 \\
{\cal O}(\lambda ^4) & {\cal O}(\lambda ^2) & {\cal O}(\lambda ^2) 
\end{array} 
\label{TRI3} 
\eeq 
\beq 
\begin{array}{ccc} 
V^*(td)V(cd) + &V^*(ts)V(cs) + &V^*(tb) V(cb) = 
\delta _{ct}=0 \\
{\cal O}(\lambda ^4) & {\cal O}(\lambda ^2) & {\cal O}(\lambda ^2) 
\end{array} 
\label{TRI4} 
\eeq 
\beq 
\begin{array}{ccc} 
V^*(ud)V(ub) + &V^*(cd)V(cb) + &V^*(td) V(tb) = 
\delta _{db}=0 \\
{\cal O}(\lambda ^3) & {\cal O}(\lambda ^3) & {\cal O}(\lambda ^3) 
\end{array} 
\label{TRI5} 
\eeq 
\beq 
\begin{array}{ccc} 
V^*(td)V(ud) + &V^*(ts)V(us) + &V^*(tb) V(ub) = 
\delta _{ut}=0 \\
{\cal O}(\lambda ^3) & {\cal O}(\lambda ^3) & {\cal O}(\lambda ^3) 
\end{array}
\label{TRI6}  
\eeq 
where below each product of matrix elements I have noted 
their size in powers of $\lambda $. 

We see that the six triangles fall into three categories: 
\begin{enumerate}
\item 
The first two triangles are extremely `squashed': two sides are 
of order $\lambda $, the third one of order $\lambda ^5$ and their 
ratio of order $\lambda ^4 \simeq 2.3 \cdot 10^{-3}$; 
Eq.(\ref{TRI1}) and Eq.(\ref{TRI2}) control the situation in 
strange and charm decays; the relevant weak phases there 
are obviously tiny. 
\item 
The third and fourth triangles are still rather squashed, yet less so: 
two sides are of order $\lambda ^2$ and the third one of order 
$\lambda ^4$. 
\item 
The last two triangles have sides that are all of the same 
order, namely $\lambda ^3$. All their angles are therefore 
naturally large, i.e. $\sim$ several $\times$ $10$ degrees! Since to 
leading order in $\lambda$ one has 
\beq 
V(ud) \simeq V(tb) \; , \; V(cd) \simeq - V(us) \; , \; 
V(ts) \simeq - V(cb) 
\eeq 
we see that the triangles of Eqs.(\ref{TRI5}, \ref{TRI6}) 
actually coincide to that order. 
\end{enumerate} 
The sides of this triangle having naturally large angles are 
given by $\lambda \cdot V(cb)$, $V(ub)$ and 
$V^*(td)$; these are all quantities that control important 
aspects of $B$ decays, namely CKM favoured and disfavoured 
$B$ decays and $B_d - \bar B_d$ oscillations! 

%%%%%%%%%%%%%%%%%
\subsubsection{Different, Yet Coherent Amplitudes!}
%%%%%%%%%%%%%%%
$B^0 - \bar B^0$ oscillations provide us with two different 
amplitudes that by their very nature have to be coherent: 
\beq 
B^0 \Rightarrow \bar B^0 \ra f \leftarrow B^0 
\eeq 
On general grounds one expects oscillations to be speedy for 
$B^0 - \bar B^0$ (like for $K^0 - \bar K^0$), yet slow for 
$D^0 - \bar D^0$ 
\footnote{$T^0 - \bar T^0$ oscillations cannot 
occur since top quarks decay before they hadronize 
\cite{RAPALLO}.}. 
Experimentally one indeed finds 
\beq 
\frac{\Delta m(B_d)}{\Gamma (B_d)} = 0.71 \pm 0.06 
\label{OSCBD}
\eeq  
\beq 
\frac{\Delta m(B_s)}{\Gamma (B_s)} \geq 10  
\label{OSCBS}
\eeq
While Eq.(\ref{OSCBD}) describes an almost optimal 
situation the overly rapid pace of $B_s - \bar B_s$ 
oscillations will presumably cause experimental 
problems. 

The conditions are quite favourable also for {\em direct} 
CP violation to surface. Consider a transition amplitude 
\beq 
T(B \ra f) = {\cal M}_1 + {\cal M}_2 = 
e^{i\phi _1}e^{i\alpha _1}|{\cal M}_1| + 
e^{i\phi _2}e^{i\alpha _2}|{\cal M}_2| \; . 
\eeq    
The two partial amplitudes ${\cal M}_1$ and ${\cal M}_2$ are 
distinguished by, say, their isospin -- as it was the case for 
$K \ra (\pi \pi )_{I=0,2}$ discussed before; $\phi _1$, $\phi _2$ 
denote the phases in the {\em weak} couplings and 
$\alpha _1$, $\alpha _2$ the phase shifts due to 
{\em strong} final state interactions. For the CP conjugate reaction 
one obtains 
\beq 
T(\bar B \ra \bar f) =  
e^{-i\phi _1}e^{i\alpha _1}|{\cal M}_1| + 
e^{-i\phi _2}e^{i\alpha _2}|{\cal M}_2| \; . 
\eeq 
since under CP the weak parameters change into their 
complex conjugate values whereas the phase shifts remain 
the same; for the strong forces driving final state 
interactions conserve CP. The rate difference is then given by 
$$ 
\Gamma (B \ra f) - \Gamma (\bar B \ra \bar f) \propto 
|T(B \ra f)|^2 - |T( \bar B \ra \bar f)|^2 = 
$$
\beq 
= - 4 {\rm sin}(\phi _1 - \phi _2) \cdot 
{\rm sin}(\alpha _1 - \alpha _2) \cdot 
{\cal M}_1 \otimes {\cal M}_2  
\eeq 
For an asymmetry to arise in this way two conditions need to be 
satisfied simultaneously, namely 
\beq 
\begin{array}{l} 
\phi _1 \neq \phi _2 \\ 
\alpha _1 \neq \alpha _2
\end{array}
\eeq 
I.e., the two amplitudes ${\cal M}_1$ and ${\cal M}_2$ have to 
differ both in their weak and strong forces! The first condition 
implies (within the Standard Model) that the reaction has to 
be KM suppressed, whereas the second one require the intervention 
of nontrivial final state interactions. 

There is a large number of KM suppressed channels in $B$ 
decays that are suitable in this context: they receive significant 
contributions from weak couplings with large phases -- 
like $V(ub)$ in the Wolfenstein representation -- and there 
is no reason why the phase shifts should be small in general 
(although that could happen in some cases). 

%%%%%%%%%%%%%%%
\subsubsection{Resume}
%%%%%%%%%
Let me summarize the discussion just given and anticipate the 
results to be presented below. 
\begin{itemize}
\item 
{\em Large} CP asymmetries are {\em pre}dicted {\em with} 
confidence to occur in $B$ decays. If they are not found, there is 
no plausible deniability for the KM ansatz. 
\item 
Some of these predictions can be made with high 
{\em parametric} reliability. 
\item 
New theoretical technologies have emerged that will allow us to 
translate this {\em parametric} reliability into 
{\em numerical} precision. 
\item 
Some of the observables exhibit a high and unambiguous 
sensitivity to the presence of New Physics since we are 
dealing with coherent processes with observables depending  
{\em linearly} on New Physics amplitudes and where the 
CKM `background' is (or can be brought)  
under theoretical control. 
\end{itemize}

%%%%%%%%%%%%%%
\subsection{General Phenomenology}
%%%%%%%%%%%%
Decay rates for CP conjugate channels can be expressed as follows: 
\beq  
\begin{array} {l} 
{\rm rate} (B(t) \ra f) = e^{-\Gamma _Bt}G_f(t) \\  
{\rm rate} (\bar B(t) \ra \bar f) = 
e^{-\Gamma _Bt}\bar G_{\bar f}(t)  
\end{array} 
\label{DECGEN}
\eeq 
where CPT invariance has been invoked to assign the same lifetime 
$\Gamma _B^{-1}$ to $B$ and $\bar B$ hadrons. Obviously if 
\beq
\frac{G_f(t)}{\bar G_{\bar f}(t)} \neq 1 
\eeq 
is observed, CP violation has been found. Yet one should 
keep in mind that this can manifest itself in two (or three) 
qualitatively different ways: 
\begin{enumerate} 
\item 
\beq 
\frac{G_f(t)}{\bar G_{\bar f}(t)} \neq 1 
\; \; {\rm with} \; \; 
\frac{d}{dt}\frac{G_f(t)}{\bar G_{\bar f}(t)} =0 \; ; 
\label{DIRECTCP1}
\eeq   
i.e., the {\em asymmetry} is the same for all times of decay. This 
is true for {\em direct} CP violation; yet, as explained later, it also 
holds for CP violation {\em in the oscillations}.  
\item 
\beq 
\frac{G_f(t)}{\bar G_{\bar f}(t)} \neq 1 
\; \; {\rm with} \; \; 
\frac{d}{dt}\frac{G_f(t)}{\bar G_{\bar f}(t)} \neq 0 \; ; 
\label{DIRECTCP2}
\eeq   
here the asymmetry varies as a function of the time of decay. 
This can be referred to as CP violation {\em involving 
oscillations} \footnote{This nomenclature falls well short of 
Shakespearean standards.}. 
\end{enumerate} 

Quantum mechanics with its linear superposition principle makes 
very specific statements about the possible time dependance of 
$G_f(t)$ and $\bar G_{\bar f}(t)$; yet before going into that 
I want to pose another homework problem: 
\begin{center} 
$\spadesuit \; \; \; \spadesuit \; \; \; \spadesuit $ \\ 
{\em Homework Problem \# 4}: 
\end{center}
Consider the reaction 
\beq 
e^+ e^- \ra \phi \ra 
(\pi ^+\pi ^-)_K  (\pi ^+\pi ^-)_K 
\eeq 
Its occurrance requires CP violation. For the {\em initial} state -- 
$\phi $ -- carries {\em even} CP parity whereas the 
{\em final} state with the two $(\pi ^+\pi ^-)$ combinations 
forming a P wave must be CP {\em odd}: 
$(+1)^2 (-1)^l = -1$! Yet Bose statistics requiring identical 
states to be in a symmetric configuration would appear to 
veto this reaction; for it places the two $(\pi ^+\pi ^-)$ states 
into a P wave which is antisymmetric. What is the flaw in this 
reasoning? The same puzzle can be formulated in terms of 
\beq 
e^+ e^- \ra \Upsilon (4S) \ra B_d \bar B_d \ra 
(\psi K_S)_B (\psi K_S)_B \; . 
\eeq 
\begin{center} 
$\spadesuit \; \; \; \spadesuit \; \; \; \spadesuit $
\end{center} 
A straightforward application of quantum mechanics yields  
the general expressions 
\cite{CARTER,BS,CECILIABOOK}:
\beq 
\begin{array}{l}
G_f(t) = |T_f|^2 
\left[ 
\left( 1 + \left| \frac{q}{p}\right| ^2|\bar \rho _f|^2 \right) + 
\left( 1 - \left| \frac{q}{p}\right| ^2|\bar \rho _f|^2 \right) 
{\rm cos}\Delta m_Bt 
+ 2 ({\rm sin}\Delta m_Bt) {\rm Im}\frac{q}{p} \bar \rho _f 
\right]  \\ 
\bar G_{\bar f}(t) = |\bar T_{\bar f}|^2 
\left[ 
\left( 1 + \left| \frac{p}{q}\right| ^2|\rho _{\bar f}|^2 \right) + 
\left( 1 - \left| \frac{p}{q}\right| ^2|\rho _{\bar f}|^2 \right) 
{\rm cos}\Delta m_Bt 
+ 2 ({\rm sin}\Delta m_Bt) {\rm Im}\frac{p}{q} \rho _{\bar f}  
\right]  
\end{array}
\eeq 
The amplitudes for the instantaneous $\Delta B=1$ 
transition into a 
final state $f$ are denoted by 
$T_f = T(B \ra f)$ and $\bar T_f = T(\bar B \ra f)$ and  
\beq 
\bar \rho _f = \frac{\bar T_f}{T_f} \; \; , 
\rho _{\bar f} = \frac{T_{\bar f}}{\bar T_{\bar f}} \; \; , 
\frac{q}{p} = \sqrt{\frac{M_{12}^* - \frac{i}{2} \Gamma _{12}^*}
{M_{12} - \frac{i}{2} \Gamma _{12}}}
\eeq 
Staring at the general expression is not always very illuminating; 
let us therefore consider three very simplified limiting cases: 
\begin{itemize}
\item
$\Delta m_B = 0$, i.e. {\em no} $B^0- \bar B^0$ oscillations: 
\beq 
G_f(t) = 2|T_f|^2 \; \; , \; \; 
\bar G_{\bar f}(t) = 2|\bar T_{\bar f}|^2 
\leadsto \frac{\bar G_{\bar f}(t)}{G_{ f}(t)} = 
\left|
\frac{\bar T_{\bar f}}{T_{ f}}
\right|^2 \; \; , \frac{d}{dt}G_f (t) \equiv 0 \equiv 
\frac{d}{dt}\bar G_{\bar f} (t) 
\eeq 
This is explicitely what was referred to above as {\em direct} 
CP violation. 
\item 
$\Delta m_B \neq  0$   
and $f$ a flavour-{\em specific} final state with {\em no} 
direct CP violation; i.e., 
$T_{f} = 0 = \bar T_{\bar f}$ and $\bar T_f = T_{\bar f}$   
\footnote{For a flavour-specific mode one has in general 
$T_f \cdot T_{\bar f} =0$; the more intriguing case arises  
when one considers a transition that requires oscillations 
to take place.}: 
\beq 
\begin{array} {c} 
G_f (t) = \left| \frac{q}{p}\right| ^2 |\bar T_f|^2 
(1 - {\rm cos}\Delta m_Bt )\; \; , \; \; 
\bar G_{\bar f} (t) = \left| \frac{p}{q}\right| ^2 |T_{\bar f}|^2 
(1 - {\rm cos}\Delta m_Bt) \\ 
\leadsto 
\frac{\bar G_{\bar f}(t)}{G_{ f}(t)} = \left| \frac{q}{p}\right| ^4 
\; \; , \; \; \frac{d}{dt} \frac{\bar G_{\bar f}(t)}{G_{ f}(t)} \equiv 0  
\; \; , \; \; \frac{d}{dt} \bar G_{\bar f}(t) \neq 0 \neq 
\frac{d}{dt} G_ f(t)
\end{array} 
\eeq 
This constitutes CP violation {\em in the 
oscillations}. For the CP conserving decay into the flavour-specific 
final state is used merely to track the flavour identity of the 
decaying meson. This situation can therefore be denoted also 
in the following way: 
\beq 
\frac{{\rm Prob}(B^0 \Rightarrow  \bar B^0; t) - 
{\rm Prob}(\bar B^0 \Rightarrow  B^0; t)}
{{\rm Prob}(B^0 \Rightarrow \bar B^0; t) + 
{\rm Prob}(\bar B^0 \Rightarrow  B^0; t)} = 
\frac{|q/p|^2 - |p/q|^2}{|q/p|^2 + |p/q|^2} = 
\frac{1- |p/q|^4}{1+ |p/q|^4} 
\eeq 

\item 
$\Delta m_B \neq  0$ with $f$ now being a  
flavour-{\em non}specific final state -- a final state {\em common} 
to $B^0$ and $\bar B^0$ decays -- of a special nature, namely 
a CP eigenstate -- $|\bar f\rangle = {\bf CP}|f\rangle = 
\pm |f\rangle $ -- {\em without} direct CP violation --  
$|\bar \rho _f| = 1 = |\rho _{\bar f}| $: 
\beq 
\begin{array} {c} 
G_f(t) = 2 |T_f|^2 
\left[ 1 + ({\rm sin}\Delta m_Bt) \cdot 
{\rm Im} \frac{q}{p} \bar \rho _f 
\right] \\  
\bar G_f(t) = 2 |T_f|^2 
\left[ 1 - ({\rm sin}\Delta m_Bt )\cdot 
{\rm Im} \frac{q}{p} \bar \rho _f 
\right] \\ 
\leadsto 
\frac{d}{dt} \frac{\bar G_f(t) }{G_f(t)} \neq 0
\end{array} 
\eeq 
is the concrete realization of what was called CP violation 
{\em involving oscillations}. 
\end{itemize} 

%%%%%%%%%%
\subsubsection{CP Violation in Oscillations}
%%%%%%%%%%
Using the convention blessed by the PDG 
\beq 
B = [\bar b q] \; \; , \; \; \bar B = [\bar q b] 
\eeq
we have 
\beq 
\begin{array} {c} 
T(B \ra l^- X) = 0 = T(\bar B \ra l^+ X) \\ 
T_{SL} \equiv  T(B \ra l^+ X) = T(\bar B \ra l^- X) 
\end{array} 
\eeq  
with the last equality enforced by CPT invariance. The 
so-called Kabir test can then be realized as follows: 
$$  
\frac
{{\rm Prob}(B^0 \Rightarrow \bar B^0; t) - 
{\rm Prob}(\bar B^0 \Rightarrow B^0; t)} 
{{\rm Prob}(B^0 \Rightarrow \bar B^0; t) + 
{\rm Prob}(\bar B^0 \Rightarrow B^0; t)} = 
$$ 
\beq 
= \frac
{{\rm Prob}(B^0 \Rightarrow \bar B^0 \ra l^-X; t) - 
{\rm Prob}(\bar B^0 \Rightarrow B^0 \ra l^+X; t)} 
{{\rm Prob}(B^0 \Rightarrow \bar B^0 \ra l^-X; t) + 
{\rm Prob}(\bar B^0 \Rightarrow B^0 \ra l^+X; t)} = 
\frac{1 - |q/p|^4}{1+|q/p|^4} 
\eeq 
Without going into details I merely state the results here 
\cite{CECILIABOOK}: 
\beq 
1 - \left|  \frac{q}{p} \right| \simeq 
\frac{1}{2} {\rm Im}\left( \frac{\Gamma _{12}}
{M_{12}} \right) \sim 
\left\{ 
\begin{array}{ccc} 
10^{-3} & \; {\rm for} \; & B_d=(\bar bd) \\ 
10^{-4} & \; {\rm for} \; & B_s =(\bar bs) \\ 
\end{array} 
\right. 
\eeq
i.e.,  
\beq 
a_{SL} (B^0) \equiv \frac{\Gamma (\bar B^0(t) \ra l^+ \nu X) - 
\Gamma ( B^0(t) \ra l^- \bar \nu X)}
{\Gamma (\bar B^0(t) \ra l^+ \nu X) +  
\Gamma ( B^0(t) \ra l^- \bar \nu X)} \simeq 
\left\{ 
\begin{array}{ccc} 
{\cal O}(10^{-3}) & \; {\rm for} \; & B_d \\ 
{\cal O}(10^{-4}) & \; {\rm for} \; & B_s \\ 
\end{array} 
\right. 
\eeq  
The smallness of the quantity $1-|q/p|$ is primarily due to 
$|\Gamma _{12}| \ll |M_{12}|$ or $\Delta \Gamma _B \ll 
\Delta m_B$. Within the Standard Model this hierarchy 
is understood (semi-quantitatively at leaast) as due to the 
hierarchy in the GIM factors of the box diagram 
expressions for $\Gamma _{12}$ and $M_{12}$, namely 
$m_c^2/M_W^2 \ll m_t^2/M_W^2$. 

For $B_s$ mesons the phase between $\Gamma _{12}$ and 
$M_{12}$ is further (Cabibbo) suppressed for reasons that 
are peculiar to the KM ansatz: for to leading order in the 
KM parameters quarks of the second and third family only 
contribute and therefore 
arg$(\Gamma _{12}/M_{12}) = 0$ to that order. If New 
Physics intervenes in $B^0 - \bar B^0$ oscillations, it would 
quite naturally generate a new phase between 
$\Gamma _{12}$ and $M_{12}$; it could also reduce 
$M_{12}$. Altogether this CP asymmetry could get 
enhanced very considerably: 
\beq 
a_{SL}^{New \; Physics} (B^0) \sim 1 \% 
\eeq 
Therefore one would be ill-advised to accept the somewhat 
pessimistic KM predictions as gospel. 

Since this CP asymmetry does not vary with the time of decay, 
a signal is not diluted by integrating over all times. It is, 
however, essential to `flavour tag' the decaying meson; i.e., 
determine whether it was {\em produced} as  a 
$B^0$ or $\bar B^0$. This can be achieved in several ways 
as discussed later.

%%%%%%%%%%%
\subsubsection{Direct CP Violation}
%%%%%%%%%%

Sizeable direct CP asymmetries arise rather naturally in 
$B$ decays. Consider 
\beq 
b \ra s \bar u u 
\eeq 
Three different processes contribute to it, namely 
\begin{itemize}
\item 
the tree process 
\beq 
b \ra u W^* \ra u (\bar u s)_W  \; , 
\eeq 
\item 
the penguin process with an internal top quark which is 
purely local (since $m_t > m_b$) 
\beq 
b \ra s g^* \ra s u \bar u \; , 
\eeq 
\item 
the penguin reaction with an internal charm quark. Since 
$m_b > 2m_c + m_s$, this last operator is {\em not} 
local: it contains an absorptive part that amounts to a 
final state interaction including a phase shift. 
\end{itemize}
One then arrives at a guestimate \cite{SONI,CECILIABOOK} 
\beq 
\frac{\Gamma (b \ra s u \bar u) - 
\Gamma (\bar b \ra \bar s u \bar u)}
{\Gamma (b \ra s u \bar u) + 
\Gamma (\bar b \ra \bar s u \bar u)} 
\sim {\cal O}(\% ) 
\eeq  
Invoking quark-hadron duality one can expect  
(or at least hope) that this quark level analysis -- rather 
than being washed out by hadronisation -- yields 
some average asymmetry or describes the 
asymmetry for some inclusive subclass of nonleptonic 
channels. I would like to draw the following lessons 
from these considerations: 
\begin{itemize}
\item 
According to the KM ansatz the natural scale for direct 
CP asymmetries in the decays of beauty hadrons 
(neutral or charged mesons or baryons) is the 
$10^{-2}$ level -- not $10^{-6} \div 10^{-5}$ as in 
strange decays! 
\item 
The size of the asymmetry in {\em individual} channels -- 
like $B\ra K \pi$ -- is shaped by the strong final state 
interactions operating there. Those are likely to differ 
considerably from channel to channel, and at present we 
are unable to predict them since they reflect long-distance 
dynamics. 
\item 
Observation of such an asymmetry (or lack thereof) will not provide 
us with reliable numerical information on the parameters of the 
microscopic theory, like the KM ansatz. 
\item 
Nevertheless comprehensive and detailed studies are an 
absolute must!
\end{itemize} 
Later I will describe examples where the relevant long-distance 
parameters -- phase shifts etc. -- can be {\em measured} 
independantly. 

%%%%%%%%%%%%
\subsubsection{CP Violation Involving Oscillations}
%%%%%%%%%%%
The essential feature that a final state in this category has to 
satisfy is that it can be fed both by $B^0$ and $\bar B^0$ decays 
\footnote{Obviously no such common channels can exist for 
charged mesons or for baryons.}. However for convenience reasons 
I will concentrate on a special subclass of such modes, namely 
when the final state is a CP eigenstate. A more comprehensive 
discussion can be found in \cite{CECILIABOOK,BOOK}. 

Three qualitative observations have to be made here: 
\begin{itemize}
\item 
Since the final state is shared by $B^0$ and $\bar B^0$ decays 
one cannot even define a CP asymmetry unless one acquires 
{\em independant} information on the decaying meson: was it 
a $B^0$ or $\bar B^0$ or -- more to the point -- was it originally 
produced as a $B^0$ or $\bar B^0$? There are several 
scenarios for achieving such {\em flavour tagging}: 
\begin{itemize} 
\item 
Nature could do the trick for us by providing us with 
$B^0$ - $\bar B^0$ production asymmetries through, say, 
associated production in hadronic collisions or the use of 
polarized beams in $e^+ e^-$ annihilation. Those production 
asymmetries could be tracked through decays that are 
necessarily CP conserving -- like $\bar B_d \ra \psi K^- \pi ^+$ vs.  
$B_d \ra \psi K^+ \pi ^-$. It seems unlikely, though, that such 
a scenario could ever be realized with sufficient statistics. 

\item 
{\em Same Side Tagging}: One undertakes to repeat the success 
of the $D^*$ tag for charm mesons -- $D^{+*} \ra D^0 \pi ^+$ 
vs. $D^{-*} \ra \bar D^0 \pi ^-$ -- through finding a conveniently 
placed nearby resonance -- $B^{-**} \ra \bar B_d \pi ^-$  vs. 
$B^{+**} \ra B_d \pi ^+$ -- or through employing correlations 
between the beauty mesons and a `nearby' pion (or kaon 
for $B_s$) as pioneered by the CDF collaboration. This method can be calibrated by 
analysing how well $B^0 - \bar B^0$ oscillations are reproduced.

\item 
{\em Opposite Side Tagging}: With electromagnetic and strong 
forces conserving the beauty quantum number, one can employ 
charge correlations between the decay products (leptons and kaons) 
of the two beauty hadrons originally produced together.  

\item 
If the lifetimes of the two mass eigenstates of the neutral $B$ 
meson differ sufficiently from each other, then one can wait 
for the short-lived  component to fade away relative to the 
long-lived one and proceed in qualitative analogy to the $K_L$ 
case. Conceivably this could become feasible -- or even 
essential -- for overly fast oscillating $B_s$ mesons 
\cite{DUNIETZ}. 

\end{itemize} 
The degree to which this flavour tagging can be achieved is a crucial 
challenge each experiment has to face. 

\item 
The CP asymmetry is largest when the two interfering amplitudes 
are comparable in magnitude. With oscillations having to provide 
the second amplitude that is absent initially at time of production, 
the CP asymmetry starts out at zero for decays that occur right after 
production and builds up for later decays. The (first) maximum 
of the asymmetry 
\beq 
\left|  1 - \frac{1 - {\rm Im}\frac{q}{p}\bar \rho _f 
{\rm sin}\Delta m_Bt}
{1 + {\rm Im}\frac{q}{p}\bar \rho _f 
{\rm sin}\Delta m_Bt}
\right| 
\eeq 
is reached for 
\beq 
\frac{t}{\tau _B} = \frac{\pi }{2} \frac{\Gamma _B}{\Delta m_B} 
\simeq 2 
\eeq 
in the case of $B_d$ mesons. 

\item 
The other side of the coin is that very rapid oscillations -- 
$\Delta m_B \gg \Gamma _B$ as is the case for $B_s$ mesons -- 
will tend to wash out the asymmetry or at least will severely 
tax the experimental resolution.  

\end{itemize} 

%%%%%%%%%%%%
\subsubsection{Resume}
%%%%%%%%%%%%%
Three classes of quantities each describe the three types of CP 
violation: 
\begin{enumerate} 
\item 
\beq 
\left| \frac{q}{p} \right| \neq 1 
\eeq 
\item 
\beq 
\left| 
\frac{T(\bar B \ra \bar f)}{T( B \ra f)} 
\right| \neq 1
\eeq 

\item 
\beq 
{\rm Im} \frac{q}{p} \frac{T(\bar B \ra \bar f)}{T( B \ra f)} 
\neq 0
\eeq 
\end{enumerate}
These quantities obviously satisfy one necessary condition 
for being observables: they are insensitive to the phase convention 
adopted for the anti-state. 

%%%%%%%%%%%%%%%
\subsection{Parametric KM Predictions}
%%%%%%%%%%%%%%%%
The triangle defined by 
\beq 
\lambda V(cb) - V(ub) + V^*(td) = 0 
\eeq
to leading order controls basic features of $B$ transitions. As 
discussed before, it has naturally large angles; it usually is called 
{\em the} KM triangle. Its angles are given by KM matrix elements 
which are most concisely expressed in the Wolfenstein 
representation: 
\beq 
e^{i\phi _1} = - \frac{V(td)}{|V(td)|} \; \; ,  \; \; 
e^{i\phi _2} =  \frac{V^*(td)}{|V(td)|} \frac{|V(ub)|}{V(ub)}\; \; , \; \; 
e^{i\phi _3} =  \frac{V(ub)}{|V(ub)|} 
\eeq 
The various CP asymmetries in beauty decays are expressed in 
terms of these three angles. I will describe `typical' examples now. 

%%%%%%%%%%%%%
\subsubsection{Angle $\phi _1$}
%%%%%%%%%%%%
Consider 
\beq 
\bar B_d \ra \psi K_S \leftarrow B_d
\eeq 
where the final state is an almost pure odd CP eigenstate. On the 
quark level one has two different reactions, namely one 
describing the direct decay process 
\beq 
\bar B_d = [b \bar d] \ra [c\bar c] [s \bar d] 
\label{BDTREE}
\eeq 
and the other one involving a $B_d - \bar B_d$ oscillation: 
\beq 
\bar B_d = [b \bar d] \Rightarrow B_d = [\bar b d] 
\ra [c \bar c] [ \bar sd] 
\label{BDOSC} 
\eeq 
\begin{center} 
$\spadesuit \; \; \; \spadesuit \; \; \; \spadesuit $ \\ 
{\em Homework Problem \# 5}: 
\end{center}
How can the $[s \bar d]$ combination in 
Eq.(\ref{BDTREE}) interfere with 
$[\bar sd]$ in Eq.(\ref{BDOSC})? 
\begin{center} 
$\spadesuit \; \; \; \spadesuit \; \; \; \spadesuit $
\end{center} 
Since the final state in $B/\bar B \ra \psi K_S$ can carry 
isospin 1/2 only, we have for the {\em direct} 
transition amplitudes: 
\beq 
\begin{array}{c}
T(\bar B_d \ra \psi K_S) = V(cb)V^*(cs) 
e^{i\alpha _{1/2}} |{\cal M}_{1/2}| \\ 
T(B_d \ra \psi K_S) = V^*(cb)V(cs) 
e^{i\alpha _{1/2}} |{\cal M}_{1/2}| 
\end{array}
\eeq 
and thus 
\beq 
\bar \rho _{\psi K_S} = \frac{V(cb)V^*(cs)}{V^*(cb)V(cs)} 
\eeq 
from which the hadronic quantities, namely the 
phase shift $\alpha _{1/2}$ and the hadronic matrix 
element $|{\cal M}_{1/2}|$ -- both of which {\em cannot} 
be calculated in a reliable manner -- have dropped out. 
Therefore 
\beq 
\left| \bar \rho _{\psi K_S} \right| = 
\left|  
\frac{T(\bar B_d \ra \psi K_S}{T(B_d \ra \psi K_S}
\right|  = 1 \; ; 
\eeq 
i.e., there can be {\em no direct} CP violation in this channel. 

Since $|\Gamma _{12}| \ll |M_{12}|$ one has 
\beq 
\frac{q}{p} \simeq \sqrt{\frac{M^*_{12}}{M_{12}}} = 
\frac{M^*_{12}}{|M_{12}|} \simeq 
\frac{V^*(tb)V(td)}{V(tb)V^*(td)}
\eeq 
which is a pure phase. Altogether one obtains 
\footnote{The next-to-last (approximate) equality 
in Eq.(\ref{SIN2PHI1}) holds in the Wolfenstein 
representation, although the overall result is general.}
\beq 
{\rm Im} \frac{q}{p} \bar \rho _{\psi K_S} = 
{\rm Im} \left( 
\frac{V^*(tb)V(td)} {V(tb)V^*(td)} 
\frac{V(cb)V^*(cs)} {V^*(cb)V(cs)}  
 \right) 
\simeq {\rm Im} \frac{V^2(td)}{|V(td)|^2} = 
-{\rm sin}2\phi _1 
\label{SIN2PHI1} 
\eeq 
That means 
that to a very good approximation the observable 
Im $\frac{q}{p} \bar \rho _{\psi K_S}$, which is the amplitude 
of the oscillating CP asymmetry, is in general given by 
{\em microscopic} parameters of the theory; within 
the KM ansatz they combine to yield the angle $\phi _1$ 
\cite{BS}. 

Several other channels are predicted to exhibit a CP asymmetry 
expressed by sin$2\phi _1$, like $B_d \ra \psi K_L$ 
\footnote{Keep in mind that 
Im$\frac{q}{p}\bar \rho _{\psi K_L} =- 
{\rm Im}\frac{q}{p}\bar \rho _{\psi K_S}$ holds because 
$K_L$ is mainly CP odd and $K_S$ mainly CP even.}, 
$B_d \ra D \bar D$ etc. 

%%%%%%%%%%%%
\subsubsection{Angle $\phi _2$}
%%%%%%%%%%%%%
The situation is not quite as clean for the angle $\phi _2$. 
The asymmetry in $\bar B_d \ra \pi ^+ \pi ^-$ vs. 
$B_d \ra \pi ^+ \pi ^-$ is certainly sensitive to $\phi _2$, 
yet there are two complications:  
\begin{itemize}
\item 
The final state is described by a superposition of {\em two} 
different isospin states, namely $I = 0$ and $2$. The 
spectator process contributes to both of them. 
\item 
The Cabibbo suppressed Penguin operator 
\beq 
b \ra d g^* \ra d u \bar u 
\eeq 
will also contribute, albeit only to the $I=0$ amplitude.  
\end{itemize} 

The direct transition amplitudes are then expressed as follows: 
$$  
T(\bar B_d \ra \pi ^+ \pi ^-) = 
V(ub)V^*(ud)e^{i \alpha _2}|{\cal M}_2^{spect}| + 
$$ 
\beq 
+ e^{i \alpha _0}\left( V(ub)V^*(ud) |{\cal M}_0^{spect}| + 
V(tb)V^*(td) |{\cal M}_0^{Peng}| 
\right)  
\eeq   
$$  
T(B_d \ra \pi ^+ \pi ^-) = 
V^*(ub)V(ud)e^{i \alpha _2}|{\cal M}_2^{spect}| + 
$$ 
\beq 
+ e^{i \alpha _0}\left( V^*(ub)V(ud) |{\cal M}_0^{spect}| + 
V^*(tb)V(td) |{\cal M}_0^{Peng}| 
\right)  
\eeq  
where the phase shifts for the $I=0,2$ states have been factored 
off. 

{\em If} there were no Penguin contributions, we would have 
\beq 
{\rm Im} \frac{q}{p} \bar \rho _{\pi \pi} = 
{\rm Im} \frac{V(td)V^*(tb)V(ub)V^*(ud)}
{V^*(td)V(tb)V^*(ub)V(ud)} = 
- {\rm sin}2\phi _2 
\eeq 
without direct CP violation -- $|\bar \rho _{\pi \pi }| =1$ --  
since the two isospin amplitudes still contain the same weak 
parameters. The Penguin contribution changes the picture in 
two basic ways: 
\begin{enumerate}
\item 
The CP asymmetry no longer depends on $\phi _2$ alone: 
$$  
{\rm Im} \frac{q}{p} \bar \rho _{\pi \pi } \simeq  
- {\rm sin} 2\phi _2  + \left|  \frac{V(td)}{V(ub)} \right|  
\left[  {\rm Im}\left( e^{-i\phi _2}
\frac{{\cal M}^{Peng}}{{\cal M}^{spect}}\right)  
- {\rm Im}\left( e^{-3i\phi _2}
\frac{{\cal M}^{Peng}}{{\cal M}^{spect}}\right) 
\right] + 
$$ 
\beq 
+ {\cal O}(|{\cal M}^{Peng}|^2/|{\cal M}^{spect}|^2) 
\label{PENGPOLL} 
\eeq 
where 
\beq 
{\cal M}^{spect} = e^{i \alpha _0}|{\cal M}^{spect}_0|  + 
e^{i \alpha _2}|{\cal M}^{spect}_2| \; \; , \;  \; 
{\cal M}^{Peng} = e^{i \alpha _0}|{\cal M}^{Peng}_0| 
\eeq 
\item 
A direct CP asymmetry emerges:
\beq 
|\bar \rho _{\pi \pi}| \neq 1 
\eeq 
\end{enumerate}
Since we are dealing with a Cabibbo suppressed Penguin operator, 
we expect that its contribution is reduced relative to the 
spectator term: 
\beq 
\left| \frac{{\cal M}^{Peng}}{{\cal M}^{spect}}  
\right| 
< 1 \; , 
\eeq 
which was already used in Eq.(\ref{PENGPOLL}). 
Unfortunately this reduction might 
not be very large. This concern is based on the observation 
that the branching ratio for $\bar B_d \ra K^- \pi ^+$ appears to 
be somewhat larger than for $\bar B_d \ra \pi ^+ \pi ^-$ 
implying that the Cabibbo favoured Penguin amplitude  
is at least not smaller than the spectator amplitude. 

Various strategies have been suggested to unfold the 
Penguin contribution through a combination of additional 
or other  
measurements (of other $B \ra \pi \pi $ channels 
or of $B\ra \pi \rho$, $B \ra K\pi$ etc.) and supplemented by 
theoretical considerations like $SU(3)_{Fl}$ symmetry 
\cite{PENGTRAP}.  
I am actually hopeful that the multitude of exclusive 
nonleptonic decays (which is the other side of the 
coin of small branching ratios!) can be harnessed to 
extract a wealth of information on the strong dynamics that 
in turn will enable us to extract 
sin$2\phi _2$ with decent accuracy. 

%%%%%%%%%%%%
\subsubsection{The $\phi _3$ Saga}
%%%%%%%%%%%%
Of course it is important to determine $\phi _3$ as accurately  
as possible. This will not be easy, and one better keep 
a proper perspective. I am going to tell this saga now in 
two installments. 

{\bf (I)} {\em CP asymmetries involving $B_s - \bar B_s$ 
Oscillations}: In principle one can extract $\phi _3$ from 
KM suppressed $B_s$ decays like one does $\phi _2$ from 
$B_d$ decays, namely by measuring and analyzing the difference 
between the rates for, say, $\bar B_s(t) \ra K_S \rho ^0$ and 
$B_s(t) \ra K_S \rho ^0$: 
Im$\frac{q}{p}\bar \rho _{K_S\rho ^0} \sim {\rm sin}2\phi _3$. 
One has to face the same complication, namely that in 
addition to the spectator term a 
(Cabibbo suppressed) Penguin amplitude contributes to $\bar \rho 
_{K_S\rho ^0}$ with different weak parameters. Yet the situation 
is much more challenging due to the rapid 
pace of the $B_s - \bar B_s$ oscillations. 

\noindent A more promising way might be to compare the rates for 
$\bar B_s (t) \ra D_s^+ K^-$ with $B_s (t) \ra D_s^- K^+$ as a function 
of the time of decay $t$ since there is no Penguin contribution. 
The asymmetry depends on sin$\phi _3$ rather than 
sin$2\phi _3$ 
\footnote{Both $D_s^+K^-$ and $D_s^- K^+$ are final states common to 
$B_s$ and $\bar B_s$ decays although they are not CP 
eigenstates.}.  

{\bf (II)} {\em Direct CP Asymmetries}: The largish direct CP 
asymmetries sketched above for $B \ra K \pi$ depend on 
sin$\phi _3$ -- and on the phase shifts which in general are neither  
known nor calculable. Yet in some cases they can be determined 
experimentally -- as first described for 
$B^{\pm} \ra D_{neutral} K^{\pm}$ 
\cite{WYLER}. There are 
{\em four independant} rates that can be measured, namely 
\beq 
\Gamma (B^- \ra D^0 K^-) \; , \; 
\Gamma (B^- \ra \bar D^0 K^-) \;  , \;  
\Gamma (B^- \ra D_{\pm} K^-) \; , \; 
\Gamma (B^+ \ra D_{\pm} K^+) 
\eeq  
The {\em flavour eigenstates} $D^0$ and $\bar D^0$ are defined 
through flavour specific modes, namely 
$D^0 \ra l^+ X$ and $\bar D^0 \ra l^- X$, respectively; 
$D_{\pm}$ denote the even/odd CP eigenstates 
$D_{\pm} = (D^0 \pm \bar D^0)/\sqrt{2}$ defined by 
$D_+ \ra K^+K^-, \, \pi ^+ \pi ^-,$ etc., 
$D_- \ra K_S\pi ^0, \, K_S \eta ,$ etc. \cite{PAISSB}. 

From these four observables one can (up to a binary ambiguity) 
extract the four basic quantities, namely the moduli of the two 
independant amplitudes ($|T(B^- \ra D^0 K^-)|$, 
$|T(B^- \ra \bar D^0 K^-)|$), their strong phaseshift -- and 
sin$\phi _3$, the goal of the enterprise! 

%%%%%%%%%%%%%%%%
\subsubsection{A Zero-Background Search for New 
Physics: $B_s \ra \psi \phi , \, D_s^+ D_s^-$}
%%%%%%%%%%%%%%%%%%
The two angles $\phi _1$ and $\phi _2$ will be measured in 
the next several years with decent or even good accuracy. I 
find it unlikely that any of the direct measurements 
of $\phi _3$ sketched above will yield a more precise 
value than inferred from simple trigonometry: 
\beq 
\phi _3 = 180^o - \phi _1 - \phi _2 
\label{180}
\eeq 
Eq.(\ref{180}) holds 
within the KM ansatz; of course the real goal is to uncover 
the intervention of New Physics in $B_s$ transitions. It then 
makes eminent sense to search for it in a reaction where 
Known Physics predicts a practically zero result. 
$B_s \ra \psi \phi , \, \psi \eta , \, D_s \bar D_s$ fit this bill 
\cite{BS}: 
to leading order in the KM parameters the CP asymmetry has 
to vanish since on that level quarks of the second and third 
family only participate in $B_s - \bar B_s$ oscillations -- 
$[s \bar b] \Rightarrow t^* \bar t^* \Rightarrow [b \bar s]$ -- and 
in these direct decays -- $[b \bar s] \ra c \bar c s \bar s$. Any 
CP asymmetry is therefore Cabibbo suppressed, i.e. $\leq 4$\% . 
More specifically 
\beq 
\left. {\rm Im} \frac{q}{p}\bar \rho _{B_s \ra \psi \eta , 
\psi \phi , D_s \bar D_s} \right| _{KM} \sim 2\% 
\eeq  
Yet New Physics has a good chance to contribute to 
$B_s - \bar B_s$ oscillations; if so, there is no reason for 
it to conserve CP and asymmetries can emerge that are easily 
well in excess of 2\% . New Physics scenarios with non-minimal 
SUSY or flavour-changing neutral currents could actually 
yield asymmetries of $\sim 10 \div 30 \%$ 
\cite{GABB} -- completely 
beyond the KM reach!

%%%%%%%%%%%%%%%%%%
\subsubsection{The HERA-B Menu}
%%%%%%%%%%%%%% 
Quite often people in the US tend to believe that a restaurant that 
presents them with a long menu must be a very good one. The 
real experts -- like the French and Italians -- of course know 
better: it is the hallmark of a top cuisine to concentrate on a 
few very special dishes and prepare them in a spectacular fashion 
rather than spread one's capabilities too thinly. That is exactly the advice 
I would like to give the HERA-B collaboration, namely to focus 
on a first class menu consisting of three main dishes and one side 
dish, namely 
\begin{enumerate} 
\item 
measure $\Delta m(B_s)$ which within the Standard Model 
allows to extract $|V(td)|$ through 
\beq 
\frac{\Delta m(B_d)}{\Delta m(B_s)} \simeq 
\frac{Bf_{B_d}^2} {Bf_{B_s}^2} \left|  \frac{V(td)}{V(ts)} 
\right| ^2 \; ; 
\eeq 

\item 
determine the rates for $\bar B_d \ra \psi K_S$ and 
$B_d \ra \psi K_S$ to obtain the value of 
sin$2\phi _1$; 

\item 
compare $\bar B_s \ra \psi \phi , \, D_s \bar D_s$ with 
$B_s \ra \psi \phi , \, D_s \bar D_s$ 
as a clean search for New Physics and 

\item 
as a side dish: measure the $B_s$ lifetime separately in 
$B_s \ra l \nu D_s^{(*)}$ and $B_s \ra \psi \phi , \, D_s \bar D_s$ where 
the former yields the algebraic average of the $B_{s, short}$ 
and $B_{s,long}$ lifetimes and the latter the $B_{s, short}$ 
lifetime. One predicts for them \cite{URALTSEV}:
\beq 
\frac{\tau (B_s \ra l \nu D_s^{(*)}) - \tau (B_s \ra \psi \phi 
,\, D_s \bar D_s)}
{\tau (B_s \ra l \nu D_s^{(*)})} \simeq 0.1 \cdot 
\left(  \frac{f_{B_s}}{200\; {\rm MeV}} \right) ^2
\eeq 

\end{enumerate} 
If the HERA-B chefs succeed in preparing one of these  
main dishes, then they have achieved three star status!

%%%%%%%%%%
\subsection{Theoretical Technologies in Heavy Flavour Decays}
%%%%%%%%%%
One other intriguing and gratifying aspect of heavy flavour 
decays has become understood just over the last several 
years, namely that the decays in particular of beauty hadrons 
can be treated with a reliability and accuracy that before would 
have seemed to be unattainable. These new theoretical 
technologies can be referred to as {\em Heavy Quark Theory} 
which combines two basic elements, namely an 
asymptotic symmetry principle on one hand and a dynamical 
treatment on the other, which tells us how the asymptotic limit 
is approached. The symmetry principle is Heavy Quark 
Symmetry stating that all sufficiently heavy quarks behave 
identically under the strong interactions. The dynamical treatment 
is provided by $1/m_Q$ expansions allowing us to express 
observable transition rates through a series in inverse powers 
of the heavy quark mass. This situation is qualitatively similar 
to chiral considerations which start from the limit of chiral 
invariance and describe the deviations from it through 
chiral perturbation theory. In both cases one has succeeded in 
describing nonperturbative dynamics in special cases. 

The lessons we have learnt can be summarized as follows 
\cite{HQEREV}: we have 
\begin{itemize}
\item 
identified the sources of the non-perturbative corrections;  
\item 
found them to be smaller than they could have been; 
\item 
succeeded in relating the basic quantities of the Heavy Quark 
Theory -- KM paramters, masses and kinetic energy of heavy quarks, 
etc. -- to various a priori independant observables with a fair 
amount of  redundancy; 
\item 
developed a better understanding of incorporating perturbative 
and nonperturbative corrections without double-counting.  
\end{itemize}  
It has been shown that the heavy quark expansion has to be 
formulated in terms of short distance masses rather than 
pole masses. One finds 
\beq 
\begin{array} {c} 
m_b - m_c = 3.50 \pm 0.04 \; {\rm GeV} \\ 
m_b(1 \; {\rm GeV}) = 4.64 \pm 0.05 \; {\rm GeV} 
\end{array} 
\eeq  
This information is then used to extract $|V(cb)|$ from the 
observed semileptonic $B$ width with the result 
\beq 
|V(cb)|_{incl} = 0.0412 \cdot 
\sqrt{\frac{{\rm BR}(B\ra l X)}{0.105}} \cdot 
\sqrt{\frac{1.6 \; {\rm psec}}{\tau _B}}\cdot 
\left( 1 \pm 0.05 |_{theor} \right) 
\label{VCBIN}
\eeq 
Alternatively one can analyze the exclusive mode 
$B\ra l \nu D^*$ and extrapolate to the kinematical 
point of zero recoil to obtain 
\beq 
|F_{D^*}(0) V(cb)| = 0.0339 \pm 0.0014 
\eeq 
From Heavy Quark Theory one infers \cite{HQEREV} 
\beq 
F_{D^*}(0) = 0.91  \pm 0.06  
\eeq 
to arrive at 
\beq 
|V(cb)|_{excl} = 0.0377 \pm 0.0016|_{exp} \pm 0.002 |_{theor}
\label{VCBEX}
\eeq 
The two determinations in Eqs.(\ref{VCBIN}) and (\ref{VCBEX}) 
are systematically very different both in their experimental and 
theoretical aspects. Nevertheless they are quite consistent with 
each other with the experimental and theoretical uncertainties 
being very similar. A few years ago it would have seemed 
quite preposterous to claim such small theoretical uncertainties! 
I am actually confident that those can be reduced from the 
present 5\% level down to the 2\% level in the foreseeable future.  

$|V(ub)|$ (or $|V(ub)/V(cb)|$) is not known with an even remotely 
similar accuracy, and so far one has relied on models rather 
than QCD proper to extract it from data. Yet we can be confident 
that over the next ten years $|V(ub)|$  will be determined with a 
theoretical uncertainty below 10\% . It will be important to obtain 
it from systematically different semileptonic distributions and 
processes; Heavy Quark Theory provides us with the 
indispensable tools for combining the various analyses in a 
coherent fashion. 

This theoretical progress can embolden us to hope that in the end 
even $|V(td)|$ can be determined with good accuracy -- say 
$\sim 10 \div 15\%$ -- from 
$\Gamma (K^+ \ra \pi ^+ \nu \bar \nu )$, 
$\Delta m(B_s)$ vs. $\Delta m(B_d)$ or 
$\Gamma (B \ra \gamma \rho /\omega )$ vs. 
$\Gamma (B \ra \gamma K^* )$ etc. 

%%%%%%%%%%%%%%
\subsection{KM Trigonometry}
%%%%%%%%%%%%%%%
One side of the triangle is exactly known since the base line can be 
normalized to unity without affecting the angles: 
\beq 
1 - \frac{V(ub)}{\lambda V(cb)} + 
\frac{V^*(td)}{\lambda V(cb)} = 0 
\eeq 
The second side is known to some degree from semileptonic $B$ 
decays: 
\beq 
\left| 
\frac{V(ub)}{V(cb)} 
\right| 
\simeq 0.08 \pm 0.03
\eeq
where the quoted uncertainty is mainly theoretical and amounts 
to little more than a guestimate. In the Wolfenstein representation 
this reads as 
\beq 
\sqrt{\rho ^2 + \eta ^2} \simeq 0.38 \pm 0.11
\eeq  
The area cannot vanish since $\epsilon _K \neq 0$. Yet at present 
not much more can be said for certain. 

In principle one would have enough observables -- namely 
$\epsilon _K$ and $\Delta m(B_d)$ in addition to 
$|V(ub)/V(cb)|$ -- to determine the two KM parameters 
$\rho$ and $\eta$ in a {\em redundant} way. In practise, though, 
there are two further unknowns, namely the size of the 
$\Delta S=2$ and $\Delta B=2$ matrix elements, as expressed through 
$B_K$ and $B_B f_B^2$. For $m_t$ sufficiently large $\epsilon _K$ 
is dominated by the top contribution:   
$d \bar s \Rightarrow  t^* \bar t^* \Rightarrow s \bar d$. The same holds 
always for $\Delta m(B_d)$. In that case things are simpler: 
\beq 
\frac{|\epsilon _K|}{\Delta m(B_d)} \propto 
{\rm sin}2\phi _1 \simeq 
0.42 \cdot UNC 
\label{SIN2BETA} 
\eeq 
with the factor $UNC$ parametrising the uncertainties 
\beq 
UNC \simeq \left( \frac{0.04}{|V(cb)|}\right) 
\left( \frac{0.72}{x_d}\right) \cdot 
\left( \frac{\eta _{QCD}^{(B)}}{0.55}\right) \cdot 
\left( \frac{0.62}{\eta _{QCD}^{(K)}}\right) \cdot 
\left( \frac{2B_B}{3B_K}\right) \cdot 
\left( \frac{f_B}{160\, {\rm MeV}}\right) ^2  
\eeq
where   
$x_d \equiv \Delta m(B_d)/\Gamma (B_d)$; 
$\eta _{QCD}^{(B)}$ and $\eta _{QCD}^{(K)}$ denote the 
QCD radiative corrections for ${\cal H}(\Delta B=2)$ and 
${\cal H}(\Delta S=2)$, respectively; $B_B$ and $B_K$ 
express the expectation value of ${\cal H}(\Delta B=2)$ or 
${\cal H}(\Delta S=2)$ in units of the `vacuum saturation' 
result which is given in terms of the decay constants 
$f_B$ and $f_K$ (where the latter is known). The main 
uncertainty is obviously of a theoretical nature related to 
the hadronic parameters $B_B$, $B_K$ and $f_B$; as discussed 
before, state-of-the-art theoretical technologies yield 
$B_B \simeq  1$, $B_K \simeq 0.8 \pm 0.2$ and 
$f_B \simeq 180 \pm 30 \, {\rm MeV}$ where the latter range 
might turn out to be anything but conservative! 
Eq.(\ref{SIN2BETA}) represents an explicit illustration that some CP 
asymmetries in $B^0$ decays are huge. 

For $m_t \simeq 180$ GeV the $c \bar c$ and $c\bar t + t \bar c$ 
contributions to $\epsilon _K$ are still sizeable; nevertheless 
Eq.(\ref{SIN2BETA}) provides a good approximation. Furthermore 
sin$2\phi _1$ can still be expressed reliably as a function of the 
hadronic matrix elements: 
\beq 
{\rm sin}2\phi _1 = f(B_Bf_B^2/B_K) 
\eeq  
It will become obvious why this is relevant. 

The general idea is, of course, to construct the triangle as accurately 
as possible and then probe it; i.e. search for inconsistencies that 
would signal the intervention of New Physics. A few remarks on that 
will have to suffice here. 

As indicated before we can expect the value of 
$|V(ub)/V(cb)|$ to be known to better than 10\% and hope for 
$|V(td)|$ to be determined with decent accuracy as well. 
The triangle 
will then be well determined or even overdetermined. Once the 
first asymmetry in $B$ decays that can be interpreted reliably -- 
say in $B_d \ra \psi K_S$ -- has been measured and $\phi _1$ 
been determined, the triangle is fully constructed from 
$B$ decays alone. Furthermore one has arrived at the first 
sensitive consistency check of the triangle: one  
compares the measured value of sin$2\phi _1$ with 
Eq.(\ref{SIN2BETA}) to infer which value of 
$B_Bf_B^2$ is thus required; this value is inserted into the 
Standard Model expression for $\Delta m(B_d)$ together 
with $m_t$ to see whether the experimental result is 
reproduced. 

A host of other tests can be performed that are highly sensitive to 
\begin{itemize}
\item 
the presence of New Physics and 
\item 
to some of their salient dynamical features. 
\end{itemize}
Details can be found in the ample literature on that subject.

%%%%%%%%%%
\section{Oscillations and CP Violation in Charm Decays -- 
The Underdog's Chance for Fame}
%%%%%%%%%%%%
It is certainly true that 
\begin{itemize}
\item 
$D^0-\bar D^0$ oscillations proceed very slowly in the 
Standard Model and 
\item 
CP asymmetries in $D$ decays are small or even tiny within 
the KM ansatz. 
\end{itemize}
Yet the relevant question quantitatively is: how slow and how small? 

%%%%%%%%%%%%%
\subsection{$D^0-\bar D^0$ Oscillations}
%%%%%%%%%%%%%%
Bounds on $D^0 - \bar D^0$ oscillations are most cleanly   
expressed through `wrong-sign' semileptonic decays: 
\beq 
r_D = \frac{\Gamma (D^0 \ra l^-X)}{\Gamma (D^0 \ra l^+X)} 
\simeq \frac{1}{2} \left( x_D^2 + y_D^2\right)  
\eeq 
with $x_D = \Delta m_D/\Gamma _D$, 
$y_D = \Delta \Gamma _D/2\Gamma _D$. It is often stated that  
the Standard Model predicts 
\beq 
r_D \leq 10^{-7} \; \; \hat = \; \; 
x_D, \, y_D \leq 3 \cdot 10^{-4} 
\eeq 
I myself am somewhat flabbergasted by the boldness of such 
predictions. For one should keep the following in mind for 
proper perspective: there are quite a few channels that can drive 
$D^0 - \bar D^0$ 
oscillations -- like $D^0 \Rightarrow K\bar K , \; \pi \pi \Rightarrow  
\bar D^0$ or $D^0 \Rightarrow K^- \pi ^+ \Rightarrow 
\bar D^0$ -- and 
they branching ratios on the $({\rm few})\times 10^{-3}$ 
level 
\footnote{For the $K^- \pi ^+$ mode this represents the average of 
its Cabibbo allowed and doubly Cabibbo suppressed incarnations.}. 
In the limit of $SU(3)_{Fl}$ symmetry all these contributions have 
to cancel of course. Yet there are sizeable violations of $SU(3)_{Fl}$ 
invariance in $D$ decays, and one should have little confidence in an 
imperfect symmetry to ensure that a host of channels with branching 
ratios of order few$\times 10^{-3}$ will cancel as to render 
$x_D, \, y_D \leq 3\cdot 10^{-4}$. To say it differently: 
The relevant question in this context is {\em not} whether 
$r_D \sim 10^{-7} \div 10^{-6}$ is a possible or even reasonable 
Standard Model estimate, but whether 
$10^{-6} \leq r_D \leq 10^{-4}$ can {\em reliably be ruled out}! I 
cannot see how anyone could make such a claim with the 
required confidence. 

The present experimental bound is 
\beq 
r_D |_{exp} \leq 3.4 \cdot 10^{-3} \; \; \hat = \; \; 
x_D, \, y_D \leq 0.1 
\eeq 
to be compared with a {\em conservative} Standard Model bound 
\beq 
r_D |_{SM} < 10^{-4} \; \; \hat = \; \; y_D, \, x_D|_{SM}
\leq 10^{-2} \; 
\eeq 
New Physics on the other hand can enhance $\Delta m_D$ 
(though not 
$\Delta \Gamma _D$) very considerably up to 
\beq 
x_D |_{NP} \sim 0.1 \; , 
\eeq 
i.e. the present experimental bound. 

%%%%%%%%%%%%
\subsection{CP Violation involving $D^0 - \bar D^0$ Oscillations}
%%%%%%%%%%%%%
One can discuss this topic in close qualitative analogy to 
$B$ decays. First one considers final states that are CP 
eigenstates like $K^+K^-$ or $\pi ^+ \pi ^-$ 
\cite{BSDDBAR}: 
$$  
{\rm rate}(D^0(t) \ra K^+ K^-) \propto e^{-\Gamma _D t} 
\left(  
1+ ({\rm sin}\Delta m_Dt) \cdot {\rm Im}\frac{q}{p}
\bar \rho _{K^+K^-}   
\right) 
\simeq 
$$
\beq 
\simeq e^{-\Gamma _D t} 
\left(  
1+ \frac{\Delta m_Dt}{\Gamma _D}\cdot 
\frac{t}{\tau _D} 
\cdot {\rm Im}\frac{q}{p} \bar \rho _{K^+K^-}   
\right) 
\eeq 
With $x_D|_{SM} \leq 10^{-2}$ and 
Im$\frac{q}{p} \bar \rho _{K^+K^-}|_{KM} \sim {\cal O}(10^{-3})$ one 
arrives at an asymmetry of around $10^{-5}$, i.e. for all practical 
purposes zero, since it presents the product of two 
very small numbers. 
Yet with New Physics one conceivably has $x_D|_{NP} \leq 0.1$, 
Im$\frac{q}{p} \bar \rho _{K^+K^-}|_{NP} \sim {\cal O}(10^{-1})$ 
leading to an asymmetry that could be as large as of order 1\%. 
Likewise one should compare the doubly Cabibbo suppressed transitions \cite{BIGIBERK,NIR}
$$ 
{\rm rate}(D^0(t) \ra K^+ \pi ^-) \propto 
e^{-\Gamma _{D^0} t} {\rm tg}^4\theta _C|\hat \rho _{K\pi }|^2 \cdot 
$$ 
$$ 
\cdot \left[ 1 - \frac{1}{2}\Delta \Gamma _D t + 
\frac{(\Delta m_Dt)^2}{4{\rm tg}^4\theta _C|\hat \rho _{K\pi }|^2} 
+ \frac{\Delta \Gamma _Dt}
{2{\rm tg}^2\theta _C|\hat \rho _{K\pi }|}
{\rm Re}\left( 
\frac{p}{q}\frac{\hat \rho _{K\pi }}{|\hat \rho _{K\pi }|}   
\right)  
- \right. 
$$ 
\beq
\left. - \frac{\Delta m_Dt}{{\rm tg}^2\theta _C|\hat \rho _{K\pi }|} 
{\rm Im}\left( 
\frac{p}{q}\frac{\hat \rho _{K\pi }}{|\hat \rho _{K\pi }|}   
\right)  
\right] 
\eeq 
$$ 
{\rm rate}(\bar D^0(t) \ra K^- \pi ^+) \propto 
e^{-\Gamma _{D^0} t} {\rm tg}^4\theta _C|\hat{\bar \rho }_{K\pi }|^2 \cdot 
$$ 
$$ 
\cdot \left[ 1 - \frac{1}{2}\Delta \Gamma _D t + 
\frac{(\Delta m_Dt)^2}{4{\rm tg}^4\theta _C
|\hat{\bar \rho}_{K\pi }|^2} 
+ \frac{\Delta \Gamma _Dt}
{2{\rm tg}^2\theta _C|\hat{\bar \rho }_{K\pi }|}
{\rm Re}\left( 
\frac{p}{q}\frac{\hat{\bar \rho }_{K\pi }}
{|\hat{\bar \rho }_{K\pi }|}   
\right)  
+\right. 
$$ 
\beq
\left. + \frac{\Delta m_Dt}{{\rm tg}^2\theta _C
|\hat{\bar \rho }_{K\pi }|} 
{\rm Im}\left( 
\frac{p}{q}\frac{\hat{\bar \rho }_{K\pi }}{|\hat{\bar \rho }_{K\pi }|}   
\right)  
\right] 
\eeq
where 
\beq 
{\rm tg}^2\theta _C \cdot \hat \rho _{K\pi } \equiv 
\frac{T(D^0 \ra K^+ \pi ^-)}{T(D^0 \ra K^- \pi ^+)}\; \; , \; \; 
{\rm tg}^2\theta _C \cdot \hat{\bar \rho }_{K\pi } \equiv 
\frac{T(\bar D^0 \ra K^- \pi ^+)}{T(\bar D^0 \ra K^+ \pi ^-)}\; ; 
\eeq  
in such New Physics scenarios 
one would expect a considerably enhanced asymmetry 
of order $1\%/{\rm tg}^2 \theta _C \sim 20\%$ -- at the cost 
of  smaller statistics. 

Effects of that size would unequivocally signal the intervention 
of New Physics!

%%%%%%%%%%%%
\subsection{Direct CP Violation}
%%%%%%%%%%%
As explained before a direct CP asymmetry requires the presence 
of two coherent amplitudes with different weak and different 
strong phases. Within the Standard Model (and the 
KM ansatz) such effects can occur in Cabibbo suppressed 
\footnote{The effect could well reach the $10^{-3}$ 
and exceptionally the $10^{-2}$ level.}, yet not 
in Cabibbo allowed or doubly Cabibbo suppressed modes. There 
is a subtlety involved in this statement. Consider for example  
$D^+ \ra K_S \pi ^+$. At first sight it appears to be 
a Cabibbo allowed mode described by a single amplitude without 
the possibility of an asymmetry. However \cite{YAMAMOTO}  
\begin{itemize} 
\item 
due to $K^0 - \bar K^0$ mixing the final state 
can be reached also through a doubly Cabibbo suppressed 
reaction, and the two amplitudes necessarily interfere; 
\item 
because of the CP violation in the $K^0 - \bar K^0$ complex 
there is an asymmetry that can be predicted on general grounds 
$$  
\frac{\Gamma (D^+ \ra K_S \pi ^+) - \Gamma (D^- \ra K_S \pi ^+)} 
{\Gamma (D^+ \ra K_S \pi ^+) + \Gamma (D^- \ra K_S \pi ^+)} 
\simeq - 2 {\rm Re}\, \epsilon _K  \simeq - 3.3 \cdot 10^{-3} 
\simeq 
$$  
\beq 
\simeq 
\frac{\Gamma (D^+ \ra K_L \pi ^+) - \Gamma (D^- \ra K_L \pi ^+)} 
{\Gamma (D^+ \ra K_L \pi ^+) + \Gamma (D^- \ra K_L \pi ^+)} 
 \; ; 
\label{DIRECTCPEPS}
\eeq  
\item
 If New Physics contributes to the doubly Cabibbo suppressed 
amplitude $D^+ \ra K^0 \pi ^+$ (or $D^- \ra \bar K^0 \pi ^-$) then 
an asymmetry could occur quite conceivably on the few percent 
scale; 
\item 
such a manifestation of New Physics would be unequivocal; against 
the impact of $\epsilon _K$, Eq.(\ref{DIRECTCPEPS}) it 
could be distinguished not only through the size of the asymmetry, 
but also how it surfaces in $D^+ \ra K_L \pi ^+$ vs. 
$D^- \ra K_L \pi ^-$: if it is New Physics one has 
\beq 
\frac{\Gamma (D^+ \ra K_S \pi ^+) - \Gamma (D^- \ra K_S \pi ^+)} 
{\Gamma (D^+ \ra K_S \pi ^+) + \Gamma (D^- \ra K_S \pi ^+)} 
= - 
\frac{\Gamma (D^+ \ra K_L \pi ^+) - \Gamma (D^- \ra K_L \pi ^+)} 
{\Gamma (D^+ \ra K_L \pi ^+) + \Gamma (D^- \ra K_L \pi ^+)} 
\eeq 
i.e.,  the CP asymmetries in $D \ra K_S \pi$ and $D \ra K_L \pi$ 
differ in sign -- in contrast to Eq.(\ref{DIRECTCPEPS}). 
\end{itemize} 

%%%%%%%%%%%%%%%%%%%%%%%%%%%%%%%%%%%%%%%%%%%
\section{Baryogenesis  in the Universe}
%%%%%%%%%%%%%%%%
%%%%%%%%%%
\subsection{The Challenge}
%%%%%%%%%%%

One of the most intriguing aspects of big bang 
cosmology is to `understand' nucleosynthesis, i.e. to reproduce the 
abundances observed for the nuclei in the universe as 
{\em dynamically} generated rather than merely dialed as 
input values. This challenge has been met successfully, in 
particular for the light nuclei, and actually so much so that 
it is used to obtain information on dark matter in the universe, 
the number of neutrinos etc. It is natural to ask whether such 
a success could be repeated for an even more basic quantity, 
namely the baryon number density of the universe which is defined 
as the difference in the abundances of baryons and antibaryons: 
\beq 
\Delta n_{Bar} \equiv n_{Bar} - n_{\overline{Bar}} 
\eeq 
Qualitatively one can summarize the observations through 
two statements: 
\begin{itemize}
\item 
The universe is not empty. 
\item 
The universe is almost empty. 
\end{itemize}
More quantitatively one finds
\beq 
r_{Bar} \equiv \frac{\Delta n_{Bar}}{n_{\gamma}} \sim {\rm few} 
\times 10^{-10} 
\label{BNUMUNI}
\eeq 
where $n_{\gamma}$ denotes the number density of photons in the 
cosmic background radiation. Actually we know more, namely 
that at least in our corner of the universe there are 
practically no primary antibaryons: 
\beq 
n_{\overline{Bar}} \ll n_{Bar} \ll n_{\gamma}
\label{TWOHIER}
\eeq
It is conceivable that in other 
neighbourhoods antimatter dominates and that the universe 
is formed by a patchwork quilt of matter and antimatter 
dominated regions with the whole being 
matter-antimatter symmetric. Yet it is widely held 
to be quite unlikely -- primarily because no mechanism has been 
found by which a matter-antimatter symmetric universe following 
a big bang evolution can develop sufficiently large regions 
with non-vanishing baryon number. While there will be 
statistical fluctuations, they can be nowhere near large 
enough. Likewise for dynamical effects: baryon-antibaryon 
annihilation is by far not sufficiently effective to create pockets 
with the observed baryon number, Eq.(\ref{BNUMUNI}). For the 
number density of {\em surviving} baryons can be estimated as 
\cite{DOLGOV1} 
\beq 
n_{Bar} \sim \frac{n_{\gamma}}{\sigma _{annih}m_N M_{PL}} 
\simeq 10^{-19}n_{\gamma} 
\eeq
where $\sigma _{annih}$ denotes the cross section of nucleon 
annihilation, $m_N$ and $M_{Pl}$ the nucleon and Planck mass, 
respectively. Hence we conclude for the universe as a whole 
\beq 
0 \neq \frac{n_{Bar}}{n_{\gamma}} \simeq 
\frac{\Delta n_{Bar}}{n_{\gamma}} \sim {\cal O}(10^{-10}) 
\eeq  
which makes more explicit the meaning of the statement 
quoted above that the universe has been observed to be 
almost empty, but not quite. Understanding this double 
observation is the challenge we are going to address now.  
%%%%%%%%%%%%%%
\subsection{The Ingredients}
%%%%%%%%%%%%%%
The question is: under which condition can one have a situation 
where the baryon number of the universe that vanishes at the initial time -- 
which for all practical purposes is the Planck time --  develops 
a non-zero  value later on: 
\beq 
\Delta n_{Bar}(t = t_{Pl}\simeq 0) = 0 
\; \; \stackrel{?}{\Longrightarrow} \; \; 
\Delta n_{Bar}(t = `today') 
\neq 0 
\eeq 
One can and should actually go one step further in the task one is 
setting for oneself: explaining the observed baryon number as 
dynamically generated 
{\em no matter what its initial value was!} 

In a seminal paper that appeared in 1967 Sakharov 
listed the three ingredients that are essential for the feasibility 
of such a program \cite{SAKH,DOLGOV2} : 
\begin{enumerate}
\item 
Since the final and initial baryon number differ, there have to be baryon number violating transitions: 
\beq 
{\cal L}(\Delta n_{Bar} \neq 0) \neq 0 
\eeq  
\item 
CP invariance has to be broken. Otherwise for every baryon number 
changing transition $N \ra f$ there is its CP conjugate one 
$\bar N \ra \bar f$ and no net baryon number can be generated. 
I.e., 
\beq 
\Gamma (N \; \; \stackrel{{\cal L}(\Delta n_{Bar} \neq 0)}
{\longrightarrow} \; \;  f ) 
\; \; \; \neq \; \; \; 
\Gamma (\bar N \; \; \stackrel{{\cal L}(\Delta n_{Bar} \neq 0)}
{\longrightarrow} \; \; \bar f ) 
\eeq 
is needed. 
\item 
Unless one is willing to entertain thoughts of CPT violations, the 
baryon number and CP violating transitions have to proceed 
out of thermal equilibrium. For in thermal equilibrium time 
becomes irrelevant globally and CPT invariance reduces to 
CP symmetry which has to be avoided, see above: 
\beq 
{\rm CPT \; invariance} \; \; \; 
\stackrel{thermal \; equilibrium}
{\Longrightarrow} \; \; \;  {\rm CP \; invariance} 
\eeq 
\end{enumerate}  
It is important to keep in mind that these three conditions have 
to be satisfied {\em simultaneously}. The other side of the coin is, 
however, the following: once a baryon number has been 
generated through the concurrance of these three effects, 
it can be washed out again by these same effects.

%%%%%%%%%%%%%%
\subsection{GUT Baryogenesis}
%%%%%%%%%%%%%%%%
Sakharov's paper was not noticed (except for \cite{KUZMIN}) 
for several years until the concept of Grand Unified Theories 
(=GUTs) emerged starting in 1974 \cite{PATI}; 
for those naturally provide all three necessary ingredients: 

\begin{enumerate} 
\item 
Baryon number changing reactions have to exist in GUTs. For placing 
quarks and leptons into common representations of the underlying 
gauge groups -- the hallmark of GUTs -- means that gauge interactions exist changing baryon and lepton numbers.  
Those gauge bosons are generically 
referred to as $X$ bosons and have two couplings to fermions 
that violate baryon and/or lepton number: 
\beq 
X \leftrightarrow qq \;  , \; \;  q \bar l  
\label{XDEC} 
\eeq  

\item  
Those models are sufficiently complex to allow for several 
potential sources of CP violation. Since $X$ bosons have (at least) 
two decay channels open CP asymmetries can arise 
\begin{eqnarray}  
\Gamma (X \ra qq) = (1+\Delta _q) \Gamma _q 
&,& \; \; 
\Gamma (X \ra q\bar l) = (1-\Delta _l) \Gamma _l \\   
\Gamma (\bar X \ra \bar q \bar q) = 
(1-\Delta _q) \Gamma _q 
&,& \; \; 
\Gamma (\bar X \ra \bar q l) = (1+\Delta _l) \Gamma _l 
\end{eqnarray}   
where  
\begin{eqnarray}  
{\bf \rm CPT} \; \; &\Longrightarrow & \; \; 
\Delta _q \Gamma _q = \Delta _l \Gamma _l \\  
{\bf \rm CP} \; \; &\Longrightarrow & \; \; \Delta _q = 0 = \Delta _l \\  
{\bf \rm C} \; \; &\Longrightarrow & \; \; \Delta _q = 0 = \Delta _l 
\end{eqnarray}   

\item  
Grand Unification means that a phase transition takes 
place around an energy scale $M_{GUT}$. For temperatures 
$T$ well above the transition point -- $T \gg M_{GUT}$ -- 
all quanta are relativistic with a number density 
\beq 
n(T) \propto T^3
\label{NREL} 
\eeq 
For temperatures around the phase transition -- 
$T \sim M_{GUT}$ -- some of the quanta, in particular those 
gauge bosons generically referred to as $X$ bosons aquire 
a mass $M_X \sim {\cal O}(M_{GUT})$ and their 
equilibrium number 
density becomes Boltzmann suppressed: 
\beq 
n_X(T) \propto (M_XT)^{\frac{3}{2}} {\rm exp}
\left( - \frac{M_X}{T}\right) 
\label{BOLTZ} 
\eeq 
More $X$ bosons will decay according to Eq.(\ref{XDEC}) 
than be regenerated from $qq$ and $q \bar l$ collisions 
ultimately bringing the number of $X$ bosons down to the 
level described by Eq.(\ref{BOLTZ}). Yet that will take some time; 
the expansion in the big bang cosmology leads to a cooling rate 
that is so rapid that thermal equilibrium cannot be maintained 
through the phase transition. That means that $X$ bosons 
decay -- and in general interact -- out of thermal 
equilibrium \cite{DOLGOV2}. 
\end{enumerate} 
To the degree that the back production of $X$ 
bosons in $qq$ and $q\bar l$ collisions can be ignored one finds 
as an order-of-magnitude estimate  
\beq 
r_{Bar} \sim \frac{\frac{4}{3}\Delta _q \Gamma _q - 
\frac{2}{3}\Delta _l \Gamma _l}{\Gamma _{tot}} \frac{n_X}{n_0} 
= \frac{\frac{2}{3}\Delta _q \Gamma _q }{\Gamma _{tot} }\frac{n_X}{n_0}
\label{GUTRESULT} 
\eeq 
with $n_X$ denoting the initial number density of $X$ 
bosons and $n_0$ the number density of the light decay 
products 
\footnote{Due to thermalization effects one can have 
$n_0 \gg 2n_X$.}. 
The three essential conditions for baryogenesis are thus 
naturally realized around the GUT scale in big bang cosmologies, as 
can be read off from Eq.(\ref{GUTRESULT}): 
\begin{itemize} 
\item 
$\Gamma _q \neq 0$ representing baryon number violation; 
\item 
$\Delta _q \neq 0$ reflecting CP violation and 
\item 
the absence of the back reaction due to an absence of thermal 
equilibrium. 
\end{itemize}   

The fact that this problem can be formulated in GUT models 
and answers obtained that are very roughly in the right 
ballpark is a highly attractive feature of GUTs, in particular 
since this was {\em not} among the original motivations 
for constructing such theories. 

On the other hand it would be highly misleading to claim 
that baryogenesis has been understood. There are  
serious problems in any attempt to have baryogenesis 
occur at a GUT scale: 
\begin{itemize}
\item 
A baryon number generated at such high temperatures is 
in grave danger to be washed out or diluted in the subsequent 
evolution of the universe. 
\item 
Very little is known about the dynamical actors operating at 
GUT scales and their characteristics -- and that is putting it mildly. Actually 
even in the future we can only hope to obtain some 
slices of indirect information on them. 
\end{itemize} 
Of course it would be premature to write-off baryogenesis at 
GUT scales, yet it might turn out that it is best 
characterised as a proof of principle -- namely that the 
baryon number of the universe can be understood as dynamically 
generated -- rather than as a semi-quantitative realization. 
%%%%%%%%%%%%%%
\subsection{Electroweak Baryogenesis}
%%%%%%%%%%%%%%%%
Baryogenesis at the electroweak scale 
\cite{RUBAKOV} 
is the most actively analyzed scenario at present. For it possesses several highly attractive features: 
\begin{itemize}
\item 
We know that dynamical landscape fairly well.
\begin{itemize}
\item 
In particular CP violation has been found to exist there. 
\item 
A well-studied phase transition, namely the spontaneous breaking \beq 
SU(2)_L\times U(1) \; \; \Longrightarrow  U(1)_{QED}
\eeq 
takes place. 
\end{itemize}
\item 
Future experiments will certainly probe that dynamical regime 
with ever increasing sensitivity, both by searching for the 
on-shell production of new quanta -- like SUSY and/or Higgs 
states -- and the indirect impact through quantum corrections on 
rare decays and CP violation.  
\end{itemize}
However at this point the reader might wonder: "What 
about the third required ingredient, baryon number violation? At the 
electroweak scale?" It is often not appreciated that 
the electroweak forces of the Standard Model by themselves 
violate baryon number, though in a very subtle way.  
We find here what is called an anomaly: the baryon 
number current is conserved on the classical, 
yet {\em not} on the quantum level:  
\beq 
\partial _{\mu} J_{\mu}^{Bar} = 
\partial _{\mu}\sum _q (\bar q_L \gamma _{\mu}q_L) = 
\frac{g^2}{16 \pi ^2} G_{\mu \nu} \tilde G_{\mu \nu} \neq 0 
\label{ANOM} 
\eeq  
where $G_{\mu \nu}$ denotes the electroweak field strength 
tensor 
\beq 
G_{\mu \nu} = \partial _{\mu} A_{\nu} - \partial _{\nu} A_{\mu} 
+ g[A_{\mu}, A_{\nu}] 
\eeq 
and $\tilde G_{\mu \nu} $ its dual: 
\beq 
\tilde G_{\mu \nu}  = \epsilon _{\mu \nu \alpha \beta }
G_{\alpha \beta} 
\eeq 
The right hand side of Eq.(\ref{ANOM}) can be written as 
the divergence of a current 
\beq 
G_{\mu \nu} \tilde G_{\mu \nu} = \partial _{\mu} K_{\mu} \; , 
\; \; K_{\mu} = 2 \epsilon _{\mu \nu \alpha \beta } 
\left(  A_{\nu} \partial _{\alpha} A_{\beta} + 
\frac{2}{3} ig A_{\nu}A_{\alpha}A_{\beta} 
\right) 
\label{KCURRENT}  
\eeq 
A total derivative is usually unobservable since partial 
integration allows to express its contribution 
through a surface integral at infinity. The field strength 
tensor $G_{\mu \nu}$ indeed vanishes at infinity -- but 
not necessarily the gauge potential $A_{\mu}$. 
To be more specific: The field configuration at infinity is 
that of a ground state for which $G_{\mu \nu} = 0$ 
holds. Yet that property does not define it uniquely:  
ground states get differentiated by the value of their  
$K$ charge, i.e. the space integral  
of $K_0$, the zeroth component of the current $K_{\mu}$ 
constructed from their gauge field configuration. 
This integral reflects differences in the gauge topology 
of the ground states and therefore is called the 
{\em topological charge}. 
While this charge is irrelevant for abelian gauge theories 
where the 
last term in Eq.(\ref{KCURRENT}) necessarily vanishes,  
it becomes relevant for non-abelian theories.  
We have encountered 
this phenonemon already in our discussion of the 
Strong CP Problem that is driven by the axial quark current 
not being conserved in the strong interactions of QCD. It is often 
referred to as `Chiral' Anomaly since it breaks chiral invariance, 
or `Triangle' Anomaly since it is produced by a triangular fermion 
loop diagram or `Adler-Bell-Jackiw' Anomaly named after the authors who discovered it. 

The concrete impact of the triangle anomaly on the physics 
depends on the specifics of the theory: here 
because of the chiral 
nature of the weak interactions it induces baryon number 
violation. 
Eq.(\ref{ANOM}) and Eq.(\ref{KCURRENT}) show that the difference 
$J_{\mu}^{Bar} - K_{\mu}$ is conserved. The transition from one 
ground state to another which represents a tunneling 
phenomenon is thus accompanied by a change in 
baryon number. Elementary quantum mechanics tells 
us that this baryon number violation is described as a 
barrier penetration and exponentially suppressed at low 
temperatures or energies \cite{THOOFTBAR}: 
$Prob(\Delta n_{Bar}\neq 0) \propto 
{\rm exp}(-16 \pi ^2/g^2) \sim {\cal O}(10^{-160})$ --   
a suppression that reflects the tiny size of the 
weak coupling.  

There is a corresponding anomaly for the lepton number 
current implying that lepton number is violated as 
well with the selection rule 
\beq 
\Delta n_{Bar} - \Delta n_{lept} = 0 \; . 
\label{B-L}
\eeq
This is usually referred to by saying that $B-L$, the difference between baryon and lepton number, is still conserved. 

At sufficiently high energies this huge suppression of baryon 
number changing transition rates will evaporate since 
the transition between different ground states can be 
achieved classically through a motion {\em over} 
the barrier. The question then is at which energy scale this 
will happen and how quickly baryon number violation will 
become operative. Some semi-quantitative observations can 
be offered and answers given 
\cite{RUBAKOV2,DOLGOV2}. 

There are special field configurations -- called sphalerons -- 
that carry the topological $K$ charge. In the Standard Model they 
induce effective multistate interactions among left-handed 
fermions that change baryon and lepton number by three 
units each: 
\beq 
\Delta n_{Bar} = \Delta n_{lept} = 3 
\eeq 
At high energies where the weak bosons $W$ and $Z$ 
are massless, the height of the transition barrier between 
different groundstates vanishes likewise and the change 
of baryon number can proceed in an unimpeded way and 
presumably faster than the universe expands. Thermal 
equilibrium is then maintained and any baryon 
asymmetry existing before this era is actually washed out 
\footnote{To be more precise, only $B+L$ is erased 
within the Standard Model whereas $B-L$ remains 
unchanged.}! 
Rather than generate a baryon number sphalerons act to 
drive the universe back to matter-antimatter 
symmetry at this point in its evolution. 

At energies below the phase transition, i.e. in the broken phase 
of $SU(2)_L \times U(1)$ baryon number is conserved for 
all practical purposes as pointed out above. 

The value of 
$\Delta n_{Bar}$ as observed today can thus be generated 
only in the transition from the unbroken high energy to the 
broken low energy phase. With $\Delta n_{Bar}\neq 0$ 
processes operating there the issue now turns to the 
strength of the phase transition: is it relatively smooth 
like a second order phase transition or violent like a 
first order one? Only the latter scenario can support 
baryogenesis. 

A large amount of interesting theoretical work has been  
on the thermodynamics of the Standard Model in an 
expanding universe. Employing perturbation theory 
and lattice studies one has arrived at the following result: 
for light Higgs masses up to around 70 GeV, the phase 
transition is first order, for larger masses it is second 
order 
\cite{SHAP}. Since no such light Higgs states have been observed 
at LEP, one infers that the phase transition is second order 
thus apparently foreclosing baryogenesis occurring at the 
electroweak scale. 

We have concentrated here on the questions of thermal 
equilibrium and baryon number while 
taking CP violation for 
granted since it is known to operate at the electroweak scale. 
Yet most authors -- with the exception of some notable 
heretics -- agree that the KM ansatz is not at all 
up to {\em this} task: it fails by several orders of 
magnitude. On the other hand New Physics 
scenarios of CP violation -- in particular 
of the Higgs variety -- can reasonably be called upon to perform 
the task. 

%%%%%%%%%%%%%%
\subsection{Leptogenesis Driving Baryogenesis}
%%%%%%%%%%%%%%
If the electroweak phase transition is indeed a 
second order one, 
sphaleron mediated reactions cannot drive baryogenesis 
as just discussed and they will wipe out any 
pre-existing $B+L$ number. Yet if at some high 
energy scales a lepton number is generated the very 
efficiency of these sphaleron processes can 
{\em communicate} this asymmetry to the baryon 
sector through them maintaining conservation of 
$B-L$. 

There are various ways in which such scenarios can 
be realized. The simplest one is to just add 
{\em heavy right-handed Majorana} neutrinos 
to the Standard Model. This is highly attractive 
in any case since it enables us 
to implement the see-saw mechanism for explaining 
why the observed neutrinos are (practically) massless; 
it is also easily embedded into $SO(10)$ GUTs. 

The basic idea is the following \cite{FUKUGITA}: 
\begin{itemize} 
\item  
A primordial lepton asymmetry is generated at high 
energies well above the electroweak phase transition: 
\begin{itemize} 
\item 
Since a Majorana neutrino $N$ is its own CPT mirror image, 
its dynamics necessarily violate lepton number. It will 
possess at least the following classes of decay channels: 
\beq 
N \ra l \bar H \; , \; \; \bar l H
\eeq 
with $l$ and $\bar l$ denoting a light charged or neutral lepton 
or anti-lepton and $H$ and $\bar H$ a Higgs or anti-Higgs field, 
respectively. 
\item 
A CP asymmetry will in general arise  
\beq 
\Gamma (N \ra l \bar H) \neq 
\Gamma (\bar N \ra \bar l H)
\eeq 
through a KM analogue in the neutrino mass 
matrix (which can be quite different from the 
mass matrix for charged leptons).  
\item 
These neutrino decays are sufficiently slow as to occur 
out of thermal equilibrium around the energy scale 
where the Majorana masses emerge. 
\end{itemize} 
\item 
The resulting lepton asymmetry is transferred into a 
baryon number through sphaleron mediated processes 
in the unbroken high energy phase of 
$SU(2)_L \times U(1)$: 
\beq 
\langle \Delta n_{lept}\rangle = 
\frac{1}{2}\langle \Delta n_{lept} + \Delta n_{Bar}\rangle + 
\frac{1}{2}\langle \Delta n_{lept} - \Delta n_{Bar}\rangle 
\Longrightarrow 
\frac{1}{2}\langle \Delta n_{lept} - \Delta n_{Bar}\rangle 
\eeq 
\item 
The baryon number thus generated survives through the 
subsequent evolution of the universe. 
\end{itemize}

%%%%%%%%%%%
\subsection{Wisdom -- Conventional and Otherwise}
%%%%%%%%%%
We understand how nuclei were formed in the universe 
given protons and neutrons. Obviously it would be 
even more fascinating if we could understand how 
these baryons were generated in the first place. 
We do not possess a specific and 
quantitative theory successfully describing baryogenesis. 
However leaving it at that statement would -- we believe -- 
miss the main point. We have learnt which kinds of 
dynamical ingredients are neccessary for baryogenesis to 
occur in the universe. We have seen that these ingredients 
can be realized naturally: 
\begin{itemize} 
\item 
GUT scenarios for baryogenesis 
provide us with a proof of principle that such a program 
can be realized. In practical terms however they suffer from 
various shortcomings: 
\begin{itemize} 
\item 
Since the baryon number is generated 
at the GUT scales, very little is and not much more might 
ever be known about that dynamics. 
\item 
It 
appears quite likely that a baryon number produced at such 
high scales is  subsequently washed out. 
\end{itemize} 
\item 
The highly fascinating proposal of baryogenesis at the electroweak 
phase transition has attracted a large degree of attention -- 
and deservedly so: 
\begin{itemize} 
\item 
A baryon number emerging from 
this phase transition would be in no danger of being 
diluted substantially. 
\item 
The dynamics involved here 
is known to a considerable degree and will be probed 
even more with ever increasing sensitivity over the 
coming years. 
\end{itemize} 
However it seems that the electroweak phase transition is 
of second order and thus not 
sufficiently violent. 
\item 
A very intriguing variant turns some of the vices of sphaleron 
dynamics into virtues by attempting to understand 
the baryon number of the universe as a reflection of a 
{\em primary} lepton asymmetry. 
The required new dynamical entities  
-- Majorana neutrinos and their decays -- obviously would 
impact on the universe in other ways as well.  
\end{itemize} 
The challenge to understand baryogenesis has already inspired 
our imagination, 
prompted the development of some very intriguing 
scenarios and thus has initiated many 
fruitful studies -- and in the end we might even be 
successful in meeting it!

%%%%%%%%%%%
\section{The Cathedral Builders' Paradigm}
%%%%%%%%%%%%
%%%%%%%%%%%%%
\subsection{The Paradigm}
%%%%%%%%%%%%

The dynamical ingredients for numerous and multi-layered 
manifestations of CP and T violations do exist or are likely to exist. Accordingly one searches 
for them in many phenomena, namely in  
\begin{itemize}
\item 
the neutron electric dipole moment probed with ultracold 
neutrons at ILL in Grenoble, France; 
\item 
the electric dipole moment of electrons studied through the 
dipole moment of atoms at Seattle, Berkeley and Amherst in the US; 
\item 
the transverse polarization of muons in 
$K^- \ra \mu ^- \bar \nu \pi ^0$ at KEK in Japan; 
\item 
$\epsilon ^{\prime}/\epsilon _K$ as obtained from $K_L$ 
decays at FNAL and CERN and soon at DA$\Phi$NE in Italy; 
\item 
in decay distributions of hyperons at FNAL; 
\item 
likewise for $\tau$ leptons at CERN, the beauty factories and BES 
in Beijing; 
\item 
CP violation in the decays of charm hadrons produced 
at FNAL and the beauty factories; 
\item 
CP asymmetries in beauty decays at DESY, at the beauty 
factories at Cornell, SLAC and KEK, at the FNAL collider and 
ultimately at the LHC. 

\end{itemize} 
A quick glance at this list already makes it clear 
that frontline research on this topic 
is pursued at all high energy labs in the world -- and then some; 
techniques from several different branches of physics -- 
atomic, nuclear and high energy physics -- are harnessed in 
this endeavour together with a wide range of set-ups. 
Lastly, experiments are performed at the lowest temperatures 
that can be realized on earth -- ultracold neutrons -- and at the 
highest -- in collisions produced at the LHC. And all of that dedicated 
to one profound goal. 
At this point I can explain what I mean by the term 
"Cathedral Builders' Paradigm". 
The building of cathedrals required interregional collaborations, 
front line technology (for the period) from many different fields 
and commitment; it had to be based on solid foundations -- and 
it took time. The analogy to the ways and needs of high energy 
physics are obvious -- but it goes deeper than that. 
At first sight a cathedral looks 
like a very complicated and confusing structure with something 
here and something there. Yet further scrutiny reveals that 
a cathedral is more appropriately 
characterized as a complex rather than a complicated 
structure, one that is multi-faceted and multi-layered -- 
with a coherent theme! One cannot (at least for 
first rate cathedrals) remove any of its elements 
without diluting (or even destroying) its technical soundness and 
intellectual message. Neither can one in our efforts to come to grips 
with CP violation!  

%%%%%%%%%%%%%
\subsection{Summary}
%%%%%%%%%%%%%
\begin{itemize} 
\item 
We know that CP symmetry is not exact in nature since 
$K_L \ra \pi \pi $ proceeds and presumably because we  
exist, i.e. because the baryon number of the universe does 
{\em not} vanish. 
\item 
If the KM mechanism is a significant actor in $K_L \ra \pi \pi$ 
transitions then there must be large CP asymmetries in the decays 
of beauty hadrons. In $B^0$ decays they 
are naturally measured in units of 10 \%!  
\item 
Some of these asymmetries are predicted with high parametric 
reliability. 
\item 
New theoretical technologies will allow us to translate such 
parametric reliability into quantitative accuracy. 
\item 
Any significant difference between certain KM predictions for the 
asymmetries and the data reveals the intervention of New Physics. 
There will be no `plausible deniability'.  
\item 
We can expect 10 years hence the theoretical uncertainties in 
some of the 
predictions to be reduced below 10 \% . 
\item 
I find it likely that deviations from the KM predictions will show up 
on that level. 
\item 
Yet to exploit this discovery potential to the fullest one will have to 
harness the statistical muscle provided by beauty production 
at hadronic colliders. 
\end{itemize}

%%%%%%%%%%%%
\subsection{Outlook}
%%%%%%%%%%%%
I want to start with a statement about the past: 
{\em The comprehensive study of kaon and hyperon physics 
has been instrumental in guiding us to the Standard Model.}  
\begin{itemize}
\item 
The $\tau -\theta $ puzzle led to the realization that parity is not 
conserved in nature. 
\item 
The observation that the production rate exceeded the decay rate 
by many orders of magnitude -- this was the origin of the 
name `strange particles' -- was explained through postulating 
a new quantum number -- `strangeness' -- conserved by the strong, 
though not the weak forces. This was the beginning of the second 
quark family. 
\item 
The absence of flavour-changing neutral currents was incorporated 
through the introduction of the quantum number `charm', which 
completed the second quark family. 
\item 
CP violation finally led to postulating yet another, the third 
family. 
\end{itemize}
All of these elements which are now essential pillars of the Standard 
Model were New Physics at {\em that} time! 

I take this historical 
precedent as clue that a detailed, comprehensive and thus 
neccessarily long-term program on beauty physics will lead to a 
new paradigm, a {\em new} Standard Model! 

CP violation is a fundamental as well as mysterious phenomenon 
that we have not understood yet. This is not surprising: after all 
according to the KM mechanism CP violation enters through the 
quark mass matrices; it thus relates it to three central 
mysteries of the Standard Model: 
\begin{itemize}
\item 
How are fermion masses generated? 
\footnote{Or more generally: how are masses produced in 
general? For in alternative models CP violation enters through 
the mass matrices for gauge bosons and/or Higgs bosons.}  
\item 
Why is there a family structure?
\item 
Why are there three families rather than one?
\end{itemize} 
In my judgement it would be unrealistic to expect that these 
questions can be answered through pure thinking. I strongly 
believe we have to appeal to nature through experimental efforts to 
provide us with more pieces that are surely missing in the 
puzzle. CP studies are essential in obtaining the full dynamical 
information contained in the mass matrices or -- in the language 
of v. Eichendorff's poem quoted in the beginning, "to find the 
magic word" that will decode nature's message for us. 

Considerable progress has been made in theoretical engineering 
and developing a comprehensive CP phenomenology from 
which I conclude: 
\begin{itemize}
\item 
$B$ decays constitute an almost ideal, certainly optimal and unique 
lab. Personally I believe that even if no deviation 
from the KM predictions were uncovered, we would find 
that the KM parameters, in particular the angles of the 
KM triangle, carry special values that would give us 
clues about New Physics. Some very interesting 
theoretical work is being done about how GUT dynamics in 
particular of the SUSY (or Supergravity) variety operating 
at very high scales would shape the observable 
KM parameters. 
\item 
A comprehensive analysis of charm decays with special emphasis on  
$D^0 - \bar D^0$ oscillations and CP violation is a moral 
imperative! Likewise for $\tau$ leptons. 
\item 
A vigorous research program must be pursued for light 
fermion systems, namely in the decays of kaons and hyperons 
and in electric dipole moments. After all it is conceivable 
of course that no CP asymmetries are found 
in $B$ decays on a measurable level. 
Then we would know that the KM ansatz is {\em not} a 
significant actor in $K_L \ra \pi \pi$, that New Physics drives it -- 
but what kind of New Physics would it be? Furthermore even if 
large CP asymmetries were found in $B$ decays, it could 
happen that the signals of New Physics are obscured by 
the large `KM background'. This would not be the case 
if electric dipole moments were found or a transverse 
polarization of muons in $K_{\mu 3}$ decays. 
\item Close feedback between experiment and theory will be 
essential. 

\end{itemize}
 As the final summary: insights about Nature's 
Grand Design that can be obtained from a 
comprehensive and detailed program of CP studies 
\begin{itemize}
\item 
are of fundamental importance, 
\item 
cannot be obtained any other way and 
\item 
cannot become obsolete!

\end{itemize}

\bigskip 

{\bf Acknowledgments:} \hspace{.4em} 
This work was supported by the National Science Foundation under
the grant numbers PHY 92-13313 and PHY 96-05080.


\begin{thebibliography}{99}
\bibitem{WINSTEIN} 
I have learnt this notation -- using $\Phi (\Delta S=1,2)$ rather 
than $\epsilon _K$ and $\epsilon ^{\prime}$ -- from B. Winstein.
\end{thebibliography}
\end{document}